\documentclass[preprint,12pt]{elsarticle}
\graphicspath{{./}}

\usepackage[dvipsnames]{xcolor}

\usepackage{amssymb}
\usepackage{subfigure}
\usepackage{amsmath}
\usepackage{color}
\usepackage{hyperref}

\usepackage{amsthm}
\usepackage{enumitem}
\usepackage{pgfplotstable}
\usepackage{pgfplots}
\pgfplotsset{compat=1.14}
\pgfplotsset{
	colormap={my basis colormap}{
		rgb255=(0, 114, 189);
		rgb255=(54, 106, 148);
		rgb255=(108, 98, 107);
		rgb255=(163, 91, 66);
		rgb255=(217, 83, 25);
	}
}
\pgfplotsset{
	colormap={my parula}{
		rgb255=(53.0655, 42.406, 134.9460);
		rgb255=(20.2336, 132.6061, 211.9511);
		rgb255=(55.5463, 184.8859, 157.9118);
		rgb255=(208.7187, 186.8470,  89.1966);
		rgb255=(248.9565, 250.6905, 13.7190);
	}
}
\usepackage{tikz}
\usepgfmodule{oo}
\usetikzlibrary{calc}
\usetikzlibrary{shapes}
\usetikzlibrary{plotmarks}
\usetikzlibrary{backgrounds}
\usetikzlibrary{decorations.pathmorphing}
\usetikzlibrary{external}
\usetikzlibrary{arrows,fit,matrix,positioning,shapes.geometric}
\usepackage{siunitx}

\usepackage{titlesec}

\titleformat{\paragraph}
{\normalfont\normalsize\it}{\theparagraph}{1em}{}
\titlespacing*{\paragraph}
{0pt}{3.25ex plus 1ex minus .2ex}{1.5ex plus .2ex}

\def\mathbf#1{\boldsymbol{#1}}

\def\vec#1{\boldsymbol{#1}}

\theoremstyle{plain}

\theoremstyle{remark}

\theoremstyle{definition}

\definecolor{myBlue}{rgb} {0,0.4470,0.7410}
\definecolor{myRed}{rgb} {0.8500,0.3250,0.0980}

\usepackage{marginnote}

     \allowdisplaybreaks

\begin{document}

\begin{frontmatter}

% \title{Fully nonlinear thin-shell isogeometric analysis using ASTS of arbitrary degree and handling trimmed NURBS surfaces}

% \title{Isogeometric analysis of fully nonlinear thin shells: ASTS of arbitrary degree and trimmed NURBS surfaces}

 % \title{Extending the subset of analysis-suitable T-splines: Various extraordinary points per face}
 
  \title{Removing membrane locking in quadratic NURBS-based discretizations of linear plane Kirchhoff rods: CAS elements}

% \title{Seamless integration of design and analysis of Kirchhoff-Love shells using analysis-suitable unstructured T-splines}  ASUTS

\author[label1]{Hugo Casquero\corref{cor1}}
\ead{casquero@umich.edu}
\author[label1]{Mahmoud Golestanian}
\address[label1]{Department of Mechanical Engineering, University of Michigan – Dearborn, 4901 Evergreen Road, Dearborn, MI 48128-1491, U.S.A.}

\cortext[cor1]{Corresponding author.}

% \cite{Scott2013,toshniwal2017smooth,casquero2020seamless}

\begin{abstract}

NURBS-based discretizations of the Galerkin method suffer from membrane locking when applied to primal formulations of curved thin-walled structures. We consider linear plane curved Kirchhoff rods as a model problem to study how to remove membrane locking from NURBS-based discretizations. In this work, we propose continuous-assumed-strain (CAS) elements, an assumed strain treatment that removes membrane locking from quadratic NURBS for an ample range of slenderness ratios. CAS elements take advantage of the $C^1$ inter-element continuity of the displacement vector given by quadratic NURBS to interpolate the membrane strain using linear Lagrange polynomials while preserving the $C^0$ inter-element continuity of the membrane strain. To the authors' knowledge, CAS elements are the first NURBS-based element type able to remove membrane locking for a broad range of slenderness ratios that combines the following distinctive characteristics: (1) No additional degrees of freedom are added, (2) No additional systems of algebraic equations need to be solved, and (3) The nonzero pattern of the stiffness matrix is preserved. Since the only additional computations required by the proposed element type are to evaluate the derivatives of the basis functions and the unit tangent vector at the knots, the proposed scheme barely increases the computational cost with respect to the locking-prone NURBS-based discretization of the primal formulation. The benchmark problems show that the convergence of CAS elements is independent of the slenderness ratio up to $10^4$ while the convergence of quadratic NURBS elements with full and reduced integration, local $\bar{B}$ elements, and local ANS elements depends heavily on the slenderness ratio and the error can even increase as the mesh is refined. The numerical examples also show how CAS elements remove the spurious oscillations in stress resultants caused by membrane locking while quadratic NURBS elements with full and reduced integration, local $\bar{B}$ elements, and local ANS elements suffer from large-amplitude spurious oscillations in stress resultants. In short, CAS elements are an accurate, robust, and computationally efficient numerical scheme to overcome membrane locking in quadratic NURBS-based discretizations.

% To the authors' knowledge, CAS elements are the first NURBS-based element type that combines the following distinctive characteristics: (1) Effective removal of membrane locking for the range of slenderness values found in engineering applications, (2) No additional degrees of freedom are added with respect to the standard NURBS-based discretization of the primal formulation, (3) No additional systems of algebraic equations need to be solved (neither at the global level nor at the element level), and (4) The nonzero pattern of the stiffness matrix obtained using the standard NURBS-based discretization of the primal formulation is preserved. Thus, for a given mesh, the proposed locking-free scheme is essentially as computationally efficient as the locking-prone NURBS-based discretization of the primal formulation.

% To the authors' knowledge, CAS elements are the first NURBS-based element type able to remove membrane locking  that combines the following distinctive characteristics: (1) No additional degrees of freedom are added with respect to the standard NURBS-based discretization of the primal formulation, (2) No additional systems of algebraic equations need to be solved (neither at the global level nor at the element level), and (3) The nonzero pattern of the stiffness matrix obtained using the standard NURBS-based discretization of the primal formulation is preserved. Thus, for a given mesh, the proposed scheme is essentially as computationally efficient as the locking-prone NURBS-based discretization of the primal formulation.

\end{abstract}

\begin{keyword}

Isogeometric analysis  \sep Thin-walled structures \sep  Membrane locking   \sep Assumed natural strain \sep Kirchhoff rods \sep Convergence studies  

% \sep Membrane locking in Kirchhoff-Love shells

\end{keyword}

\end{frontmatter}

%\linenumbers

\renewcommand{\thefootnote}{\fnsymbol{footnote}}

\section{Introduction}

Isogeometric analysis (IGA) \cite{1003.000, cottrell2009isogeometric} enables a seamless integration between computer-aided design (CAD) and finite element analysis (FEA) of thin-walled structures \cite{casquero2020seamless, wei2022analysis, toshniwal2017smooth, nagy2015numerical, leidinger2019explicit}. Furthermore, the higher inter-element continuity of splines enables the application of the Galerkin method to discretize the primal formulation of higher order displacements theories based on Kirchhoff assumptions. Spline discretizations of structural theories based on Kirchhoff assumptions have been developed for rods \cite{greco2013b, greco2014implicit, greco2016isogeometric} and shells \cite{Kiendl2009, Kiendl2015, casquero2017arbitrary}. These theories neglect transverse shear deformation, which is considered to be negligible as long as $R/t \geq 20$ \cite{bischoff2004models}, where $R$ is the radius of curvature, $t$ is the thickness, and $R/t$ is the slenderness ratio. Most structures used in engineering are slender enough to satisfy the above inequality. In addition to resulting in fewer degrees of freedom than the theories that take into account transverse shear deformation, Kirchhoff theories avoid shear locking. However, structural theories based on Kirchhoff assumptions with coupled membrane and bending responses still suffer from membrane locking \cite{greco2017efficient, armero2012invariant, meier2015locking, bieber2018variational, nguyen2022leveraging, greco2018reconstructed} as it is also the case for theories that take into account transverse shear deformation \cite{stolarski1983shear, bouclier2012locking, bouclier2015isogeometric, oesterle2016shear, zou2021galerkin, zou2022efficient}.

% leonetti2018efficient, leonetti2019simplified, hallquist1985implementation

In commercial FEA software \cite{lsdyna,abaqus}, the schemes that are more frequently used to treat shear locking and membrane locking when using Lagrange polynomials as basis functions are reduced \cite{zienkiewicz1971reduced, flanagan1981uniform, belytschko1984hourglass, belytschko1984explicit} and selective-reduced \cite{hughes1977simple, hughes1978reduced,  hughes1981nonlinear} integration rules and assumed natural strains (ANS) \cite{macneal1978simple, hughes1981finite, macneal1982derivation, dvorkin1984continuum}. Reduced/selective-reduced integration rules and ANS are equivalent to mixed formulations under certain conditions \cite{malkus1978mixed, simo1986variational}. Directly applying these schemes to NURBS basis functions, that is, using reduced/selective-reduced integration rules at the element level or using ANS treatments that result in discontinuous assumed strains across elements is not an effective strategy to overcome locking \cite{greco2017efficient, greco2018reconstructed, bouclier2013efficient, kim2022isogeometric}. Therefore, the development of locking treatments that take into account the higher inter-element continuity of NURBS is needed. Reduced/selective-reduced integration rules at the patch level were developed in \cite{hughes2010efficient, adam2015selective, adam2014improved, adam2015improved}. Reduced/selective-reduced integration rules at the patch level were used to alleviate locking in solid shells \cite{leonetti2018efficient} and Kirchhoff-Love shells \cite{leonetti2019simplified}. Global versions of the $\bar{B}$ method \cite{simo1986variational} were proposed for nearly incompressible solids \cite{Elguedj2008}, Timoshenko rods \cite{bouclier2012locking, zhang2018locking}, Kirchhoff rods \cite{greco2017efficient}, Kirchhoff-Love shells \cite{greco2018reconstructed}, and solid shells \cite{bouclier2013efficient}. Global versions of the discrete strain gap (DSG) method \cite{bletzinger2000unified, koschnick2005discrete} were proposed for Timoshenko rods \cite{echter2010numerical, bouclier2012locking} and Kirchhoff-Love, Reissner-Mindlin, and 7-parameter shells \cite{echter2013hierarchic}. The global versions of the $\bar{B}$ method and the DSG method avoid having discontinuous assumed strains across elements and remove locking effectively for NURBS basis functions. However, as acknowledged by the authors in \cite{greco2017efficient, greco2018reconstructed, bouclier2012locking,  bouclier2013efficient}, these solutions are not computationally efficient since (a) a global mass matrix needs to be inverted and (b) the resulting global stiffness matrix is not a spare matrix anymore, but a completely full matrix instead. Because of this, avoiding condensation of the strain variables and solving the full mixed problem directly is suggested in \cite{echter2010numerical}, but this heavily increases the size of the system of algebraic equations that needs to be solved. Another alternative is to reconstruct assumed strains at the global NURBS patch level from local $\bar{B}$ projections at the element level \cite{bouclier2013efficient, miao2018bezier, greco2017efficient, greco2018reconstructed}. Least-square-type procedures \cite{govindjee2012convergence, mitchell2011method, cardoso2014blending}, B\'ezier projection \cite{thomas2015bezier}, and $L^2$ projection  are used for the local projection at the element level in \cite{bouclier2013efficient}, \cite{miao2018bezier}, and \cite{greco2017efficient, greco2018reconstructed}, respectively. In \cite{bouclier2013efficient, miao2018bezier, greco2017efficient, greco2018reconstructed}, solving additional systems of algebraic equations at the local level is needed. In addition, even though the global stiffness matrix is no longer completely full, its bandwidth is significantly larger than the bandwidth of the global stiffness matrix obtained by applying the locking-prone NURBS-based discretization of the primal formulation. A special mention is deserved for the locking treatment proposed in \cite{zou2020isogeometric}. In \cite{zou2020isogeometric}, the starting point is a mixed formulation with independent displacements and strains as unknowns. The strain unknowns are condensed out at the element level by leveraging B\'ezier dual basis functions \cite{zou2018isogeometric, miao2020isogeometric}. However, the resulting bandwidth of the stiffness matrix increases with respect to the standard NURBS-based discretization of the primal formulation (namely, the number of nonzero entries increases by a factor of three when solving Reissner-Mindlin shell problems with this locking treatment).

% \cite{cardoso2012enhanced}

% bartovn2016gaussian, bartovn2017gauss

% hiemstra2017optimal, johannessen2017optimal

In this work, we develop an ANS treatment that successfully overcomes the membrane locking existent in quadratic NURBS-based discretizations of linear plane curved Kirchhoff rods while being almost as computationally efficient as the locking-prone NURBS-based discretization of the primal formulation since

\begin{itemize}
\item no additional degrees of freedom are added with respect to the standard NURBS-based discretization of the primal formulation,
\item no additional systems of algebraic equations need to be solved (neither at the global level nor at the element level), and
\item the nonzero pattern of the stiffness matrix obtained using the standard NURBS-based discretization of the primal formulation is preserved.
\end{itemize}

The proposed ANS treatment leverages the $C^1$ inter-element continuity of the displacement vector given by quadratic NURBS to preserve the $C^0$ continuity of the compatible strains by directly interpolating at the element level the compatible strains at the knots using linear Lagrange polynomials. Heretofore, the proposed element type to treat locking is referred to as continuous-assumed-strain (CAS) elements. Membrane locking causes not only smaller displacements and bending moments than expected, but also large-amplitude spurious oscillations of membrane forces. Thus, we study the convergence and plot the distributions of both displacements and stress resultants to show that CAS elements eliminate the spurious oscillations of membrane forces.

The paper is outlined as follows. Section 2 sets forth the mathematical theory of linear plane curved Kirchhoff rods. Section 3 describes how to solve the problem using a NURBS-based discretization of the Galerkin method. Section 4 develops CAS elements, the new element type proposed in this work to remove membrane locking while barely increasing the computational cost for a given mesh in comparison with the locking-prone NURBS-based discretization of the Galerkin method. The performance evaluation of CAS elements and comparisons with the global $\bar{B}$ method, quadratic NURBS elements with full and reduced integration, local $\bar{B}$ elements, and local ANS elements are included in Section 5. Sections 5.1, 5.2, and 5.3 consider a pinched circular ring, a clamped-clamped semi-circular arch under a distributed load, and a clamped elliptical arch under a point load at the free end, respectively. Concluding remarks and directions of future work are drawn in Section 6.

\section{Linear plane curved Kirchhoff rod model}

In this section, we consider Kirchhoff rods with infinitesimal deformations and small strains, that is, we do not consider either geometric nonlinearities or material nonlinearities. The geometry of the rod is defined by its axis and its cross section. We state the Kirchhoff rod formulation using the Lagrangian description and a curvilinear coordinate. For a full mathematical derivation of the model the reader is referred to \cite{kirchhoff1859ueber, clebsch1862theorie, winkler1867lehre, love1927mathematical, timoshenko1983history, armero2012invariant}.

\subsection{Kinematics in infinitesimal deformations}

The geometry of the axis is defined by the parametric curve $\vec r (\xi) : [0,1] \mapsto \mathbb{R}^2$, where $\xi$ is a parametric coordinate and $\mathbf{r} (\xi)$ is the position vector of a material point on the axis of the rod. The displacement vector of a material point in the axis of the rod is defined as $\vec u (\xi) : [0,1] \mapsto \mathbb{R}^2$. Both $\mathbf{r} (\xi)$ and $\vec u  (\xi)$ are defined using a global system of Cartesian coordinates. We reparametrize the axis of the rod in terms of its arc length $s$, which is done taking into account that
\begin{equation}
ds =  \Biggr\vert \Biggr\vert \frac{{\rm d} \vec{r}}{{\rm d} \xi} \Biggr\vert \Biggr\vert d \xi \text{,}
\end{equation}
\noindent where $||\cdot||$ denotes the length of a vector. Using the arc length as the parametric coordinate, the unit tangent vector to the axis of the rod is obtained by
\begin{equation}
\vec{a}_1 =  \frac{{\rm d} \vec{r}}{{\rm d} s} \text{.}
\end{equation}
The unit normal vector to the axis of the rod is obtained by
\begin{equation}
\vec{a}_2 =  \displaystyle{
	\begin{pmatrix}
		 0 & -1 \\
		 1 &  0 \\
	\end{pmatrix} } \vec{a}_1 \text{.}
\end{equation}
$\vec{a}_1$ and $\vec{a}_2$ form a counterclockwise pair. The membrane strain\footnote{To be precise, $\epsilon$ is the axial strain of the rod. Nevertheless, since $\epsilon$ has an analogous mathematical expression to the membrane strains of a linear Kirchhoff-Love shell formulation and since Kirchhoff rods are used in this work as a model problem to study membrane locking, $\epsilon$ is referred to as membrane strain throughout this manuscript.} is defined as
\begin{equation}
\epsilon =  \vec{a}_1 \cdot \frac{{\rm d} \vec{u}}{{\rm d} s} \text{.}
\end{equation}
The bending strain is defined as
\begin{equation}
\kappa =  \vec{a}_2 \cdot \frac{{\rm d^2} \vec{u}}{{\rm d} s^2}  +  \frac{{\rm d} \vec{a}_2}{{\rm d} s} \cdot \frac{{\rm d} \vec{u}}{{\rm d} s} \text{.} \label{bendingstrain}
\end{equation}

\subsection{Linear material}

\begin{figure} [t!] 
\centering
\includegraphics[width=4cm]{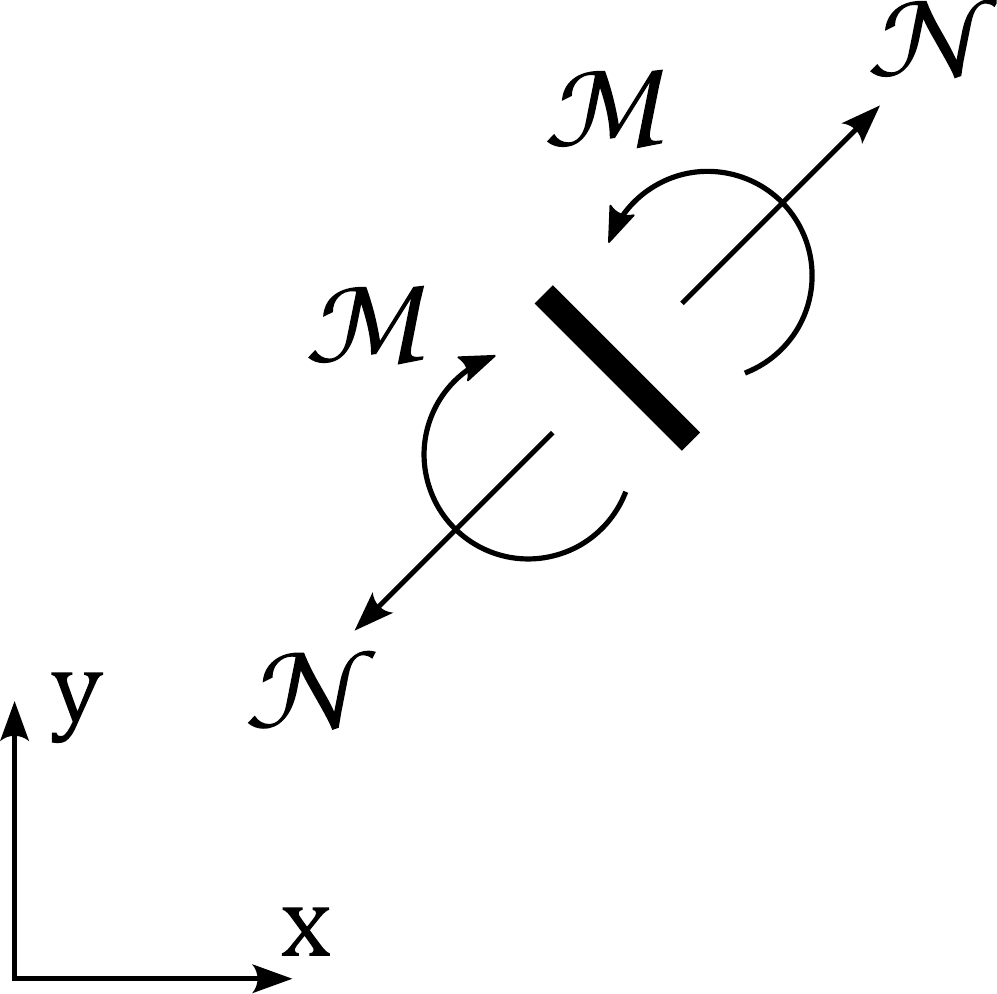}
\caption{Positive signs for the membrane force and the bending moment.} 
\label{signconvention}
\end{figure}

The membrane force and the bending moment are the stress resultants of plane curved Kirchhoff rods that are obtained from constitutive equations. As in \cite{armero2012invariant, greco2017efficient}, we use the Kirchhoff-Clebsch theory for linear elastic materials \cite{kirchhoff1859ueber, clebsch1862theorie, love1927mathematical}. For this material theory, the membrane force is defined as
\begin{equation}
\mathcal{N} =  EA\epsilon \text{.}
\end{equation}
where $E$ is the Young modulus of the material and $A$ is the area of the cross section. The bending moment is defined as
\begin{equation}
\mathcal{M} =  EI\kappa \text{.}
\end{equation}
where $I$ is the cross section's moment of inertia. The positive signs for the membrane force and the bending moment are indicated in Fig. \ref{signconvention}.

\subsection{Variational form}

The variational form can be obtained from the principle of virtual work which states that the internal virtual work ($\delta W^{int}$) must be equal to the external virtual work ($\delta W^{ext}$) for any virtual displacement ($\delta \mathbf{u}$), i.e.,
\begin{equation} 
 \delta W^{int} =  \delta W^{ext} \quad \forall \delta \mathbf{u} \text{,}  \label{virtualwork}
\end{equation}
with
\begin{align}
   \delta W^{int}&=  \int^L_{0}\left(\mathcal{N} \delta\epsilon + \mathcal{M} \delta\kappa \right) \, \mathrm ds  \text{,}\label{Wint} \\
   \delta W^{ext}&=  \int^L_{0} \vec{f} \cdot \delta \vec{u}  \, \mathrm ds + \mathbf{P}_0  \cdot   \delta \vec{u} (s=0)   + \mathbf{P}_L  \cdot   \delta \vec{u} (s=L) \text{,} \label{Wext}
\end{align}
where $L$ is the length of the rod axis, $\delta \epsilon$ and $\delta \kappa$ are the virtual membrane strain and the virtual bending strain, respectively, $\mathbf{f}$ is a distributed load acting along the rod axis, $\mathbf{P}_0$ and $\mathbf{P}_L$ are point loads acting on the ends of the rod axis.

\section{NURBS elements}

NURBS basis functions are built from a knot vector. A knot vector is a non-decreasing set of coordinates in the parametric space, written $\bold{\Xi} = \{  {\xi}_{1} ,  {\xi}_{2}, ...,  {\xi}_{n + p + 1} \}$, where ${\xi}_{i}$ is the $i$-th knot, $p$ is the polynomial degree, and $n$ is the number of NURBS basis functions. Knot values may be repeated. The continuity of the NURBS basis functions at a knot is $C^{p-m}$, where $m$ is the multiplicity of the knot. A knot vector is said to be open if its first and last knot values are repeated $p+1$ times. A knot span ${\Delta\xi}_{i}$ is the difference between two consecutive knots, i.e., ${\Delta\xi}_{i}= {\xi}_{i+1} - {\xi}_{i}$. Nonzero knot spans play the role of elements, i.e., nonzero knot spans are the regions where numerical integration is performed. 

Once the knot vector is defined, the B-spline basis functions are defined recursively starting with piecewise constants ($p=0$)
\begin{align}
M_{B,0} (\xi) & = \begin{cases}
1 \quad  \text{if} \: \: \xi_B \leq \xi < \xi_{B+1} \text{,} \\
0 \quad  \text{otherwise}  \text{.} \end{cases} 
\end{align} 
For $p = 1, 2, 3, ...$, the B-spline basis functions are defined by
\begin{equation} 
 M_{B,p} (\xi) =  \frac{\xi - \xi_B}{\xi_{B+p} - \xi_B}  M_{B,p-1} (\xi)   +   \frac{\xi_{B+p+1} - \xi}{\xi_{B+p+1} - \xi_{B+1}}  M_{B+1,p-1} (\xi) \text{,} 
\end{equation}
This is referred to as the Cox–de Boor recursion formula \cite{piegl2012nurbs, cottrell2009isogeometric}. For evaluating this formula, whenever 0/0 is obtained, 0/0 is supposed to be substituted with 0. NURBS basis functions are defined as follows
\begin{equation} 
 N_B (\xi) =  \frac{w_B M_{B,p} (\xi)}{\sum_{C=1}^{n} w_C M_{C,p} (\xi)} \text{,} 
\end{equation}
where $w_B$ are the weights. The weights are introduced to represent exactly conic curves. For further information about the properties of NURBS basis functions and how to perform $h$-refinement using the knot insertion algorithm, the reader is referred to \cite{cottrell2009isogeometric}. In this work, we use open knot vectors with no repeated interior knots and basis functions of degree $p=2$.

\begin{figure} [t!] 
\centering
\includegraphics[width=8cm]{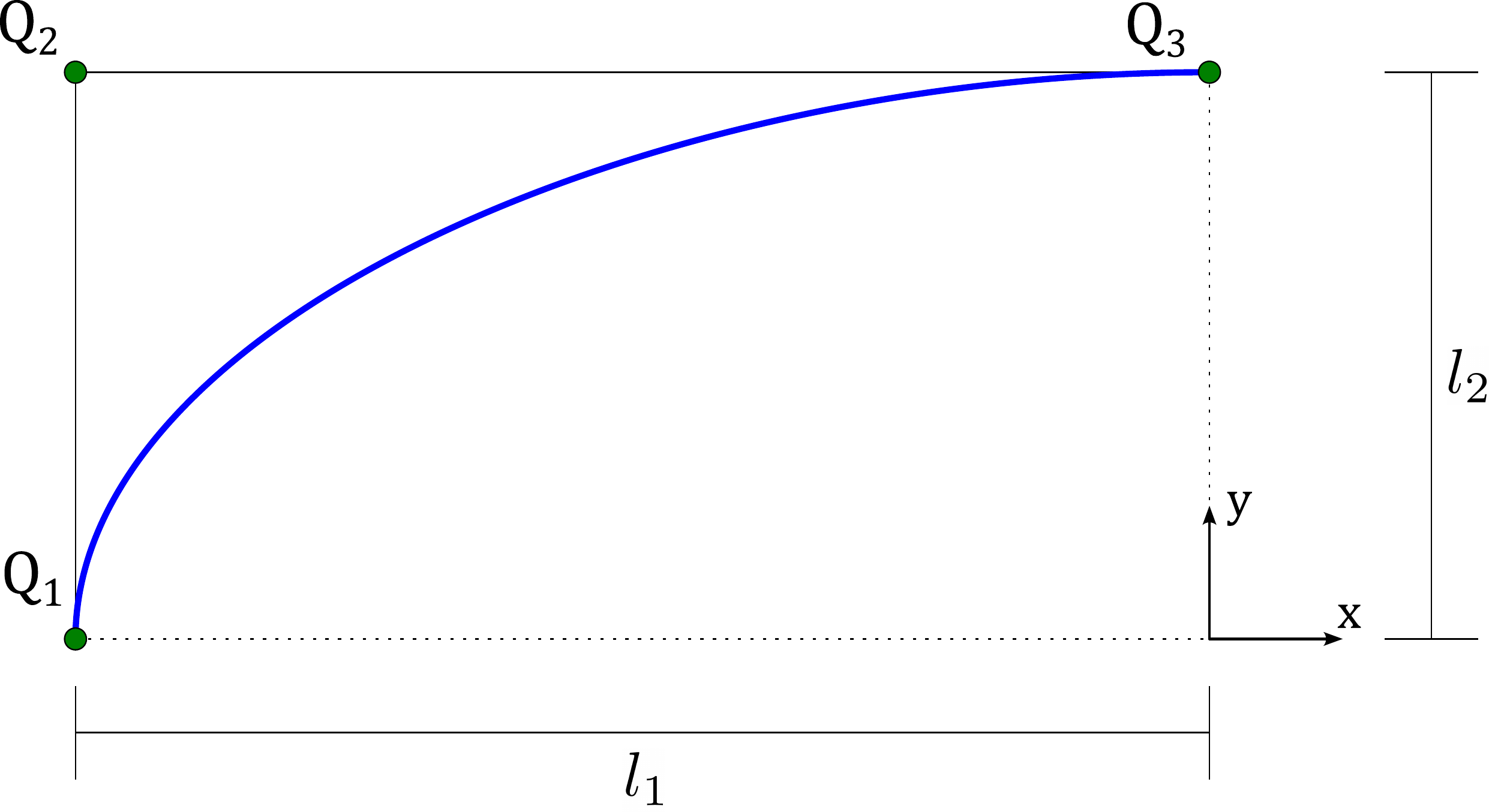}
\caption{A quarter of an ellipse exactly represented using one quadratic NURBS element. The values of the control points are $\vec{Q}_1 = (-l_1,0)$, $\vec{Q}_2 = (-l_1,l_2)$, and $\vec{Q}_3 = (0,l_2)$. The values of the weights are $w_1 = 1$, $w_2 = \sqrt{2}/2$, and $w_3 = 1$.} 
\label{ellipse}
\end{figure}

The axis of the rod is constructed by taking a linear combination of the NURBS basis functions. Thus,
\begin{equation}
\vec{r} (\xi) = \sum_{B=1}^{n} N_B (\xi) \vec{Q}_B  \text{,}
\end{equation}
where $\vec{Q}_B$ is the $B$-th control point. Fig. \ref{ellipse} shows the values of the control points and weights needed to exactly represent a quarter of an ellipse using only one quadratic NURBS element. Invoking the isoparametric concept, the displacement vector is discretized as follows
\begin{equation}
\vec{u}^h (\xi) = \sum_{B=1}^{n} N_B (\xi) \vec{U}_B  \text{,}
\end{equation}
where $\vec{U}_B$ is the $B$-th control variable of the displacement vector. In order to discretize the virtual displacements, the Bubnov-Galerkin method is used, i.e., $\delta \mathbf{u}^h (\xi) \in \text{span} \{ N_{B}(\xi)\}_{B=1}^{n}$.

As a result of the discretization explained above, the element stiffness matrix using NURBS elements is obtained as follows
\begin{equation} 
 \mathbf{k} =  \mathbf{k}_{\epsilon} + \mathbf{k}_{\kappa} \text{,} 
\end{equation}
\begin{equation} 
 \mathbf{k}_{\epsilon} =  \left[ k^{ij}_{\epsilon,bc} \right], \quad  \mathbf{k}_{\kappa} =  \left[ k^{ij}_{\kappa,bc} \right] \text{,} 
\end{equation}
\begin{equation} 
 k^{ij}_{\epsilon,bc} =  \int^{s_2^e}_{s_1^e} \left( \vec{a}_1 \cdot \frac{dN_b}{ds} \vec{e}_i \right) EA  \left( \vec{a}_1 \cdot \frac{dN_c}{ds} \vec{e}_j \right)  \, \mathrm ds   \text{,} 
\end{equation}
\begin{equation} 
 k^{ij}_{\kappa,bc} =  \int^{s_2^e}_{s_1^e} \left( \vec{a}_2 \cdot \frac{d^2N_b}{ds^2} \vec{e}_i + \frac{d \vec{a}_2}{ds} \cdot \frac{dN_b}{ds} \vec{e}_i \right) EI  \left( \vec{a}_2 \cdot \frac{d^2N_c}{ds^2} \vec{e}_j + \frac{d \vec{a}_2}{ds} \cdot \frac{dN_c}{ds} \vec{e}_j \right)  \, \mathrm ds   \text{,} 
\end{equation}
where $s_1^e$ and $s_2^e$ are the arc-length coordinates in which element $e$ starts and ends, respectively, $\vec{e}_l$ is the $l$-th versor of the global Cartesian coordinate system, $ \mathbf{k}$ is the element stiffness matrix, $\mathbf{k}_{\epsilon}$ is the element membrane stiffness matrix, and $\mathbf{k}_{\kappa}$ is the element bending stiffness matrix. Following standard FEA paraphernalia, the integrals above are computed performing change of variables twice. First, from the arc length coordinate $s$ to the parametric coordinate $\xi$ and then from the parametric coordinate $\xi$ to the parent element with coordinate $\widehat{\xi} \in [-1,1]$. The assembly of the element stiffness matrices into the global stiffness matrix is performed using conventional connectivity arrays \cite{Hughes2012, cottrell2009isogeometric}.

%%
%\begin{equation} 
% K^{ij}_{m,AB} =  \int^L_{0} \left( \vec{a}_1 \cdot \frac{dN_A}{ds} \vec{e}_i \right) EA  \left( \vec{a}_1 \cdot \frac{dN_B}{ds} \vec{e}_j \right)  \, \mathrm ds   \text{,} 
%\end{equation}
%%
%\begin{equation} 
% K^{ij}_{b,AB} =  \int^L_{0} \left( \vec{a}_2 \cdot \frac{d^2N_A}{ds^2} \vec{e}_i + \frac{d \vec{a}_2}{ds} \cdot \frac{dN_A}{ds} \vec{e}_i \right) EI  \left( \vec{a}_2 \cdot \frac{d^2N_B}{ds^2} \vec{e}_j + \frac{d \vec{a}_2}{ds} \cdot \frac{dN_B}{ds} \vec{e}_j \right)  \, \mathrm ds   \text{,} 
%\end{equation}
%%
%\begin{equation} 
% \mathbf{U} =  \left[ U^{i}_{A} \right], \quad  \mathbf{F} =  \left[ F^{i}_{A} \right] \text{,} 
%\end{equation}
%%
%\begin{equation} 
% F^{i}_{A} =  \int^L_{0} \vec{f} \cdot N_A \vec{e}_i  \, \mathrm ds + \mathbf{P_0}  \cdot   N_A (0) \vec{e}_i    + \mathbf{P_L}  \cdot   N_A (L) \vec{e}_i \text{,} 
%\end{equation}
%%

\section{CAS elements}

The membrane strain of a quadratic NURBS element has the following expression
\begin{equation}
\epsilon^h (s) = \vec{a}_1 (s) \cdot \frac{{\rm d} \vec{u}^h}{{\rm d} s} (s) \text{.}
\end{equation}
Taking advantage of the $C^1$ inter-element continuity of the geometry and the displacement vector given by quadratic NURBS, CAS elements interpolate the membrane strain at the knots using linear Lagrange polynomials resulting in a piecewise  linear mathematical expression for the membrane strain with $C^0$ inter-element continuity. Thus, the membrane strain of a CAS element is defined as follows
\begin{equation} \label{ans}
\epsilon^{\text{CAS},h} (s) =  L_1 (s) \epsilon^h (s^e_1) + L_2 (s) \epsilon^h (s^e_2) \text{,}
\end{equation}
with
\begin{equation}
L_1 (s) =  \frac{s^e_2 - s}{s^e_2 - s^e_1} \text{,}
\end{equation}
\begin{equation}
L_2 (s) =  \frac{s - s^e_1}{s^e_2 - s^e_1}  \text{,}
\end{equation}
where $s_1^e$ and $s_2^e$ are the arc-length coordinates in which element $e$ starts and ends, respectively, $L_1$ and $L_2$ are linear Lagrange polynomials.

Using the assumed natural strain proposed in Eq. \eqref{ans}, the element stiffness matrix of CAS elements is obtained as follows
\begin{equation} 
 \mathbf{k}^{\text{CAS}} =  \mathbf{k}^{\text{CAS}}_{\epsilon} + \mathbf{k}_{\kappa} \text{,} 
\end{equation}
\begin{equation} 
 \mathbf{k}^{\text{CAS}}_{\epsilon} =  \left[ k^{\text{CAS}, ij}_{\epsilon,bc} \right], \quad  \mathbf{k}_{\kappa} =  \left[ k^{ij}_{\kappa,bc} \right] \text{,} 
\end{equation}
\begin{equation} 
 k^{\text{CAS}, ij}_{\epsilon,bc} = \sum_{l=1}^{2} \sum_{m=1}^{2}  \int^{s_2^e}_{s_1^e} L_l (s) \left( \vec{a}_1 (s_l^e) \cdot \frac{dN_b}{ds} (s_l^e) \vec{e}_i \right) EA  L_m (s) \left( \vec{a}_1 (s_m^e) \cdot \frac{dN_c}{ds} (s_m^e) \vec{e}_j \right)  \, \mathrm ds   \text{,} 
\end{equation}
\begin{equation} 
 k^{ij}_{\kappa,bc} =  \int^{s_2^e}_{s_1^e} \left( \vec{a}_2 \cdot \frac{d^2N_b}{ds^2} \vec{e}_i + \frac{d \vec{a}_2}{ds} \cdot \frac{dN_b}{ds} \vec{e}_i \right) EI  \left( \vec{a}_2 \cdot \frac{d^2N_c}{ds^2} \vec{e}_j + \frac{d \vec{a}_2}{ds} \cdot \frac{dN_c}{ds} \vec{e}_j \right)  \, \mathrm ds   \text{,} 
\end{equation}
where $ \mathbf{k}^{\text{CAS}}$ is the element stiffness matrix of CAS elements and $\mathbf{k}^{\text{CAS}}_{\epsilon}$ is the element membrane stiffness matrix of CAS elements. As in Section 3, the integrals above are computed performing change of variables twice ($s \rightarrow \xi \rightarrow \widehat{\xi} \,$). In the parent element, $L_k(\widehat{\xi} \, ) = (1+ (-1)^k \widehat{\xi} \,)/2$. The assembly of the element stiffness matrices into the global stiffness matrix is performed using conventional connectivity arrays \cite{Hughes2012, cottrell2009isogeometric}.

When computing stress resultants using CAS elements, the membrane force is obtained as
\begin{equation}
\mathcal{N}^{\text{CAS},h} =  EA\epsilon^{\text{CAS},h} \text{.}
\end{equation}
The numerical experiments included in the next section will show that the assumed strain treatment proposed in this section removes the spurious oscillations of the membrane force and results in a numerical scheme whose accuracy is independent of the slenderness ratio for a wide range of values.

\section{Numerical experiments}

	\begin{figure} [t!] 
\centering
\subfigure[Pinched circular ring]{\includegraphics[scale=0.3]{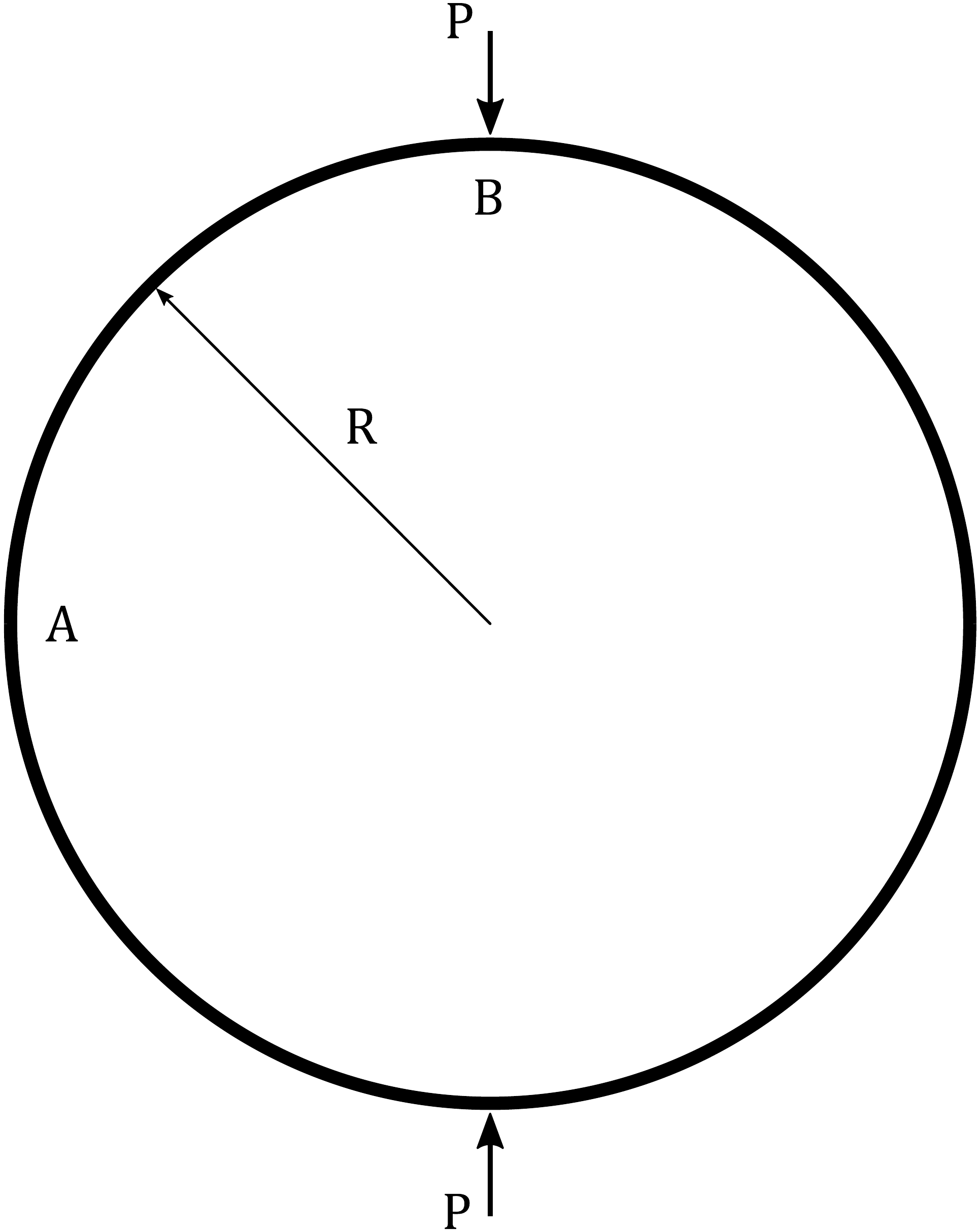}} \hspace*{+20mm}
 \subfigure[A quarter of the ring]{\includegraphics[scale=0.3]{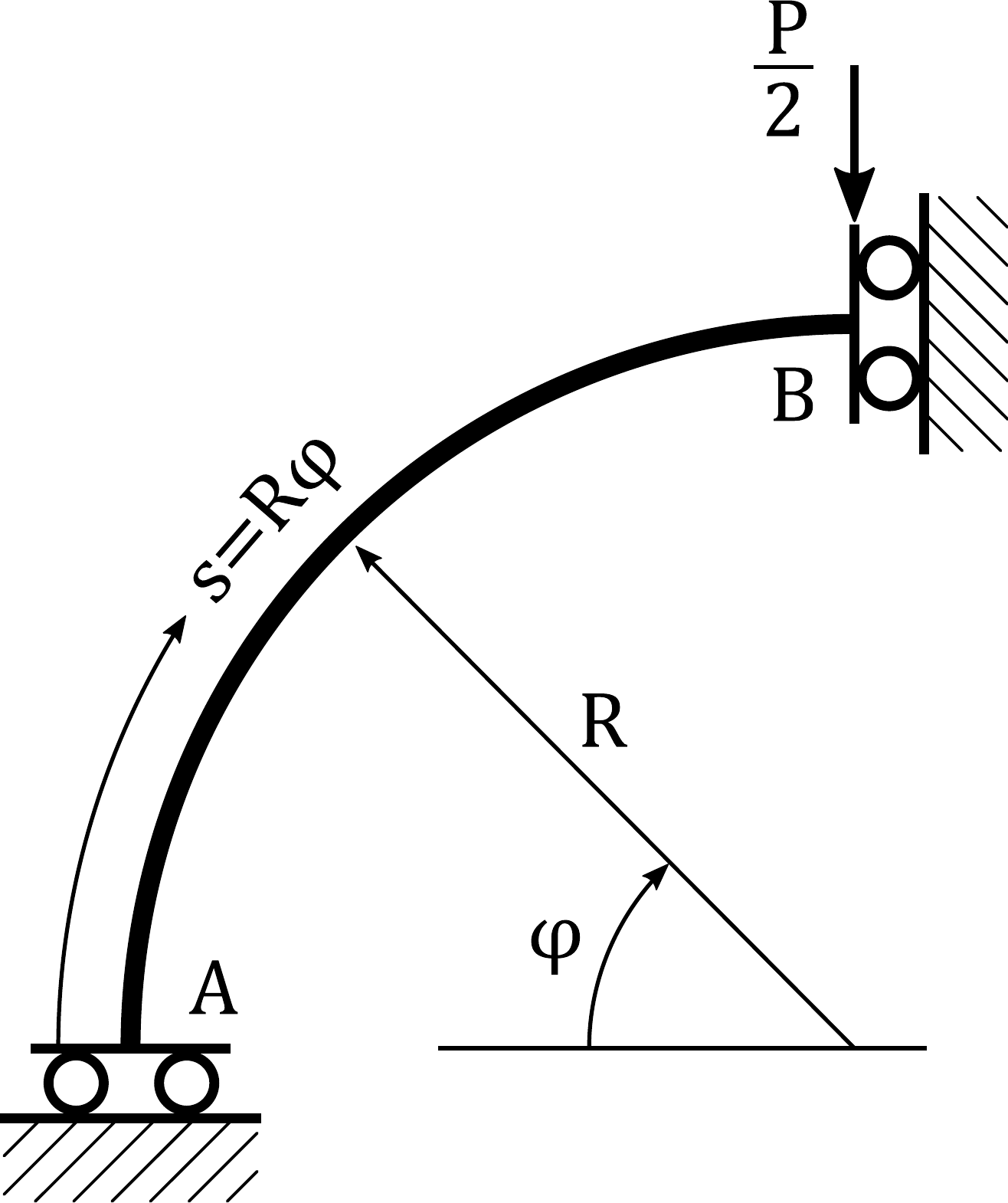}} \\
\caption{Geometry, boundary conditions, and applied load for the pinched circular ring. a) Before applying symmetry. b) After applying symmetry.} 
\label{ringgeo}
\end{figure}

In this section we perform numerical investigations using the discretizations
introduced in Sections 3 and 4 as well as the global $\bar{B}$ method \cite{Elguedj2008, bouclier2012locking, greco2017efficient, zhang2018locking}, local $\bar{B}$ elements \cite{greco2017efficient, hu2016order, antolin2020simple}, and local ANS elements \cite{caseiro2014assumed, caseiro2015assumed, greco2017efficient}. Unless mentioned otherwise, a Gauss-Legendre quadrature rule with $p+1$ integration points is used to compute all the integrals. The code used to perform these simulations has been
developed on top of the PetIGA framework \cite{dalcin2016petiga}, which adds NURBS discretization
capabilities and integration of forms to the scientific library PETSc \cite{petsc-web-page}.

In this section, we use analytical solutions to study the convergence in $L^2$ norm of the displacement vector, the membrane force, and the bending moment. In order to do so, we define the relative errors in $L^2$ norm of the displacement vector, the membrane force, and the bending moment as
\begin{align}
   e_{L^2}(\mathbf{u}^h) &=  \frac{ \sqrt{ \int^L_{0} \left( u^h_x - u_x \right)^2 \, \mathrm ds  + \int^L_{0} \left( u^h_y - u_y \right)^2 \, \mathrm ds } }{ \sqrt{ \int^L_{0}  u^2_x \, \mathrm ds + \int^L_{0} u^2_y \, \mathrm ds } }  \text{,} \\
   e_{L^2}(\mathcal{N}^h) &=  \frac{ \sqrt{\int^L_{0} \left( \mathcal{N}^h - \mathcal{N} \right)^2 \, \mathrm ds }}{ \sqrt{ \int^L_{0}  \mathcal{N}^2 \, \mathrm ds} }   \text{,} \\
   e_{L^2}(\mathcal{M}^h) &=  \frac{ \sqrt{\int^L_{0} \left( \mathcal{M}^h - \mathcal{M} \right)^2 \, \mathrm ds }}{ \sqrt{ \int^L_{0}  \mathcal{M}^2 \, \mathrm ds} } \text{,}
\end{align}
respectively. Since we are solving a fourth-order differential equation with basis functions of degree 2, the optimal convergence rates for $e_{L^2}(\mathbf{u}^h)$, $e_{L^2}(\mathcal{N}^h)$, and $e_{L^2}(\mathcal{M}^h)$ are 2, 2, and 1, respectively \cite{Hughes2012}. In engineering applications, discretization errors are acceptable in case they are smaller than the model errors (errors between reality and the mathematical model). Since $e_{L^2}(\mathbf{u}^h)$, $e_{L^2}(\mathcal{N}^h)$, and $e_{L^2}(\mathcal{M}^h)$ are relative errors, values of $e_{L^2}(\mathbf{u}^h)$, $e_{L^2}(\mathcal{N}^h)$, and $e_{L^2}(\mathcal{M}^h)$ equal to $10^{-2}$ ($1\%$ errors) are accurate enough for most engineering applications. However, values of $e_{L^2}(\mathbf{u}^h)$, $e_{L^2}(\mathcal{N}^h)$, and $e_{L^2}(\mathcal{M}^h)$ greater than 1 ($100\%$ errors) are unlikely to be acceptable in engineering applications.

\subsection{Pinched circular ring}

The first numerical investigation considers a circular ring with two opposite point loads as shown in Fig. \ref{ringgeo} a). Given the double symmetry of this problem, we solve a quarter of the ring with the appropriate symmetry boundary conditions and load shown in Fig. \ref{ringgeo} b). The next values are used in this example
\begin{equation} 
 P =  1.0,  \quad R = 1.0,  \quad EI = 1.0 \text{.} 
\end{equation}
In order to consider different values of the slenderness ratio, the values $EA=10^4$, $EA=10^6$, and $EA=10^8$ are used. As in \cite{armero2012invariant}, the cross section thickness is estimated as $t = \sqrt{EI/EA}$ in this example. Note that this thickness estimation scales down the value of the thickness by a factor of $2\sqrt{3}$ in comparison with defining a rectangular cross section as it is done in Sections 5.2 and 5.3 of this manuscript.

\begin{figure} [t!] 
 \centering
\subfigure[NURBS and CAS]{\includegraphics[scale=0.53]{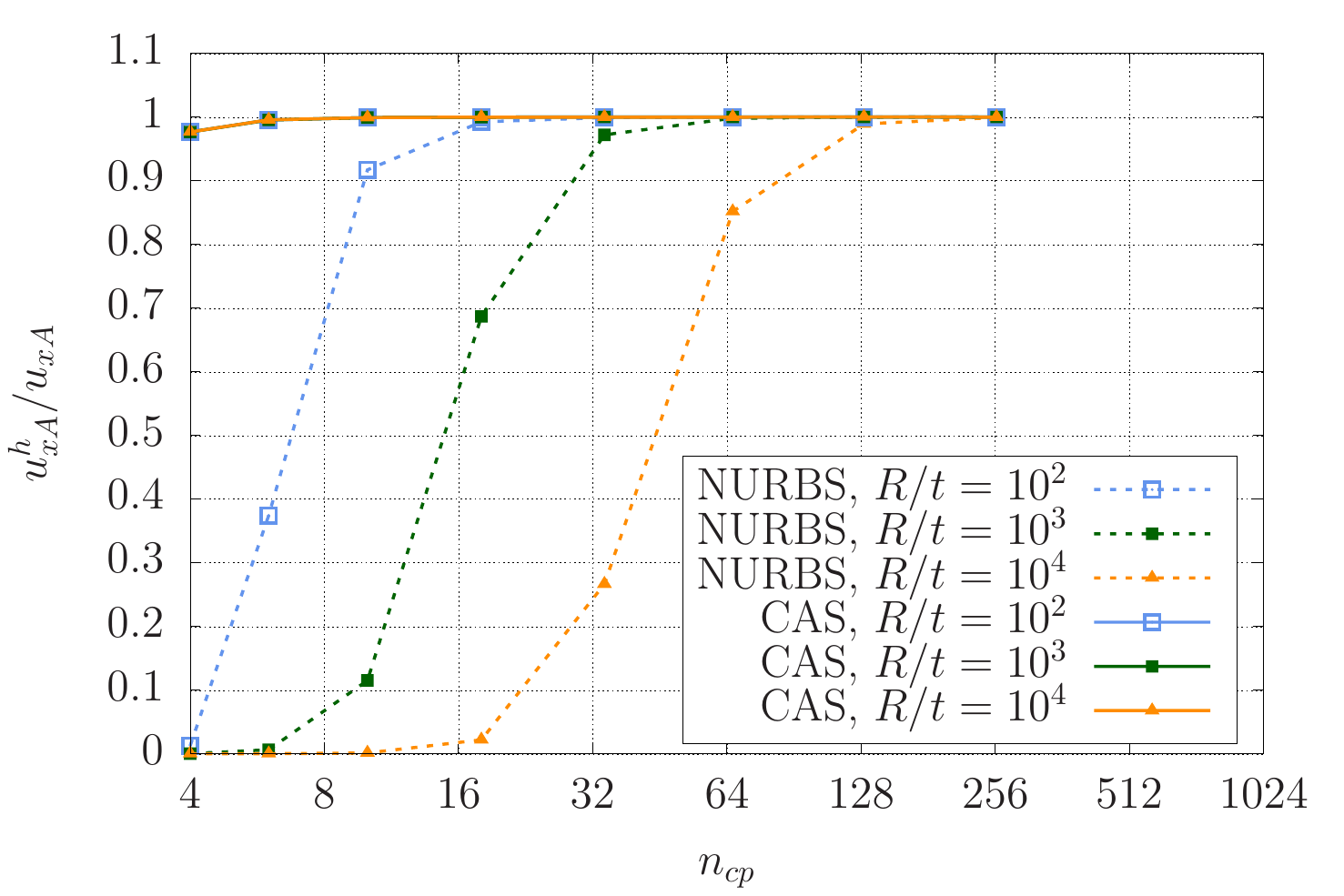}}
 \subfigure[Global $\bar{B}$ and CAS]{\includegraphics[scale=0.53]{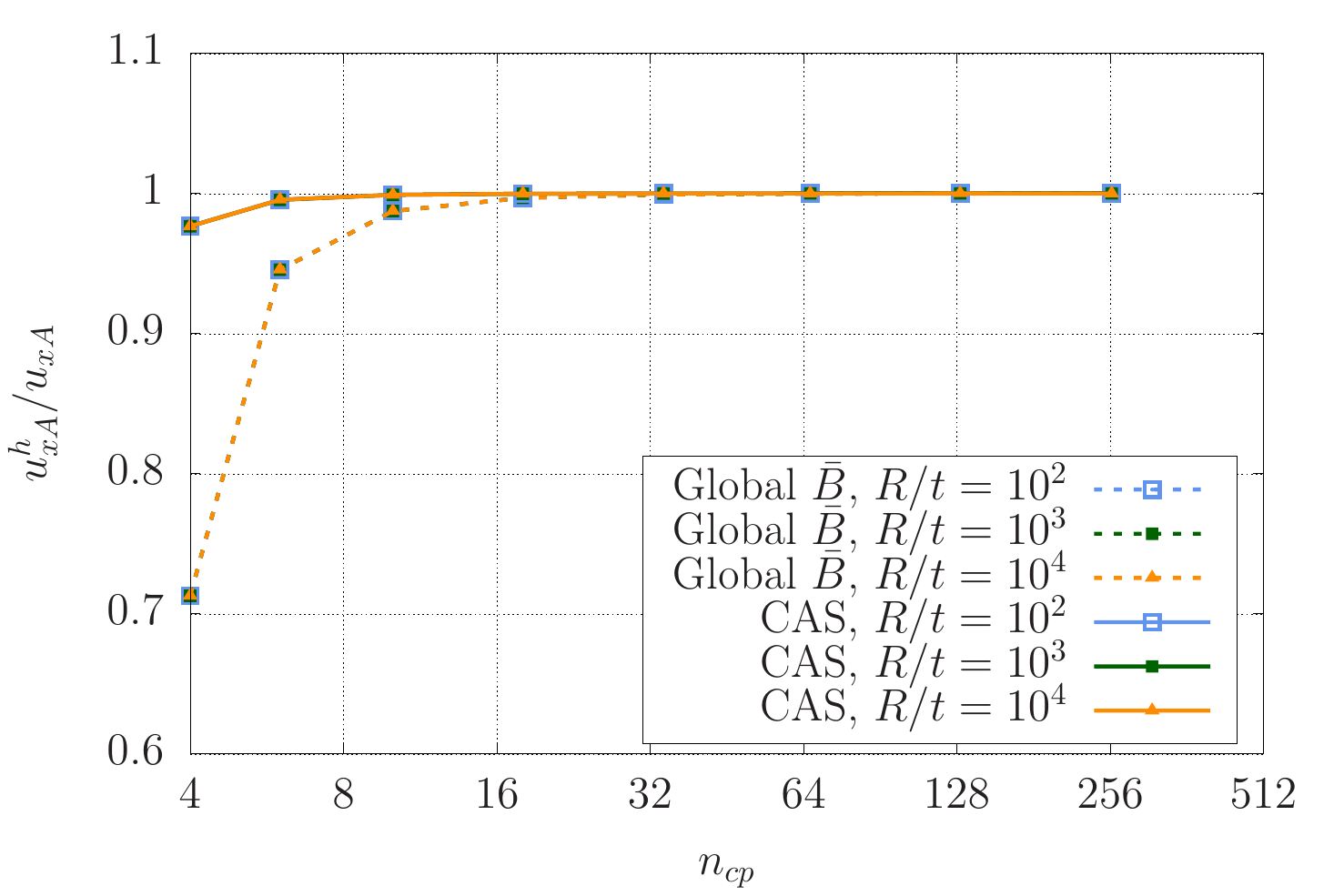}} \\
 \subfigure[NURBS and CAS]{\includegraphics[scale=0.53]{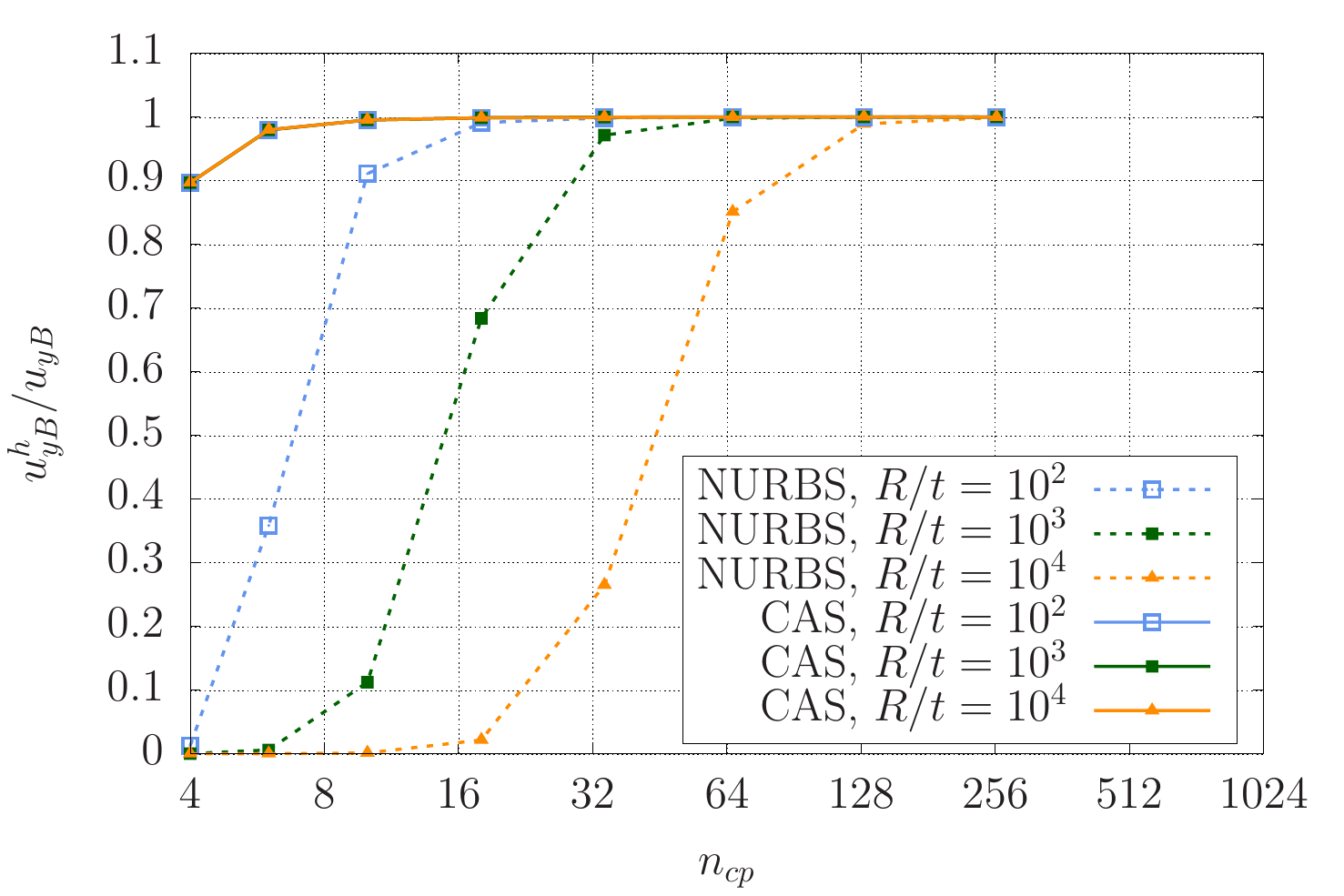}}
 \subfigure[Global $\bar{B}$ and CAS]{\includegraphics[scale=0.53]{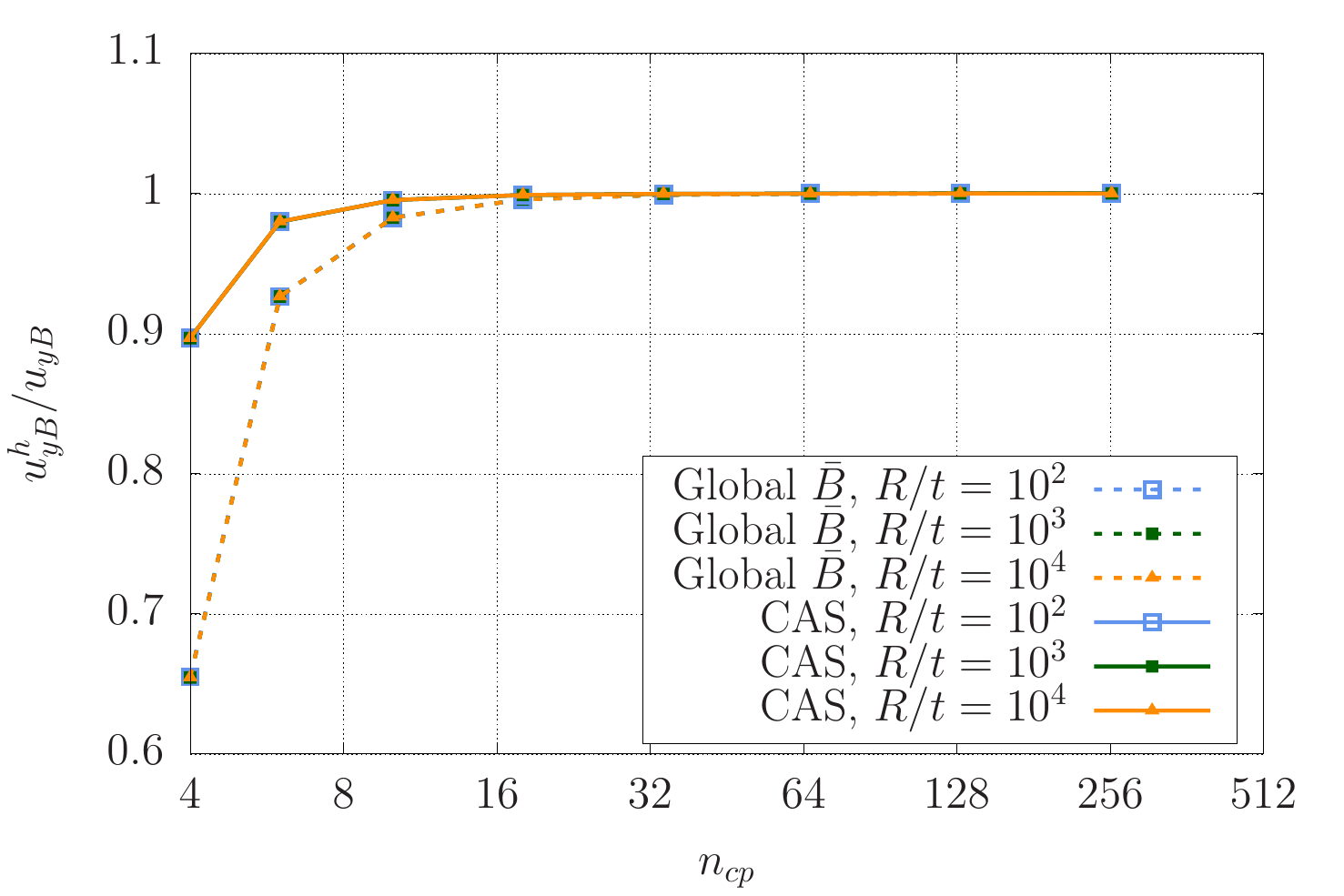}} \\
\caption{(Color online) Pinched circular ring. Convergence of the deflections at points A and B using the global $\bar{B}$ method, NURBS elements, and CAS elements. For the ample range considered, the convergence of the global $\bar{B}$ method and CAS elements are independent of the slenderness ratio.}
\label{ringdeflections}
\end{figure}

In \cite{armero2012invariant}, the exact values of the horizontal displacement of point $A$ and the vertical displacement of point $B$ are given as
\begin{align}
   u_{xA} &=  - \frac{PR^3}{EI} \left[ \frac{\pi^2 - 8}{8\pi} + \frac{\pi}{8} \left(\frac{t}{R}\right)^2 \right] \text{,} \\
   u_{yB} &=  - \frac{PR^3}{EI} \left[ \frac{4 - \pi}{4\pi} - \frac{1}{4} \left(\frac{t}{R}\right)^2 \right] \text{,}
\end{align}
respectively. Points $A$ and $B$ are shown in Fig. \ref{ringgeo}. In \cite{armero2012invariant}, the exact distribution of the membrane force and the bending moment are given as
\begin{align}
   \mathcal{N} &=  - \frac{P}{2} \cos (\varphi) \text{,} \\
   \mathcal{M} &=  \frac{PR}{2} \left[ \frac{2}{\pi} - \cos (\varphi) \right] \text{,}
\end{align}
respectively, where the angle $\varphi$ is shown in Fig. \ref{ringgeo} b).

\begin{figure} [t!] 
 \centering
\subfigure[NURBS and CAS]{\includegraphics[scale=0.53]{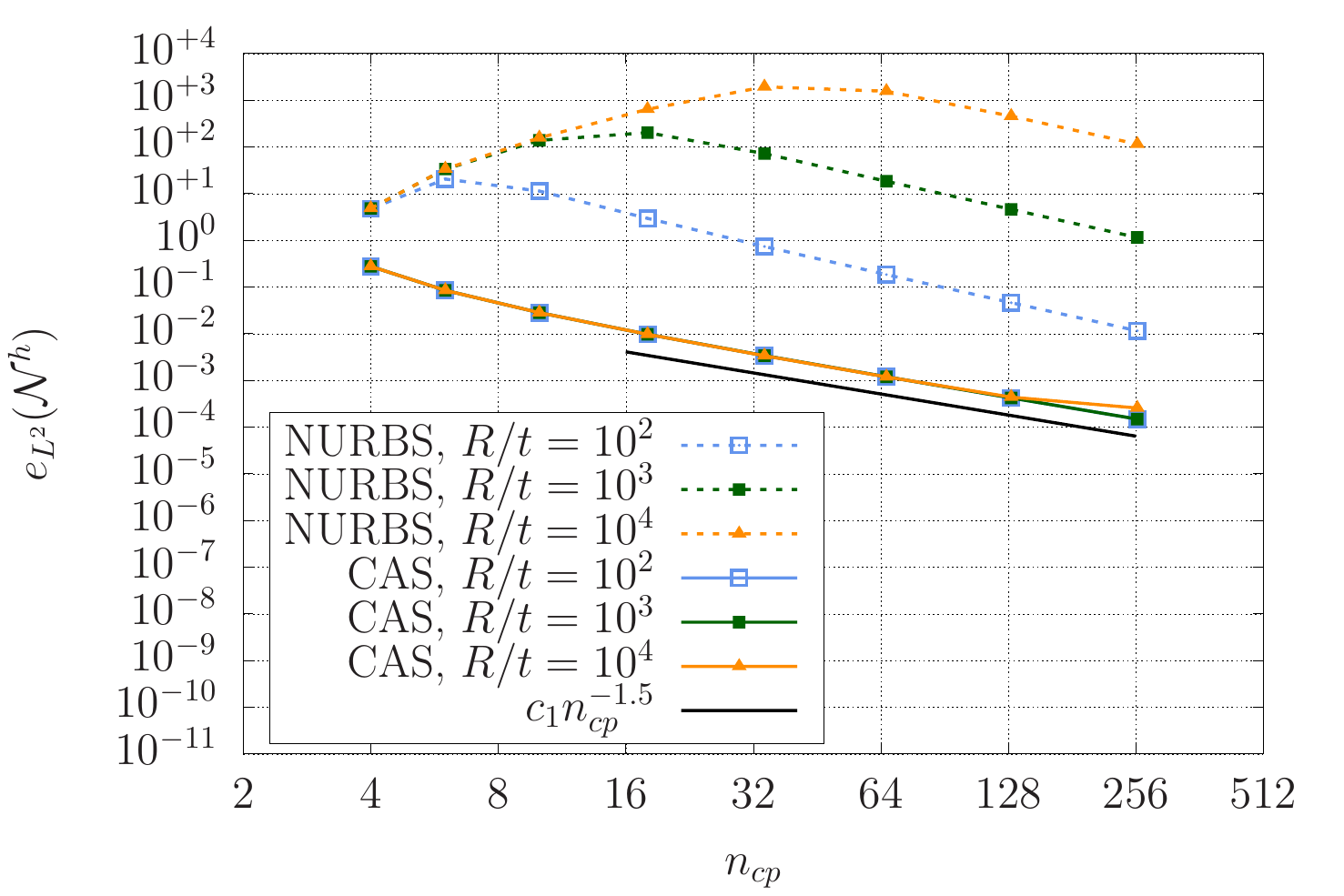}}
 \subfigure[Global $\bar{B}$ and CAS]{\includegraphics[scale=0.53]{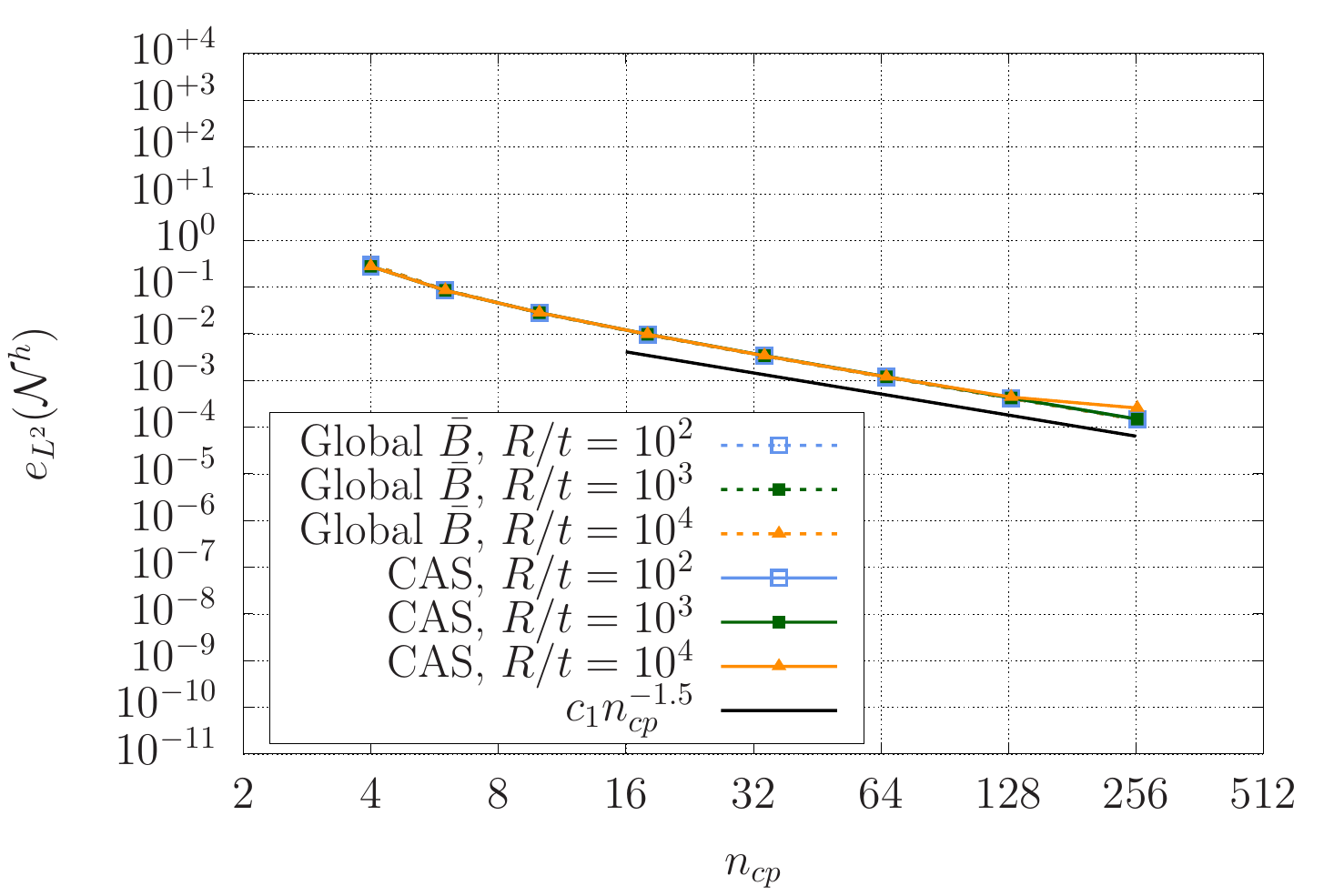}} \\
 \subfigure[NURBS and CAS]{\includegraphics[scale=0.53]{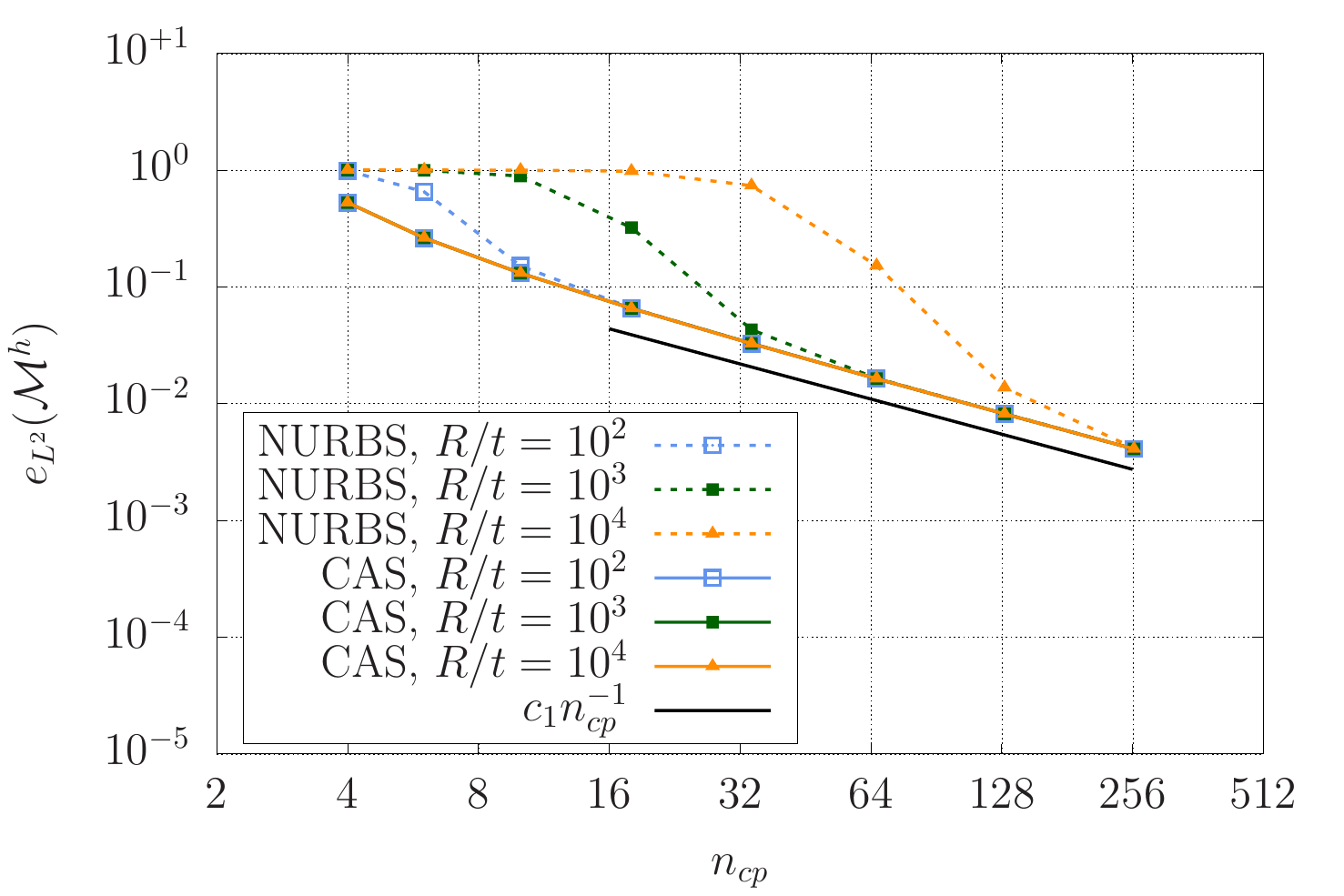}}
 \subfigure[Global $\bar{B}$ and CAS]{\includegraphics[scale=0.53]{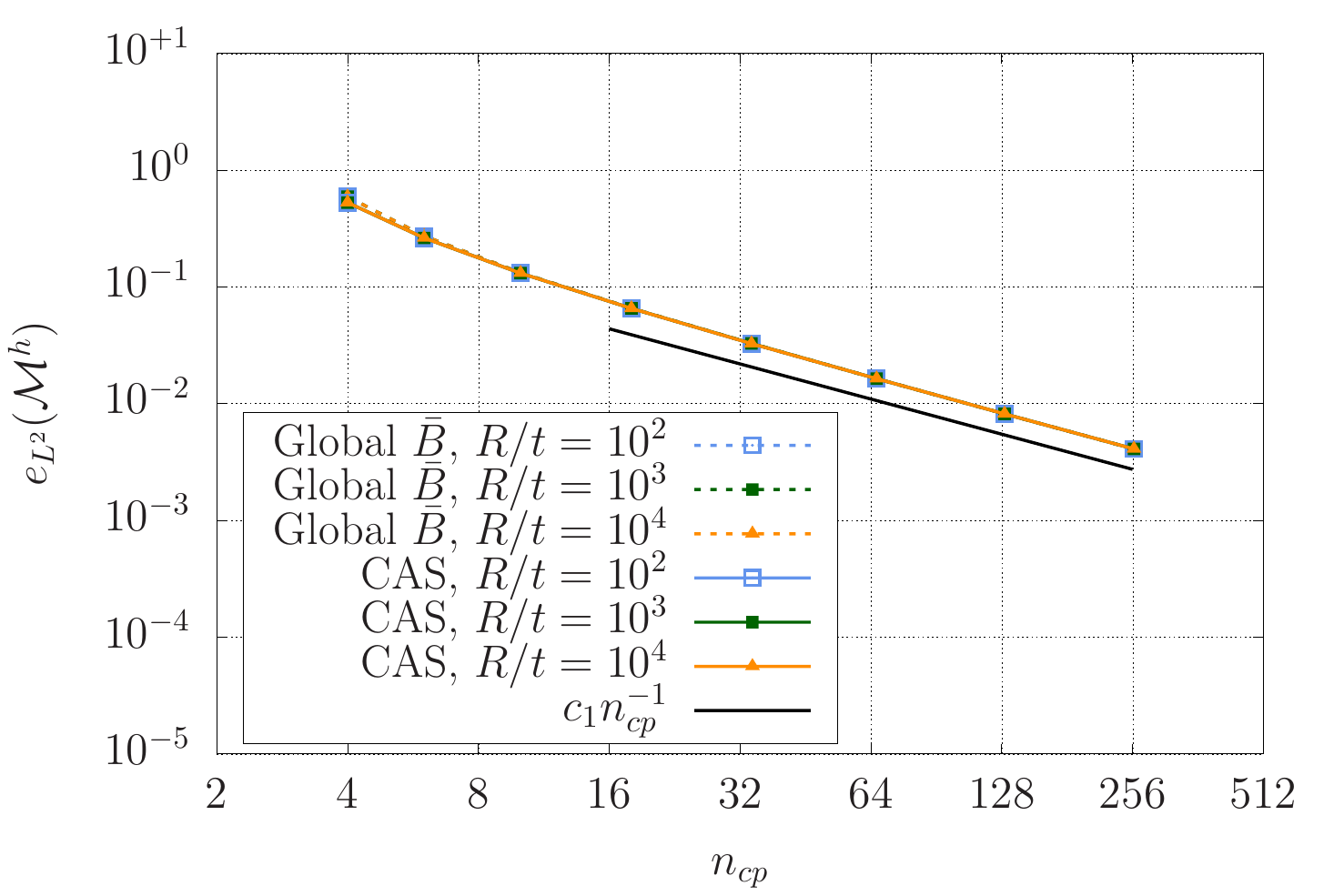}} \\
\caption{(Color online) Pinched circular ring. Convergence of the membrane force and bending moment using the global $\bar{B}$ method, NURBS elements, and CAS elements. The numerical solutions using CAS elements and the global $\bar{B}$ method overlap.}
\label{ringconvergence}
\end{figure}

\begin{figure} [t!] 
 \centering
\subfigure[8 elements]{\includegraphics[scale=0.52]{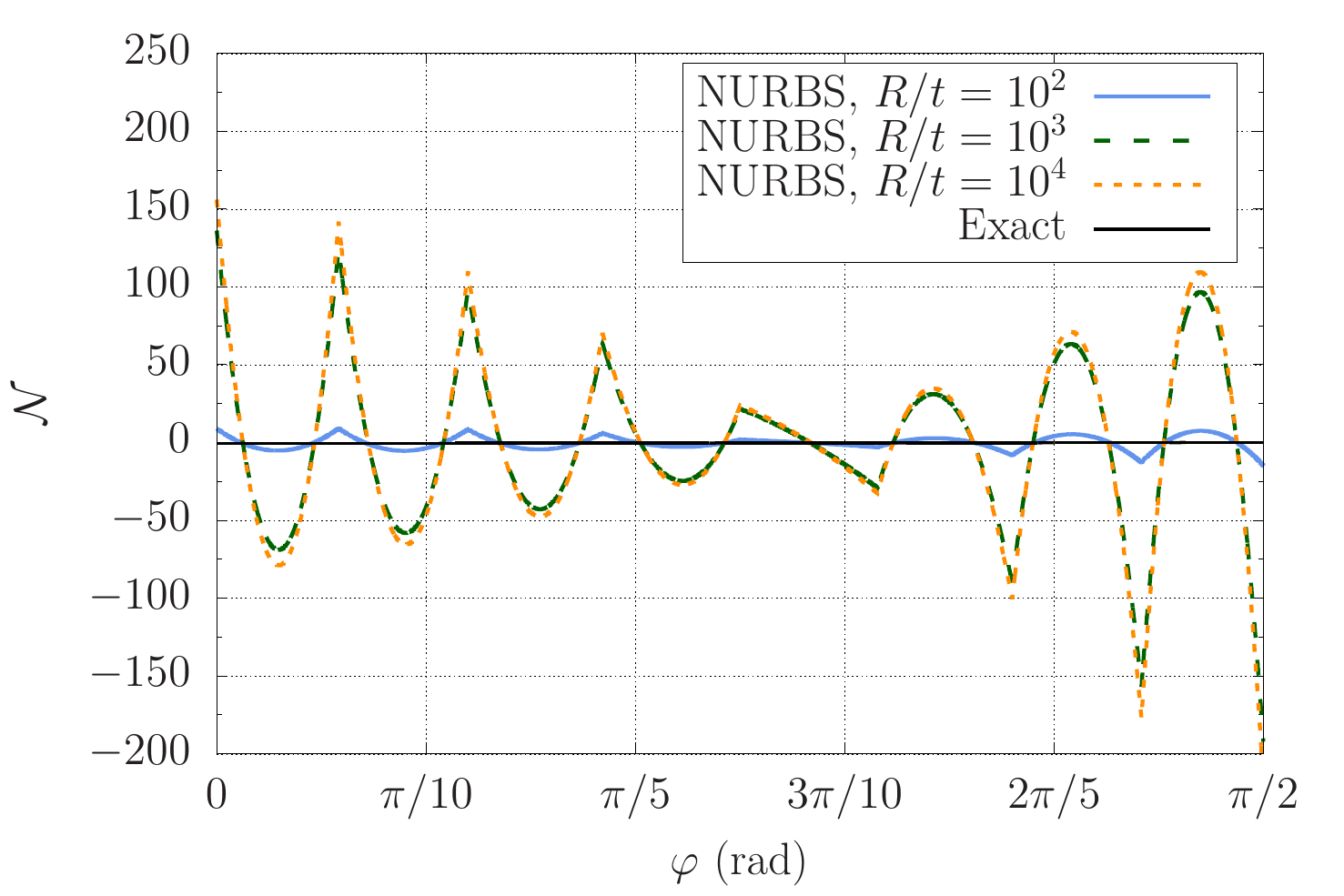}}
 \subfigure[8 elements]{\includegraphics[scale=0.52]{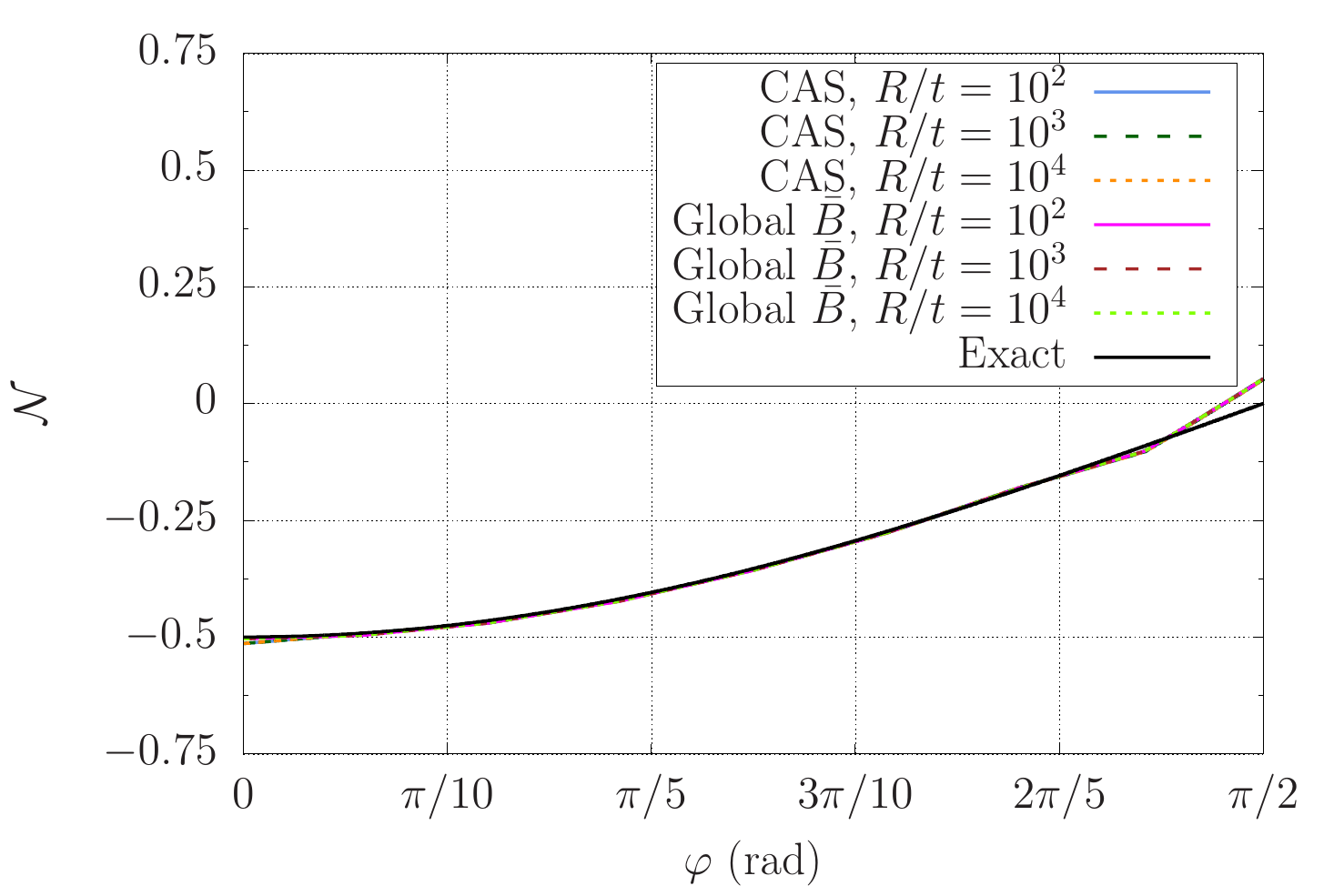}} \\
\subfigure[16 elements]{\includegraphics[scale=0.52]{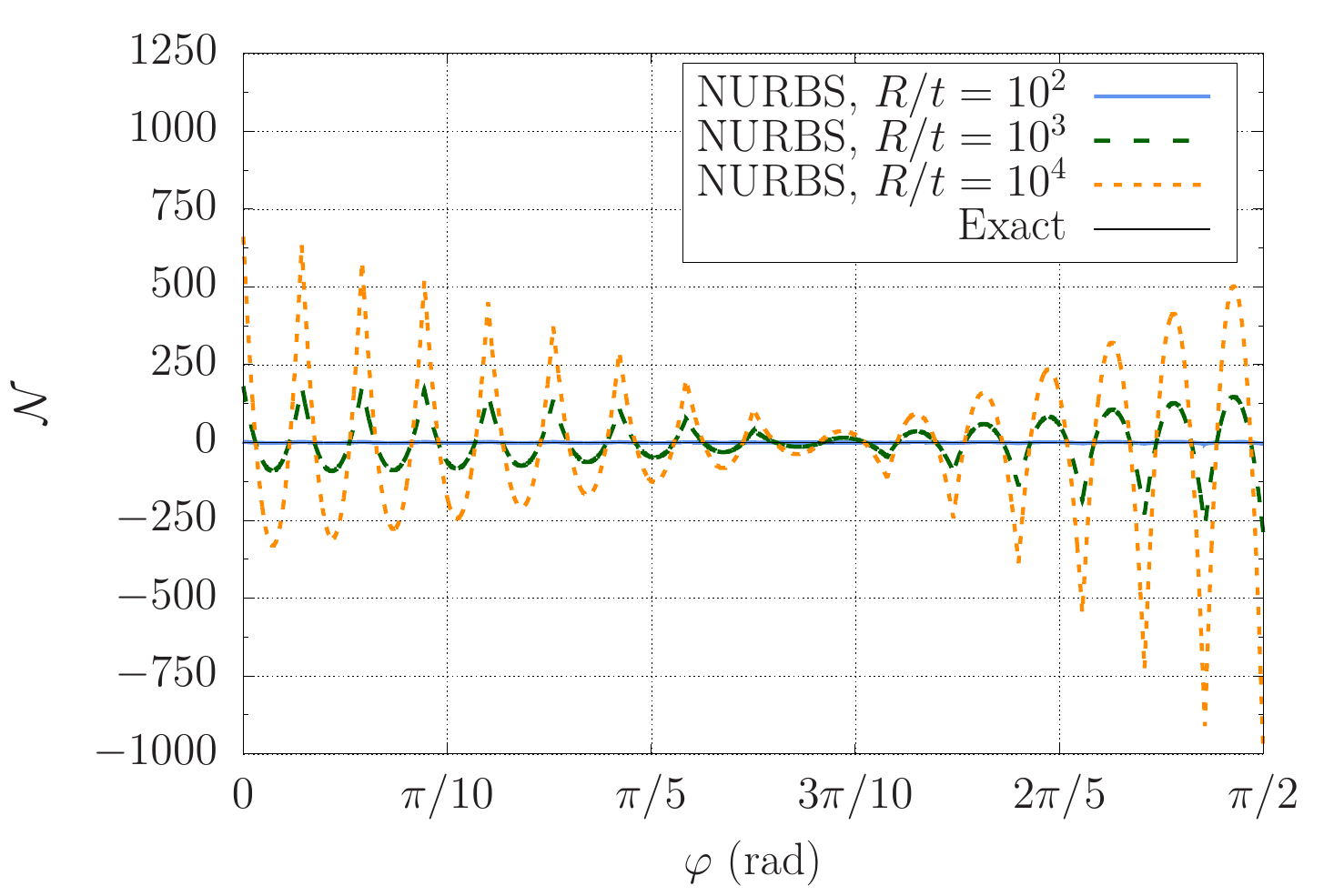}}
 \subfigure[16 elements]{\includegraphics[scale=0.52]{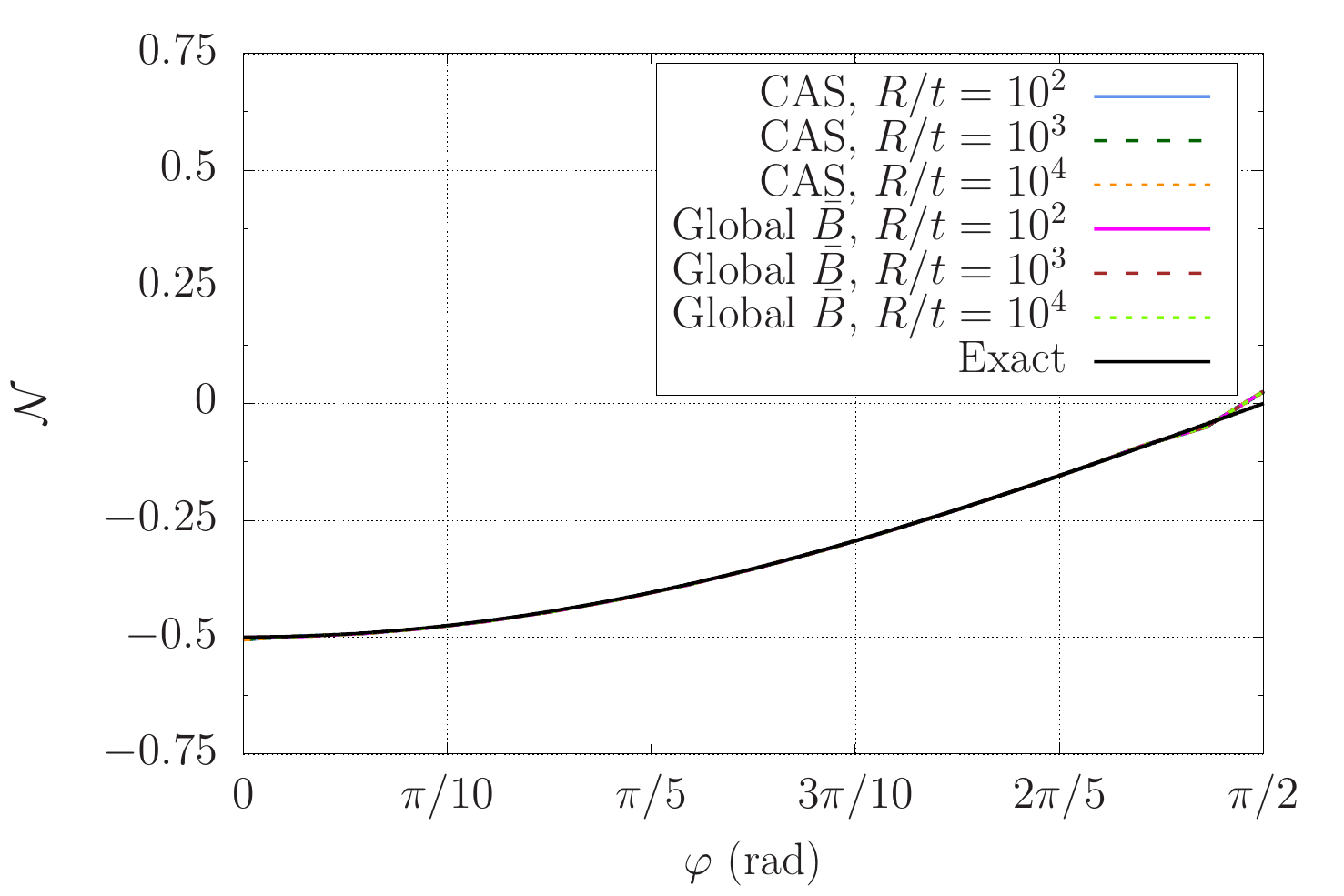}} \\
 \subfigure[32 elements]{\includegraphics[scale=0.52]{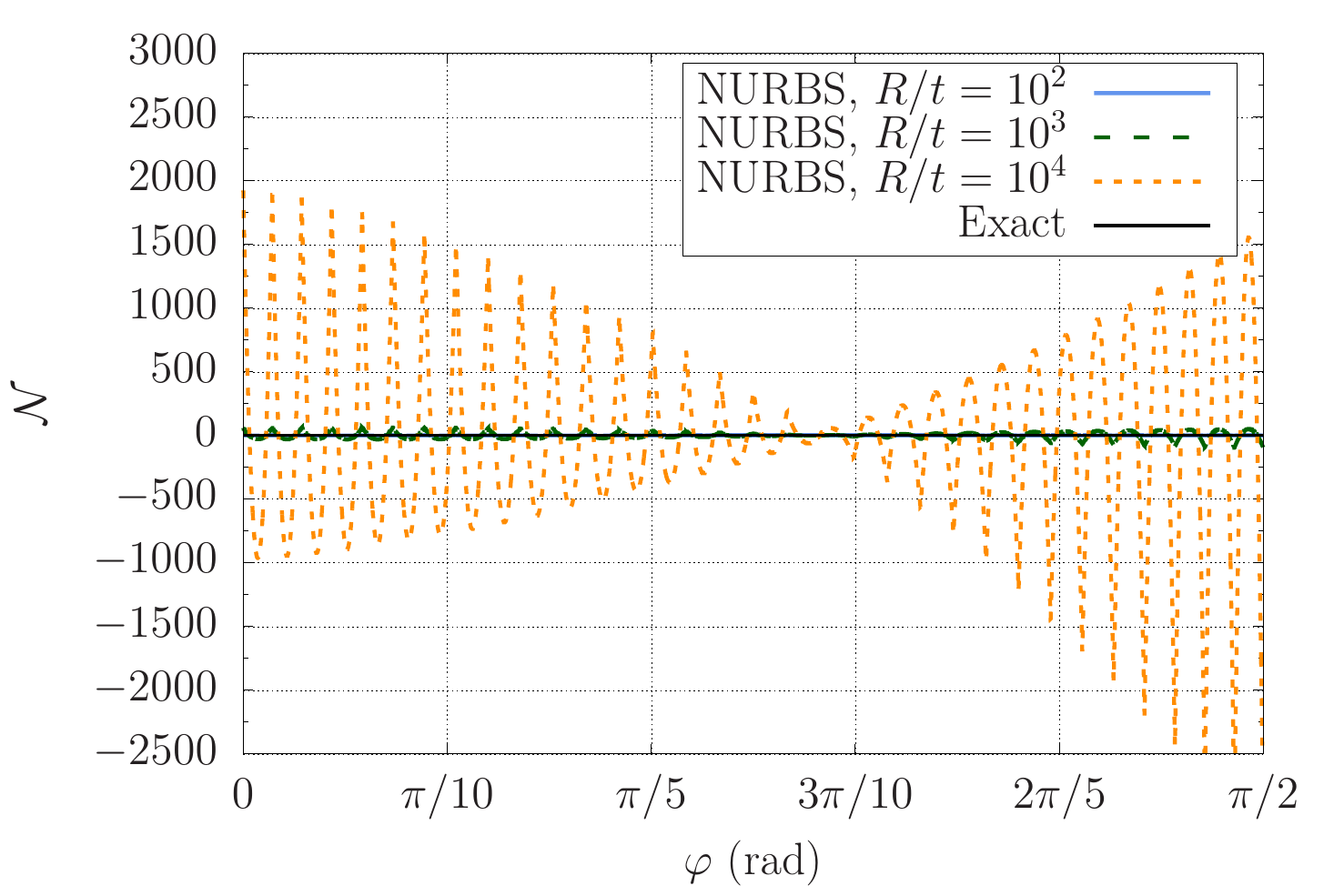}}
 \subfigure[32 elements]{\includegraphics[scale=0.52]{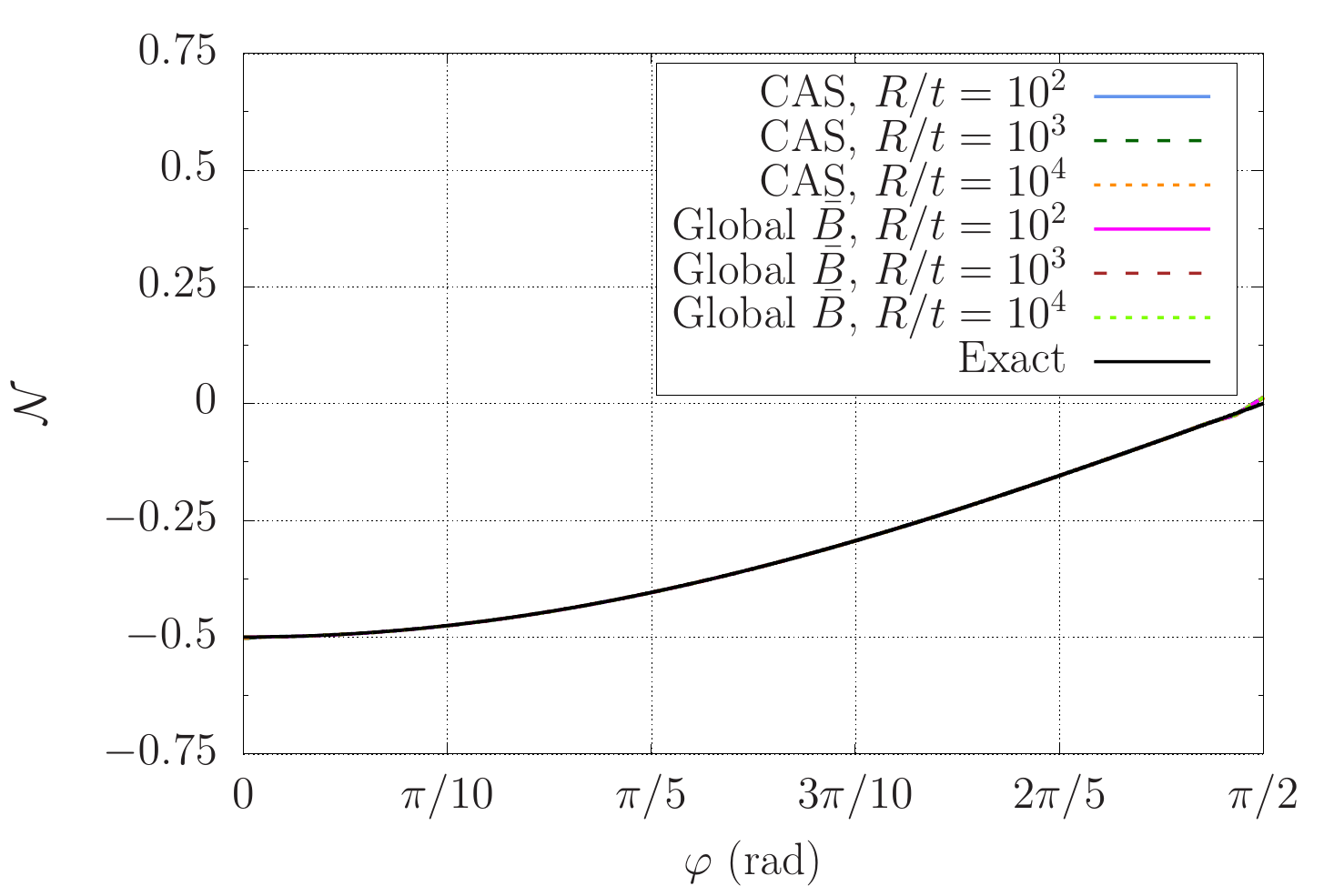}} \\
\caption{(Color online) Membrane force of the pinched circular ring for different mesh resolutions using the global $\bar{B}$ method, NURBS elements, and CAS elements. The numerical solutions using either CAS elements or the global $\bar{B}$ method overlap. The numerical solutions using NURBS lock resulting in spurious oscillations whose amplitude is orders of magnitude greater than the maximum exact membrane force of this problem. Note the different vertical scale used in each plot.}
\label{ringmfdistributions}
\end{figure}

\begin{figure} [t!] 
 \centering
\subfigure[8 elements]{\includegraphics[scale=0.52]{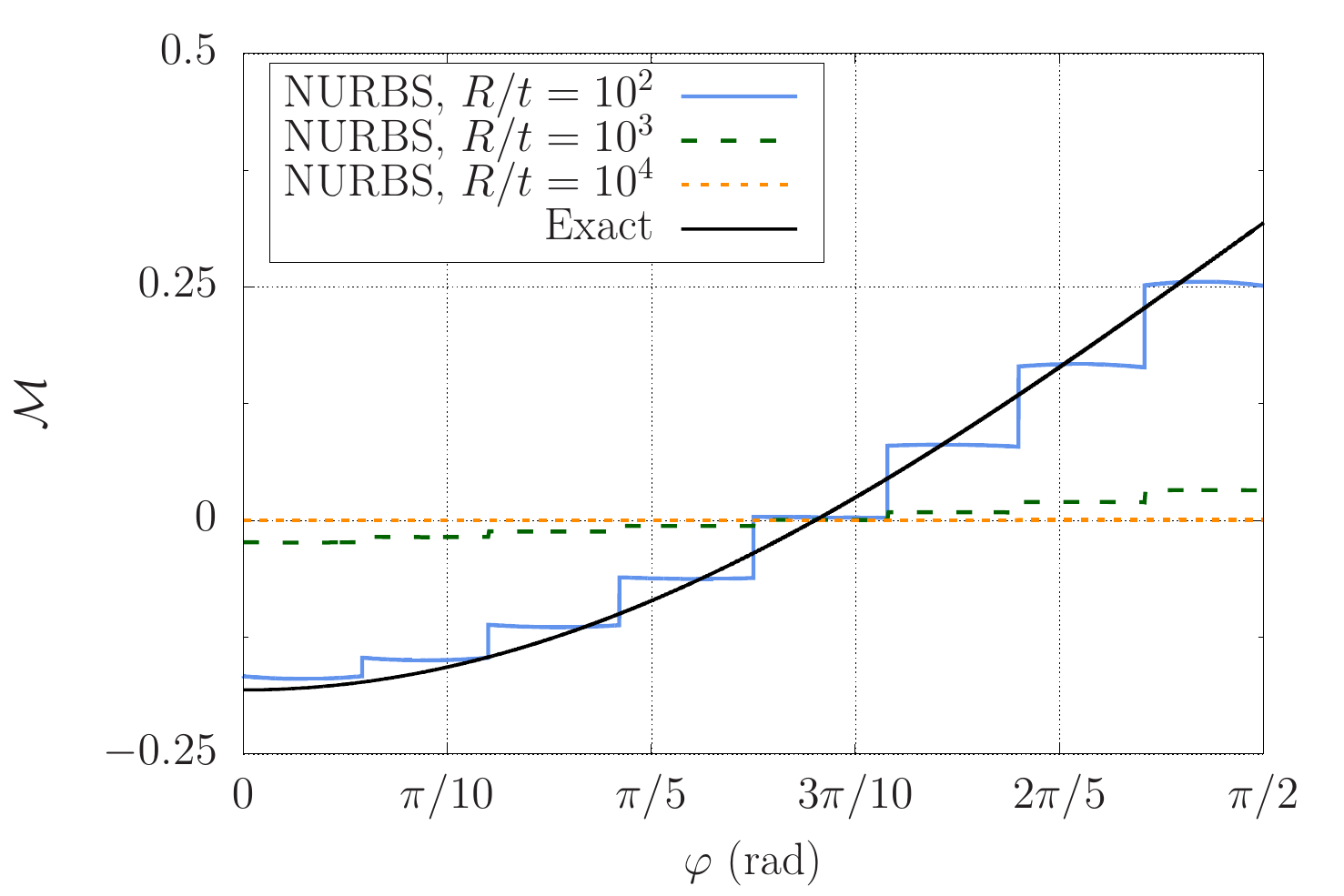}}
 \subfigure[8 elements]{\includegraphics[scale=0.52]{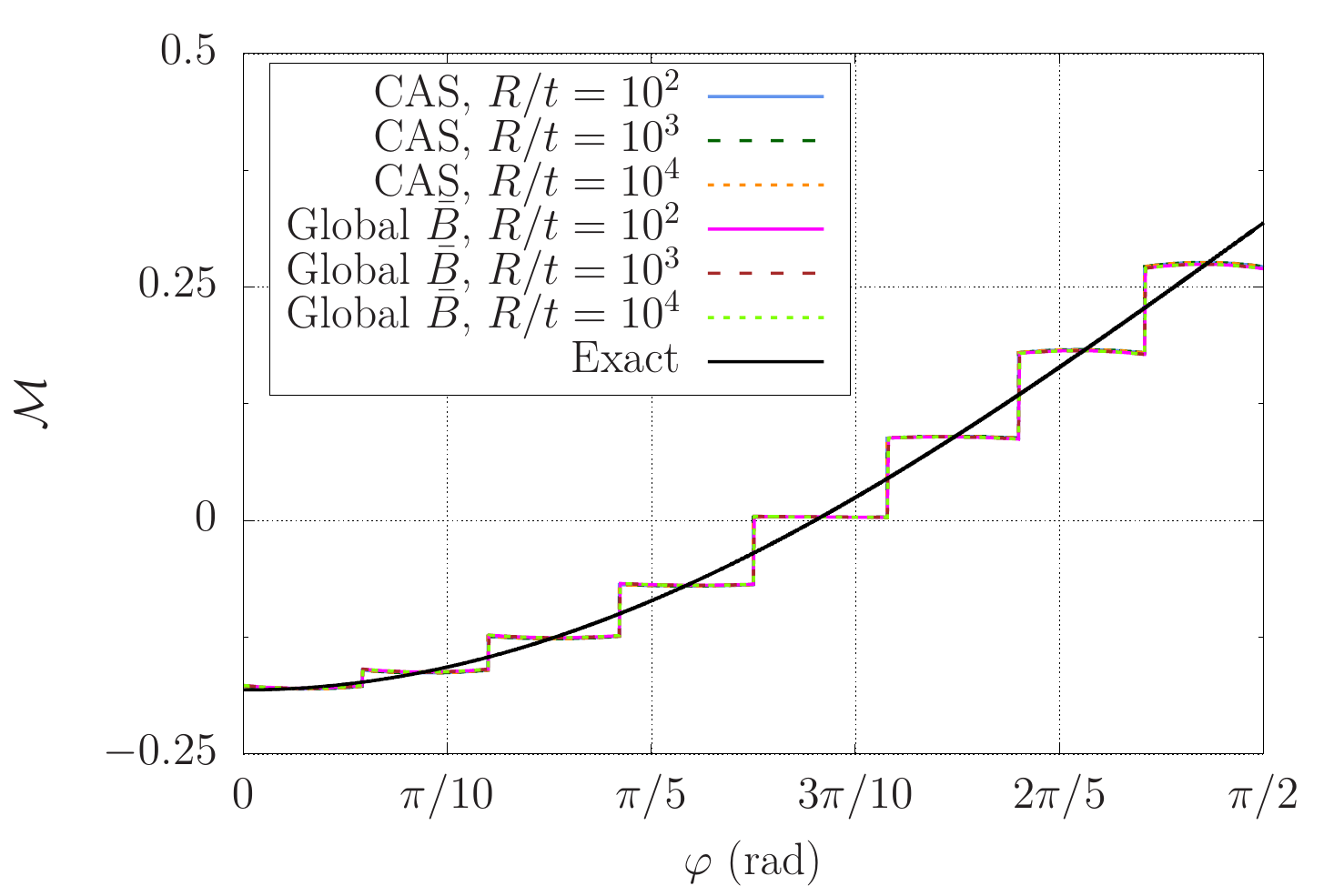}} \\
\subfigure[16 elements]{\includegraphics[scale=0.52]{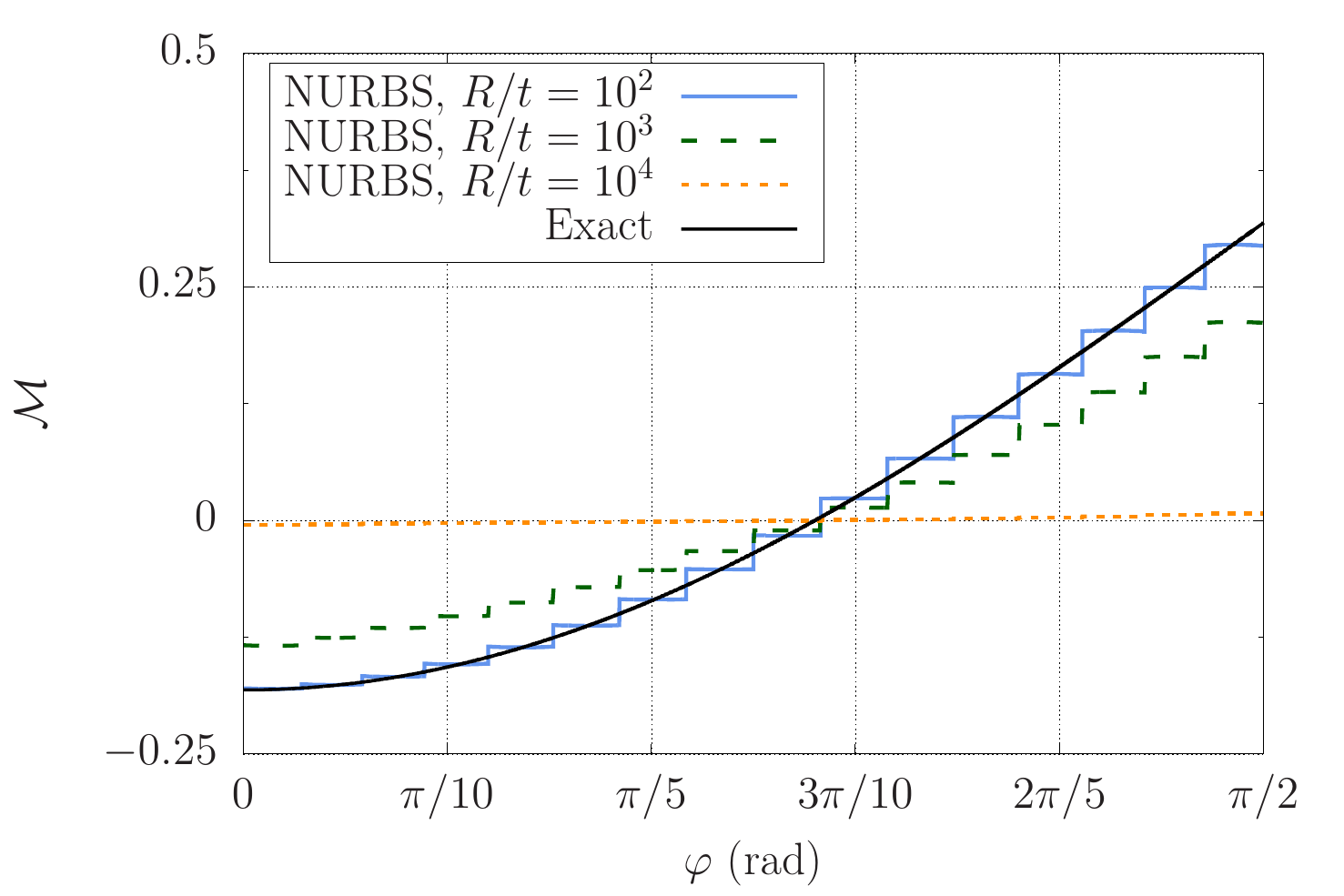}}
 \subfigure[16 elements]{\includegraphics[scale=0.52]{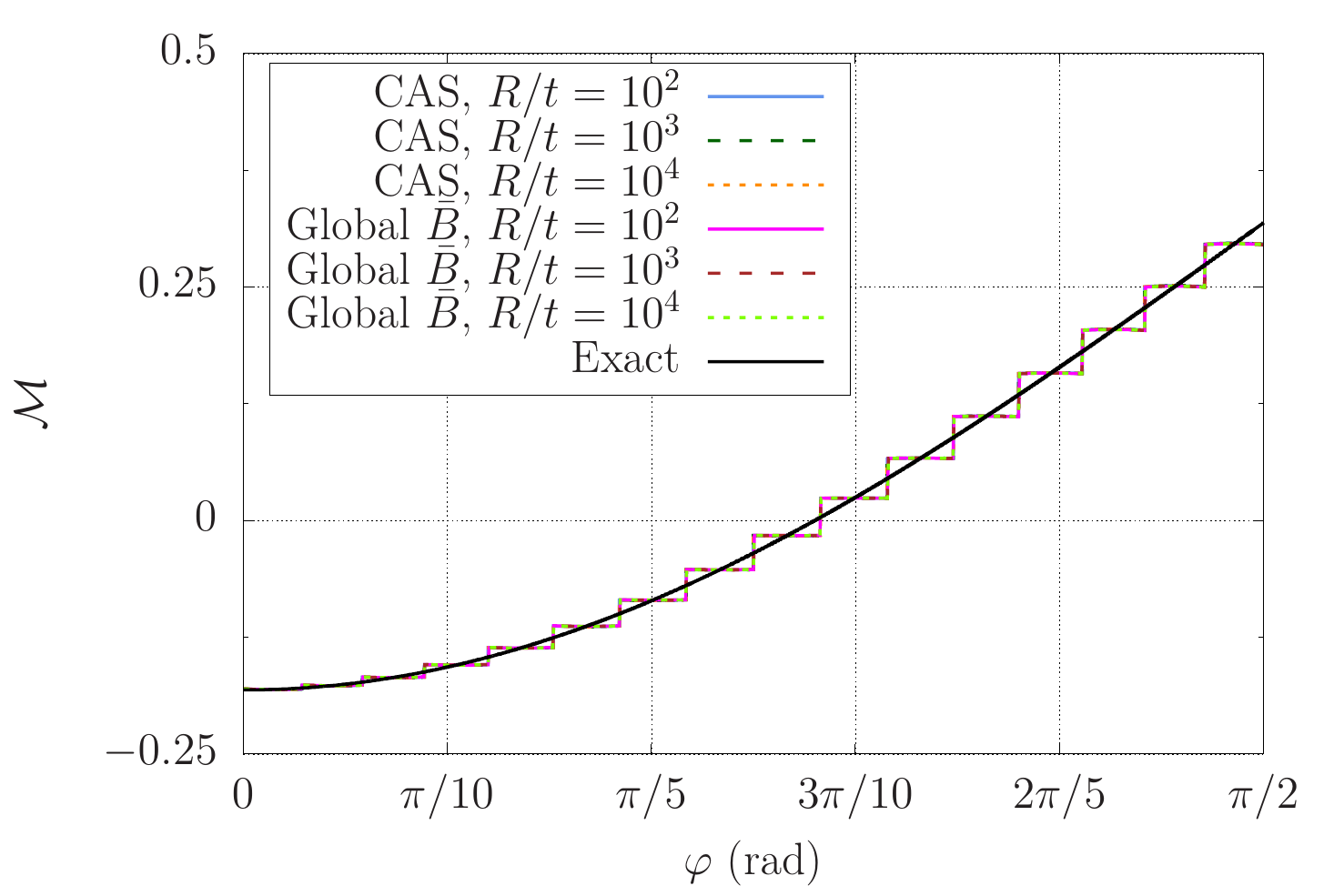}} \\
 \subfigure[32 elements]{\includegraphics[scale=0.52]{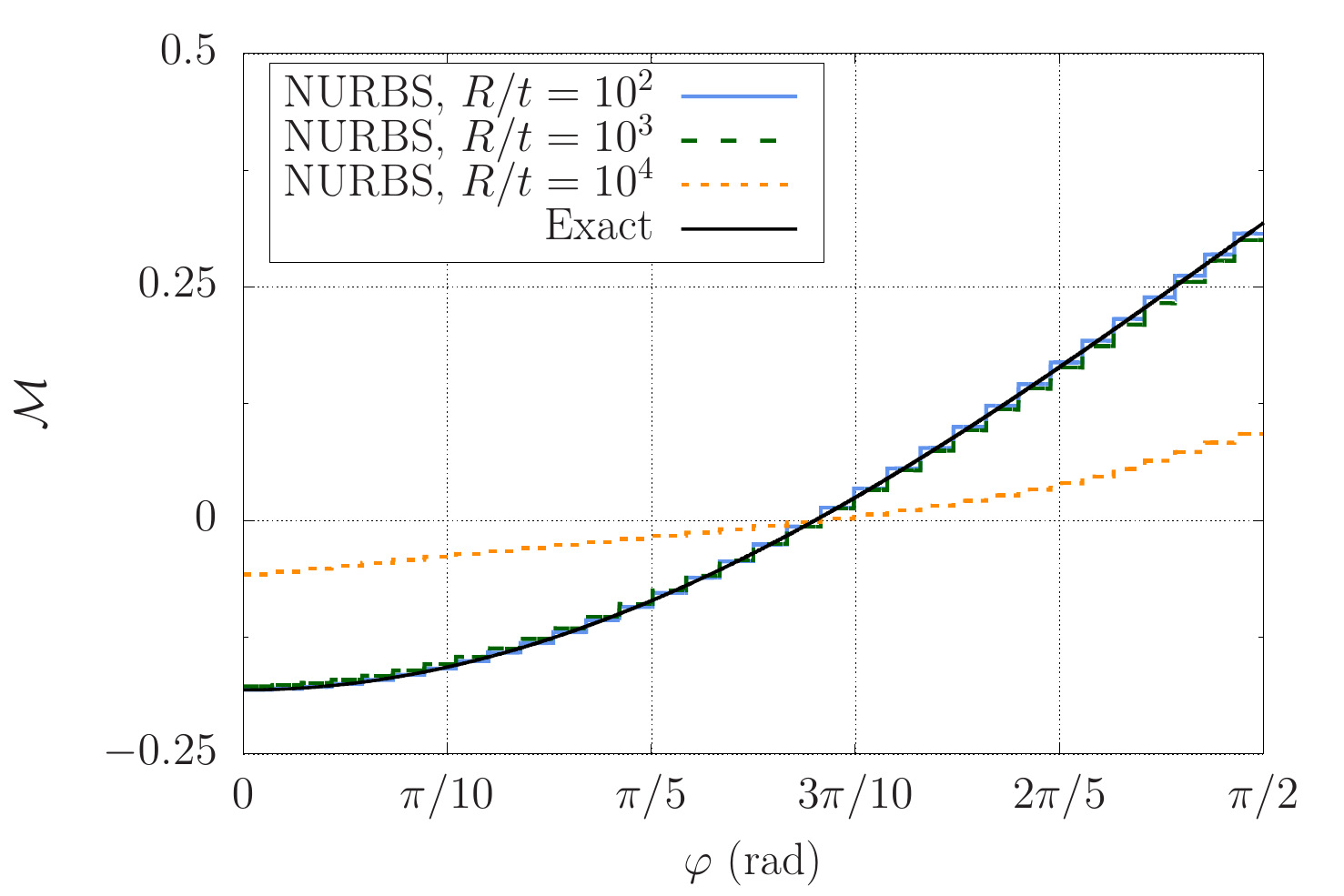}}
 \subfigure[32 elements]{\includegraphics[scale=0.52]{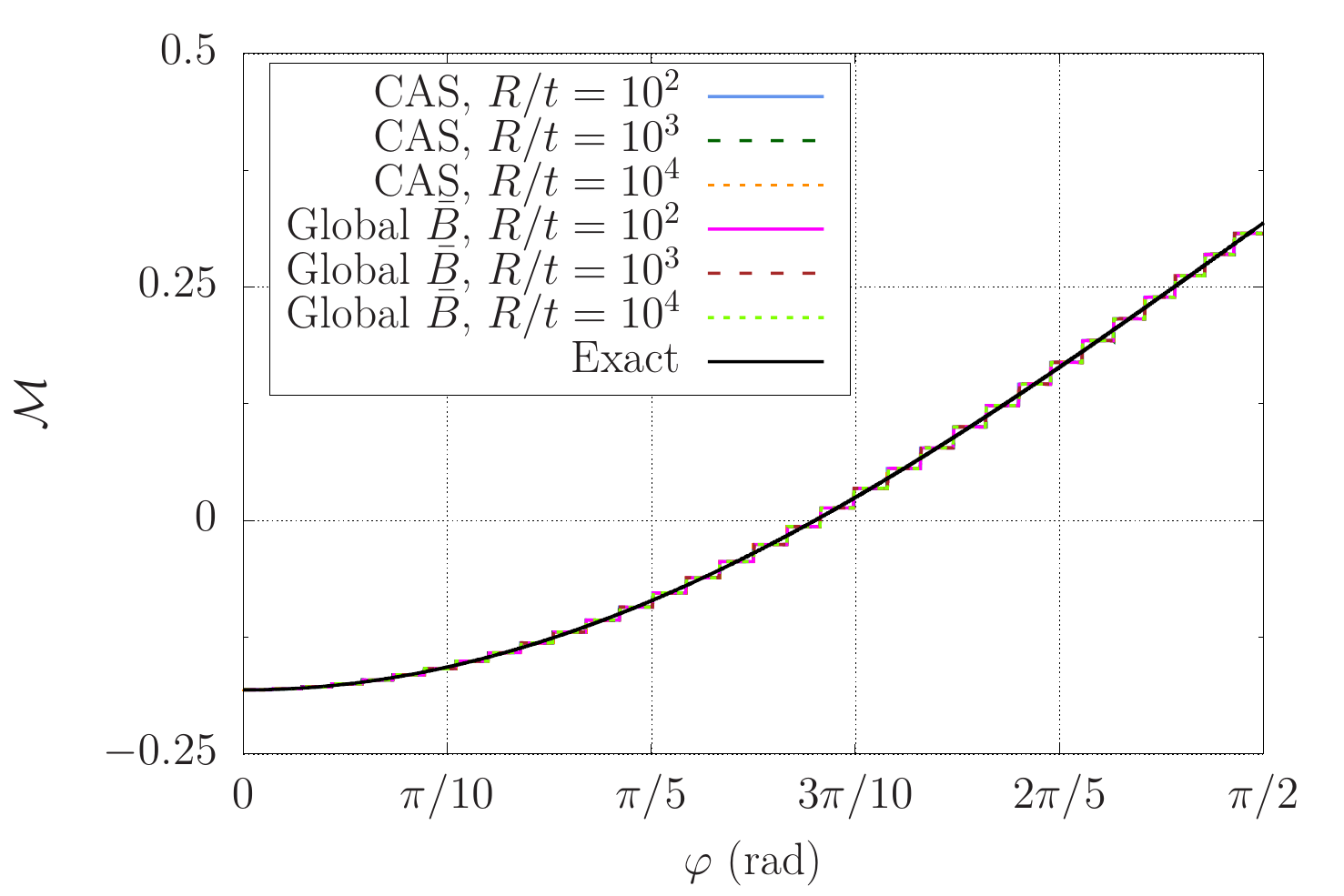}} \\
\caption{(Color online) Bending moment of the pinched circular ring for different mesh resolutions using the global $\bar{B}$ method, NURBS elements, and CAS elements. The numerical solutions using either CAS elements or the global $\bar{B}$ method overlap for the different $R/t$ ratios considered. The numerical solutions using NURBS elements lock as the $R/t$ ratio increases resulting in zero bending moment.}
\label{ringbmdistributions}
\end{figure}

We initiate our convergence study with a uniform mesh composed of two quadratic elements. The axis of the rod is represented exactly since we are using quadratic NURBS. After that, we perform uniform $h$-refinement seven times. Using the global $\bar{B}$ method \cite{Elguedj2008, bouclier2012locking, greco2017efficient, zhang2018locking}, NURBS elements, and CAS elements, Fig. \ref{ringdeflections} plots the convergence of the horizontal displacement of point $A$ and the vertical displacement of point $B$ while Fig. \ref{ringconvergence} plot the convergence in $L^2$ norm of the membrane force and the bending moment. When applied to linear plane Kirchhoff rods discretized using quadratic NURBS, the global $\bar{B}$ method performs a $L^2$ projection of the membrane strain at the patch level into the space of linear Lagrange polynomials \cite{Elguedj2008, bouclier2012locking, greco2017efficient, zhang2018locking}. In \cite{bouclier2012locking}, the global $\bar{B}$ method was applied to curved plane Timoshenko rods and shown to be at least one order of magnitude more accurate than selective-reduced integration and the global DSG method for coarse meshes. These results motivated our choice of comparing the accuracy of CAS elements with the global $\bar{B}$ method. As shown in Figs. \ref{ringdeflections} and \ref{ringconvergence}, the convergence of CAS elements and the global $\bar{B}$ method are independent of the slenderness ratio for the broad range of $R/t$ values considered while the convergence of NURBS elements heavily deteriorates as the slenderness ratio increases. Note that for coarse meshes, the displacement values obtained with CAS elements are more accurate than the displacement values obtained with the global $\bar{B}$ method. The convergence rate of the membrane force in $L^2$ norm using either CAS elements or the global $\bar{B}$ method is 1.5 instead of 2. In \cite{greco2017efficient}, both the global $\bar{B}$ method and the locally reconstructed version of the $\bar{B}$ method that preserves the continuity of the strains also resulted in the convergence rate of the membrane force in $L^2$ norm being 1.5 when applied to linear plane curved Kirchhoff rods. Regarding computational efficiency, the global $\bar{B}$ method requires to compute the inverse of a mass matrix at the patch level and the resulting stiffness matrix is not sparse anymore, but completely full instead. In contrast, the only additional cost of CAS elements in comparison with the locking-prone NURBS elements is having to compute the derivatives of the basis functions and the unit tangent vector at the knots. To measure average computational times with each numerical scheme, we solved a hundred times on a loop this problem using 128 elements with NURBS elements, CAS elements, and the global $\bar{B}$ method. The average computational time of CAS elements only increased $4\%$ with respect to NURBS elements while the average computational time of the global $\bar{B}$ method increased more than an order of magnitude with respect to NURBS elements. Even though computational times should always be taken with a grain of salt since they depend on the specific implementation of each numerical scheme, the general conclusion is that CAS elements barely increase the computational cost with respect to NURBS elements and are significantly faster than the global $\bar{B}$ method.

Fig. \ref{ringconvergence} a) reveals an anomalous behavior in the convergence of the membrane force using NURBS elements, namely, the relative error in $L^2$ norm of the membrane force increases as uniform $h$-refinement is performed multiple times (note that for most mesh resolutions and slenderness ratios the relative error of the membrane force is greater than $100\%$). This anomalous behavior caused by membrane locking has been reported using B-splines in \cite{greco2013b}. For both coarse and fine meshes, the relative error in $L^2$ norm of the membrane force obtained with CAS elements is several orders of magnitude smaller than the relative error in $L^2$ norm of the membrane force obtained with NURBS elements.

For $R/t= 10^2$, $10^3$, and $10^4$ and using 8, 16, and 32 elements, the distribution of the membrane force is plotted in Fig. \ref{ringmfdistributions}. As shown in Fig. \ref{ringmfdistributions}, NURBS elements undergo large-amplitude spurious oscillations of the membrane force which get worse as the slenderness ratio increases (the amplitude of the spurious oscillations can be up to three orders of magnitude greater than the maximum exact membrane force of this problem). In contrast, the distribution of the membrane force obtained using either CAS elements or the global $\bar{B}$ method is completely free of spurious oscillations. In addition, the curves obtained using CAS elements and the global $\bar{B}$ method overlap for $R/t= 10^2$, $10^3$, and $10^4$.

For $R/t= 10^2$, $10^3$, and $10^4$ and using 8, 16, and 32 elements, the distribution of the bending moment is plotted in Fig. \ref{ringbmdistributions}. As shown in Fig. \ref{ringbmdistributions}, NURBS elements may lock and result in flat distributions of the bending moment. This phenomenon is analogous to the essentially zero displacements obtained for those meshes in Fig. \ref{ringdeflections}. In contrast, the distribution of the bending moment obtained using either CAS elements or the global $\bar{B}$ method is insensitive to the slenderness ratio for the wide interval of $R/t$ values considered. When using $C^1$-continuous quadratic NURBS for the discretization of the displacement vector, the bending moment is discontinuous across element boundaries. Thus, the small-amplitude zigzagging shown in Fig. \ref{ringbmdistributions} is expected. Note that the mean value of the bending moment in any element obtained using either CAS elements or the global $\bar{B}$ method approximates very accurately the mean exact value of the bending moment in that element.

\subsection{Clamped-clamped semi-circular arch under a distributed load}

	\begin{figure} [h!] 
\centering
\subfigure[Before applying symmetry]{\includegraphics[scale=0.3]{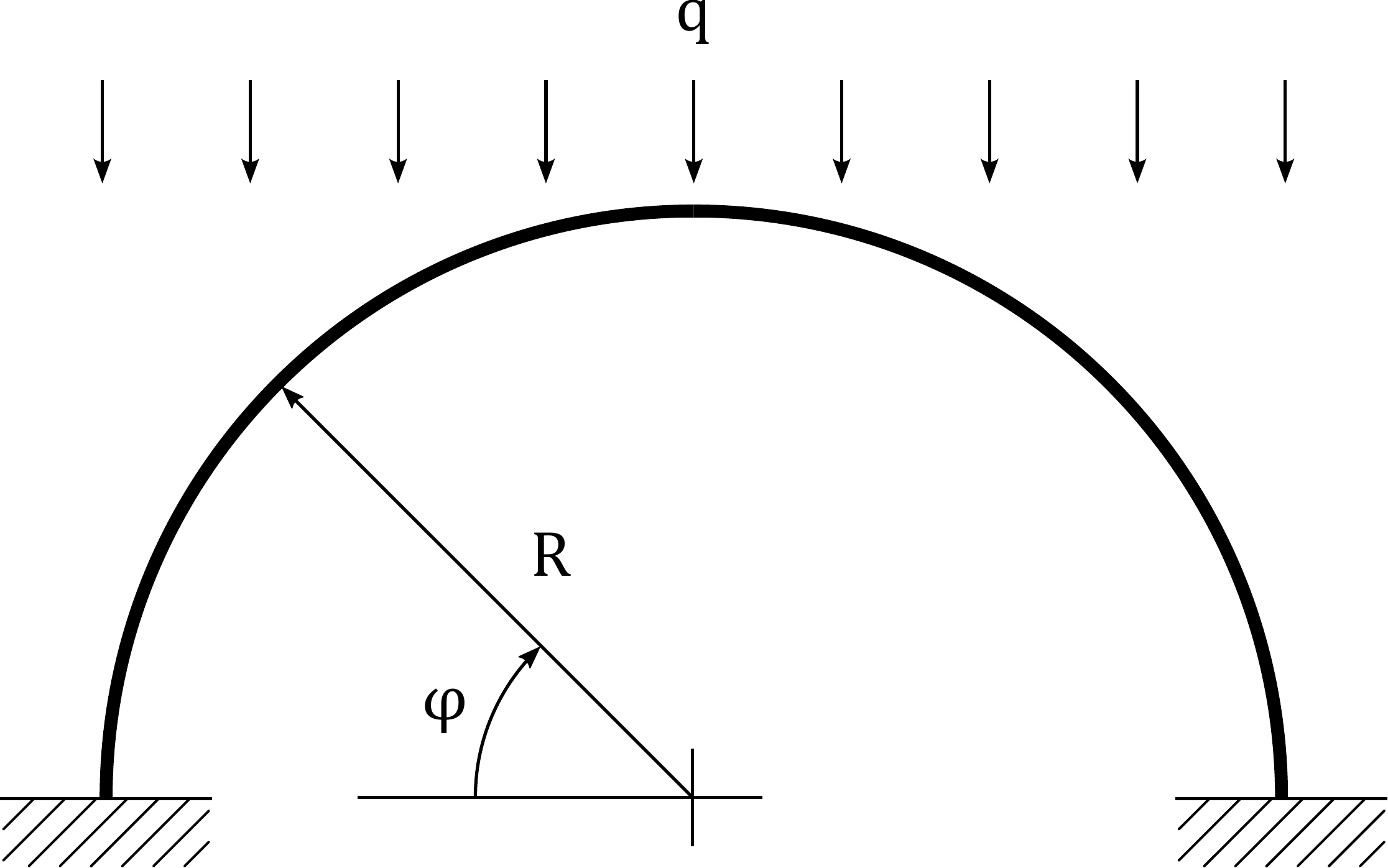}} \hspace*{+10mm}
 \subfigure[After applying symmetry]{\includegraphics[scale=0.3]{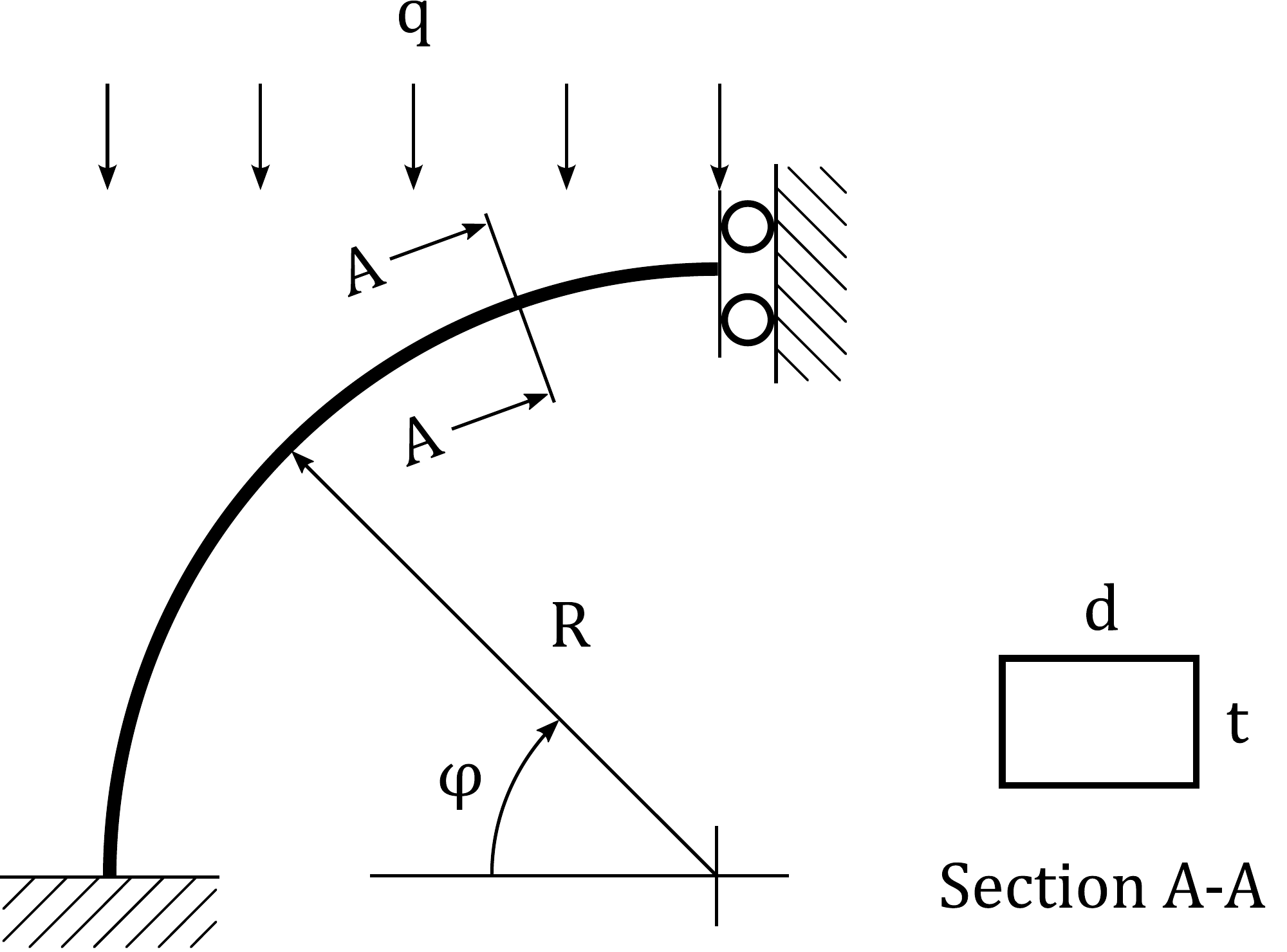}} \\
\caption{Geometry, boundary conditions, and applied load for the clamped-clamped semi-circular arch under a distributed load. a) Before applying symmetry. b) After applying symmetry.} 
\label{halfcirclegeo}
\end{figure}

The second numerical investigation considers a clamped-clamped semi-circular arch under a distributed load as shown in Fig. \ref{halfcirclegeo} a). Given the symmetry of this problem, we solve half of the arch with the appropriate symmetry boundary conditions shown in Fig. \ref{halfcirclegeo} b). The next values are used in this example:
\begin{equation} 
 q =  10^6 t^3,  \quad R = 10.0,  \quad E = 2.1 \times 10^{11},  \quad d = 0.1 \text{.} 
\end{equation}
In order to consider different values of the slenderness ratio, three values are used for the thickness in this example, namely, $t=0.1$, $t=0.01$, and $t=0.001$. Since the cross section of the rod is a rectangle, $A = td$ and $I= t^3d/12$. Note that $q$ is a distributed load per unit of horizontal length while $\vec f$ in Eq. \eqref{Wext} is a distributed load vector per unit length along the axis of the rod. Therefore, $\vec f = (0 , - q \sin (\varphi))$, where the angle $\varphi$ is shown in Fig. \ref{halfcirclegeo} b).

\begin{figure} [h!] 
 \centering
\subfigure[NURBS and CAS]{\includegraphics[scale=0.52]{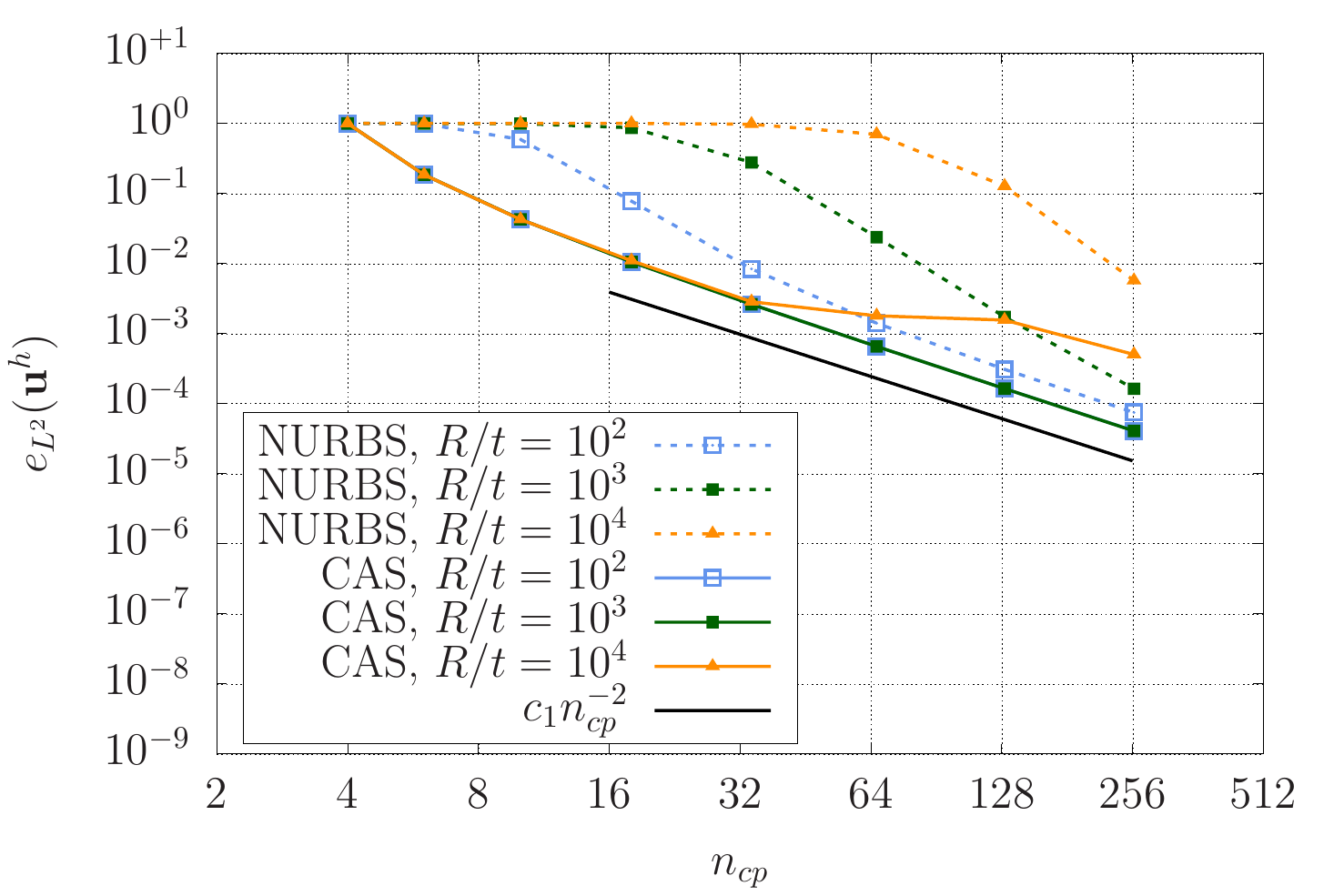}}
 \subfigure[Local $\bar{B}$ and local ANS]{\includegraphics[scale=0.52]{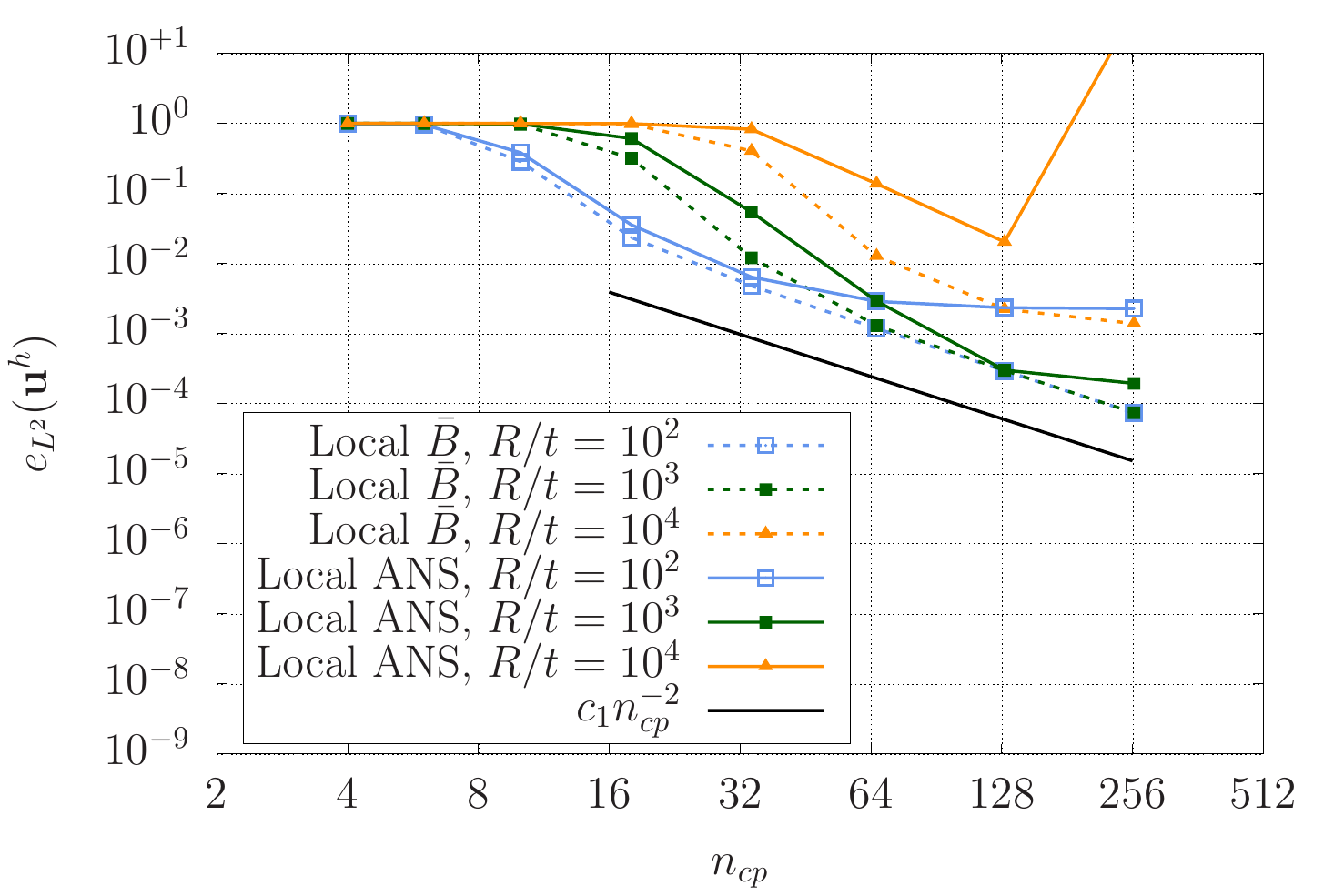}} \\
\subfigure[NURBS and CAS]{\includegraphics[scale=0.52]{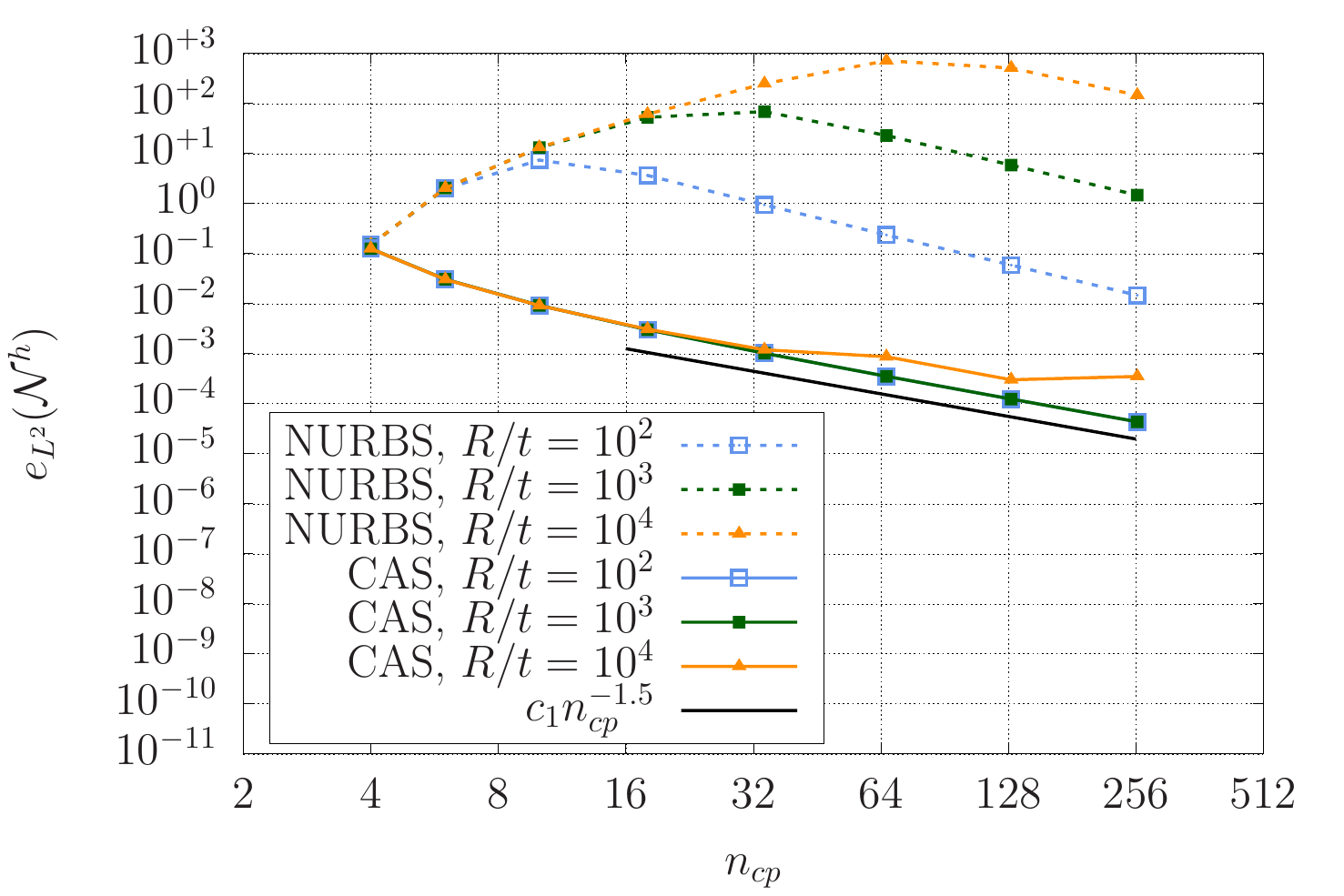}}
 \subfigure[Local $\bar{B}$ and local ANS]{\includegraphics[scale=0.52]{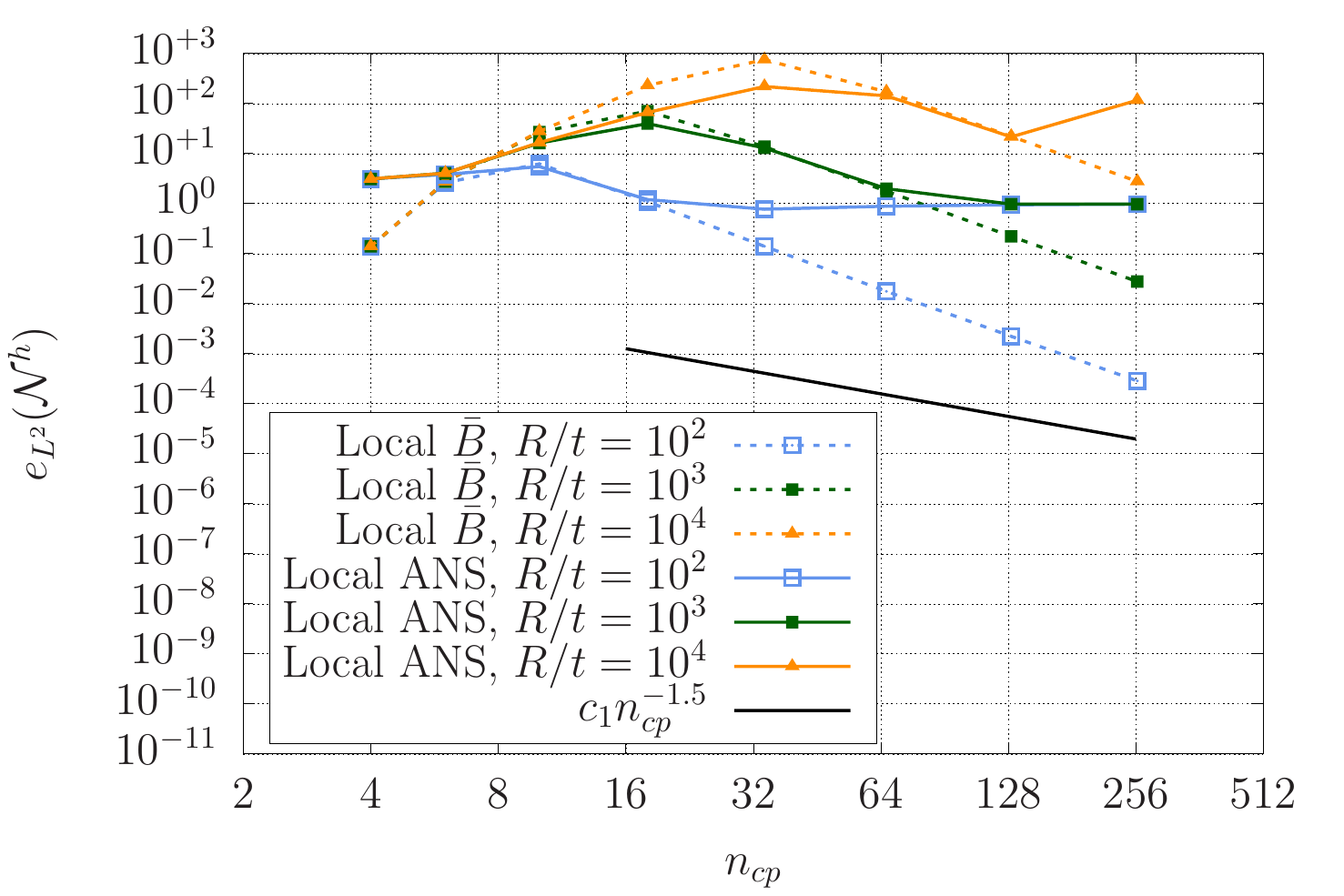}} \\
 \subfigure[NURBS and CAS]{\includegraphics[scale=0.52]{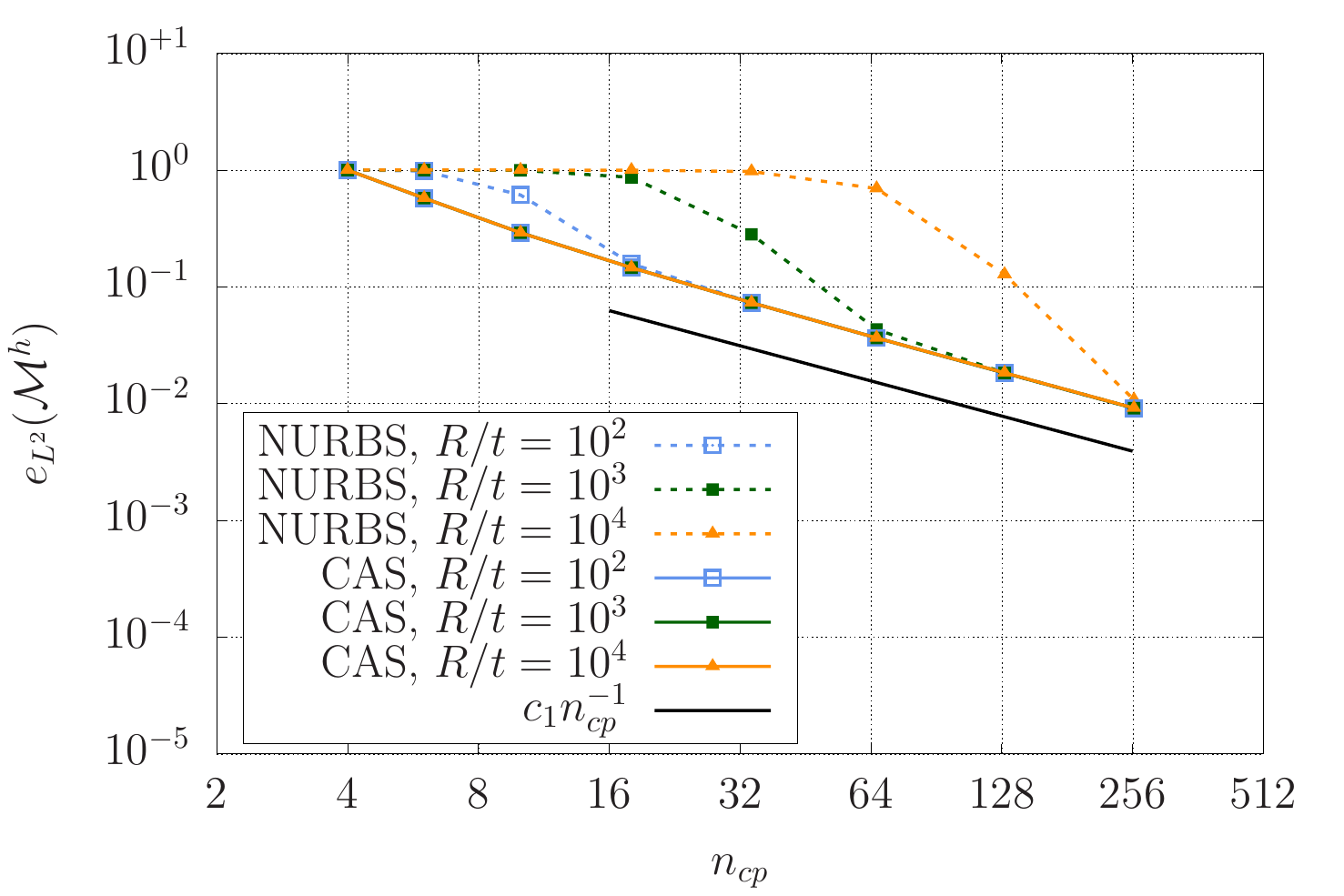}}
 \subfigure[Local $\bar{B}$ and local ANS]{\includegraphics[scale=0.52]{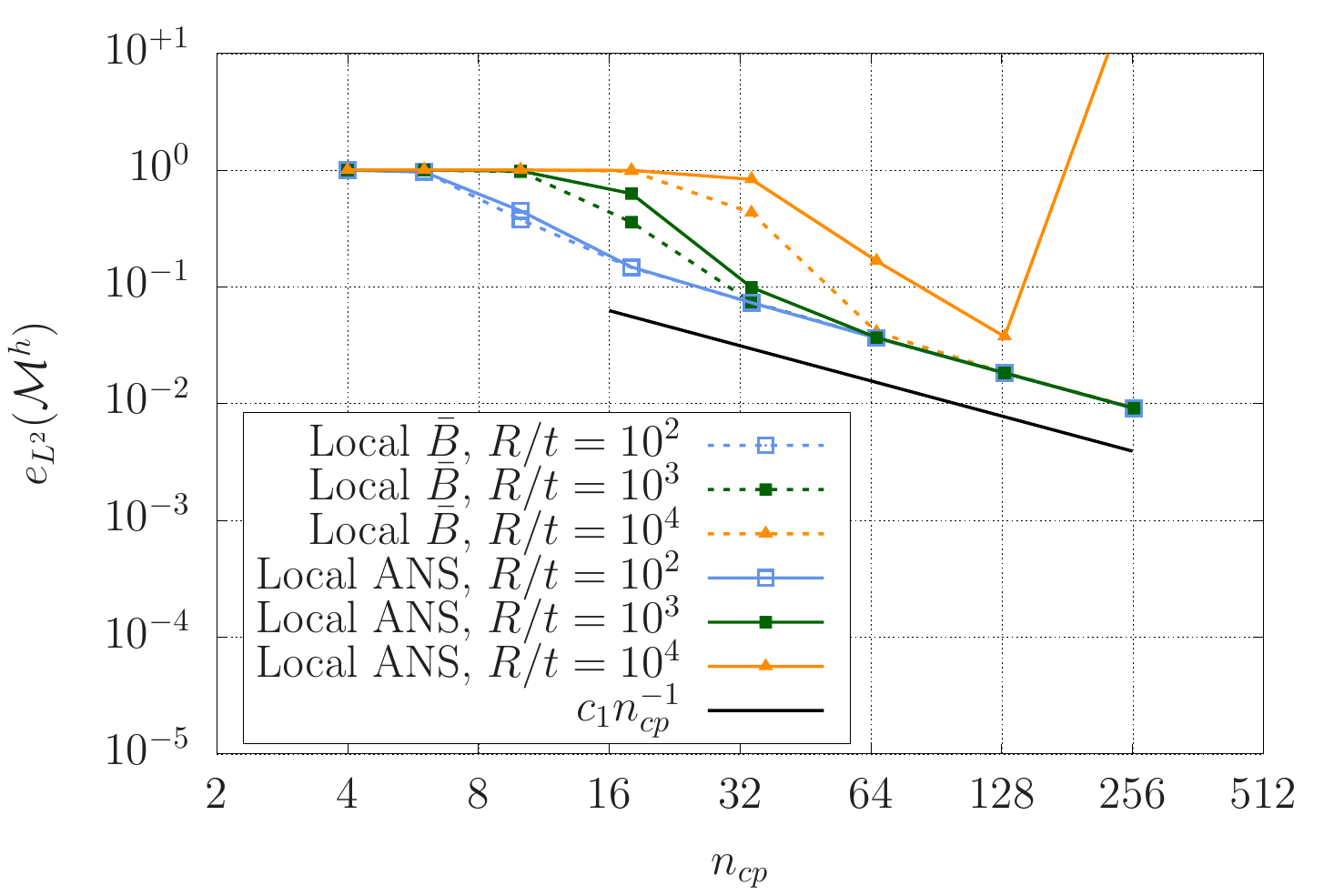}} \\
\caption{(Color online) Clamped-clamped semi-circular arch under a distributed load. Convergence of the displacement vector, the membrane force, and the bending moment using NURBS elements, CAS elements, local $\bar{B}$ elements, and local ANS elements. For any of the slenderness ratios considered, CAS elements are the only element type that overcomes locking.}
\label{halfcircleconvergence}
\end{figure}

\begin{figure} [h!] 
 \centering
\subfigure[Local $\bar{B}$ and local ANS]{\includegraphics[scale=0.52]{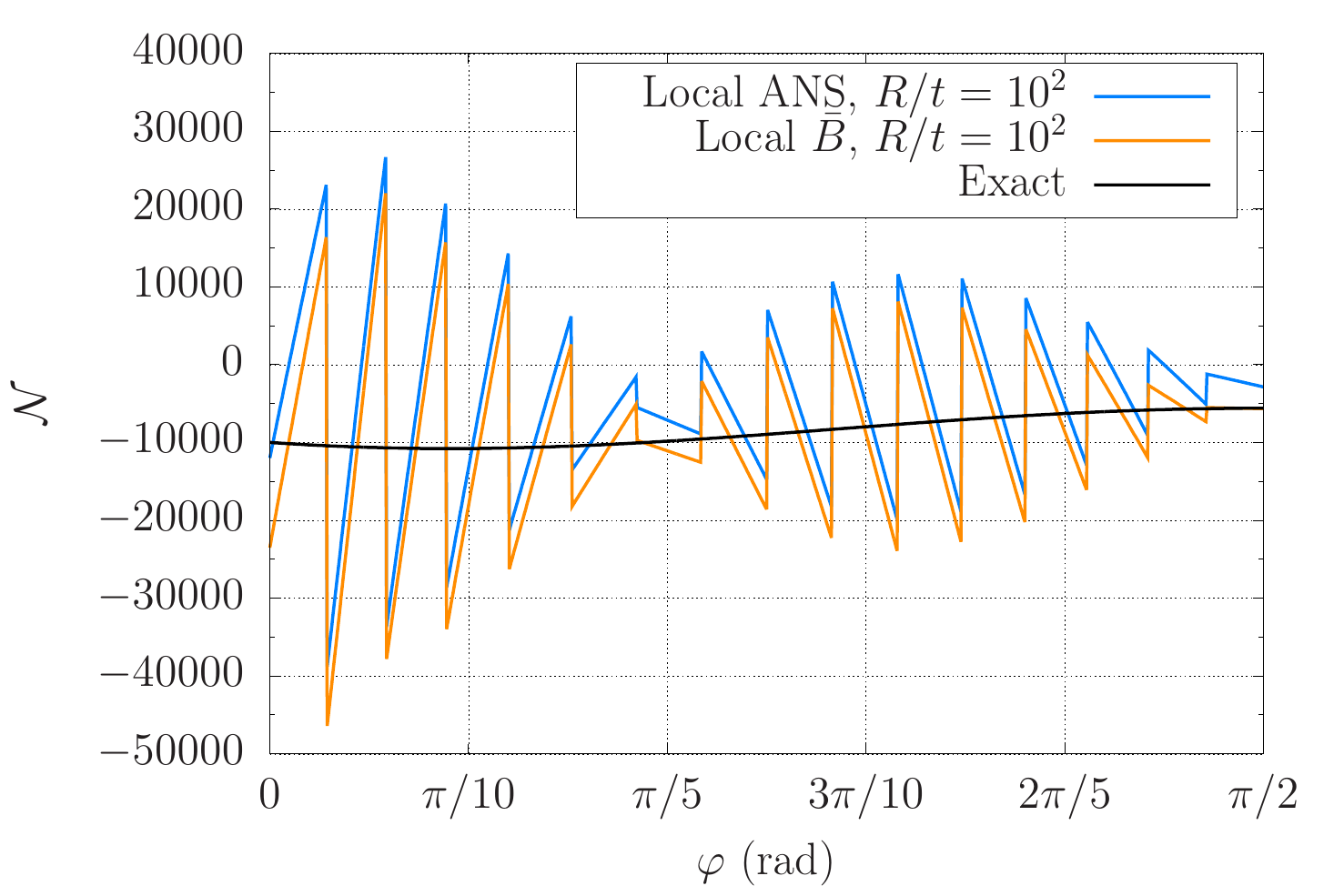}}
 \subfigure[Global $\bar{B}$ and CAS]{\includegraphics[scale=0.52]{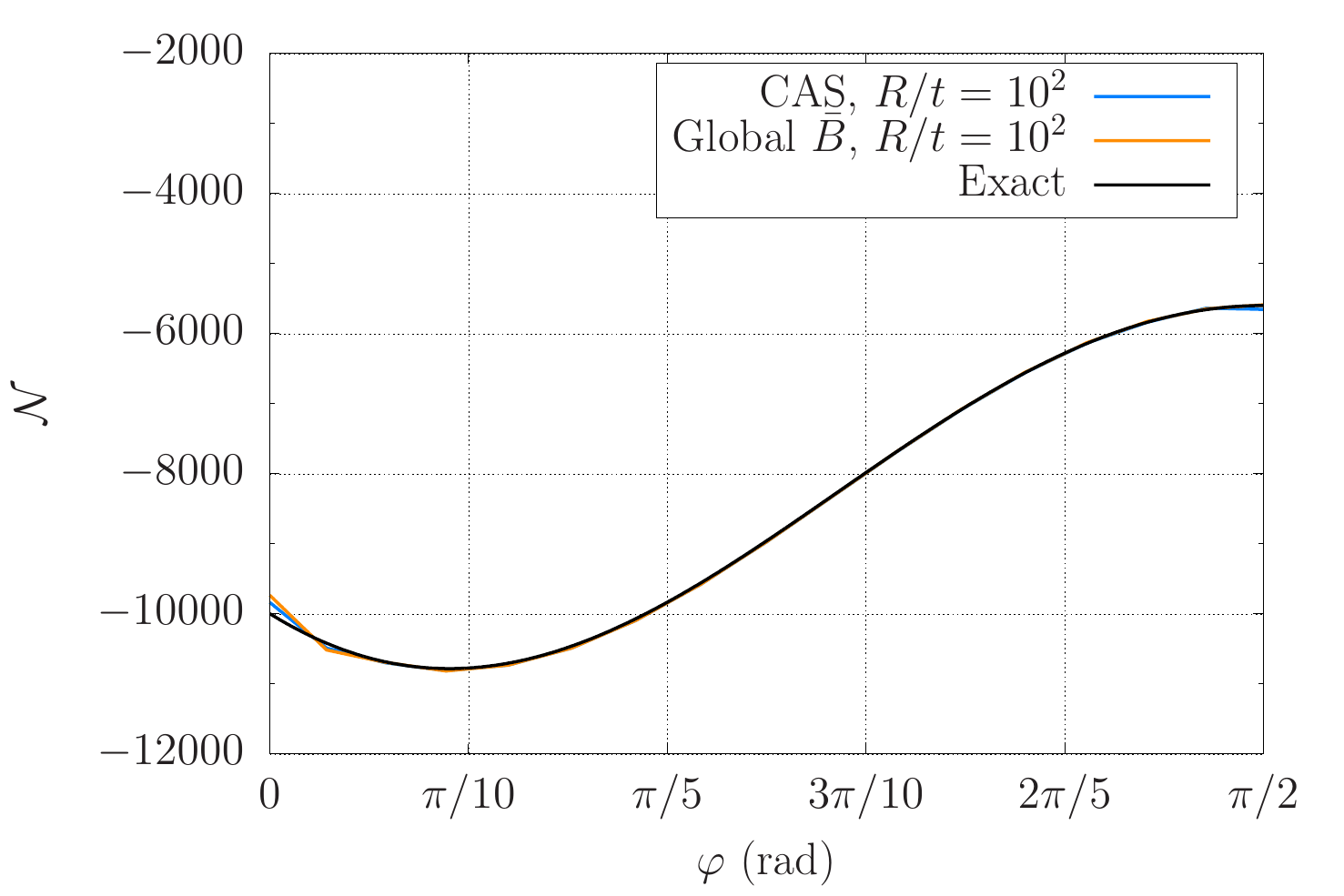}} \\
\caption{(Color online) Membrane force of the clamped-clamped semi-circular arch under a distributed load using local $\bar{B}$ elements, local ANS elements, CAS elements, and the global $\bar{B}$ method. The mesh has 16 elements and $R/t = 10^2$. The numerical solution using local $\bar{B}$ elements and local ANS elements have spurious oscillations whose amplitude is more than four times greater than the maximum exact membrane force of this problem. The numerical solution using CAS elements and the global $\bar{B}$ method overlap. Note the different vertical scale used in each plot.}
\label{halfcirclemfdistributions1}
\end{figure}

In \cite{cazzani2016isogeometric}, the exact solution to this problem is given as
\begin{align}
   u_t &=  A_{1}\left[c_{1} \varphi \sin (\varphi) - c_{3} R(1-\cos (\varphi) )\right]-A_{2} c_{3}(\varphi-\sin (\varphi))+ \nonumber \\ & A_{3} \sin (\varphi) - q R\left[\sin (2 \varphi) \left(2 / 3 c_{1}-1 / 6 c_{2}-1 / 8 c_{3} R\right)-\varphi c_{3} R / 2\right]   \text{,} \label{e1} \\
   u_n &=  A_{1}\left[c_{1}(\varphi \cos (\varphi) - \sin (\varphi) )+c_{2} \sin (\varphi) -c_{3} R \sin (\varphi) \right]-A_{2} c_{3}(1-\cos (\varphi) )+ \nonumber \\ & A_{3} \cos (\varphi) + q R\left[c_{1}-1 / 2 c_{2}+1 / 2 c_{3} R-\cos (2 \varphi) \left(1 / 3 c_{1}+1 / 6 c_{2}-1 / 4 c_{3} R\right)\right] \text{,} \label{e2} \\
   \mathcal{N} &= A_{1} \sin (\varphi) - q R \cos ^{2} (\varphi)   \text{,}  \label{e3} \\
   \mathcal{M} &=  A_{1} R \sin (\varphi) + A_{2} - q R^{2} / 2(1+1 / 2 \cos (2 \varphi))   \text{,} \label{e4}
\end{align}
\noindent with
\begin{equation}
   c_1 = \frac{1}{2}\left(\frac{R}{E A}+\frac{R^{3}}{E I}\right)  \text{,} \;
   c_2 = \frac{R^{3}}{E I}   \text{,}  \;
   c_3 = \frac{R^{2}}{E I}   \text{,}   
\end{equation}
\begin{align}
   A_1 &= \frac{8 \pi q\left(c_{1}-c_{2}\right)+3 \pi q R c_{3}}{6 \pi^{2}\left(c_{1} / R\right)-24 c_{3}}   \text{,}  \\
   A_2 &= \frac{q R^{2}}{2}-\frac{16 \pi q R\left(c_{1}-c_{2}\right)+6 \pi q R^{2} c_{3}}{6 \pi^{3}\left(c_{1} / R\right)-24 \pi c_{3}}   \text{,}  \\
   A_3 &= -\frac{2 q R\left(c_{1}-c_{2}\right)}{3}-\frac{3 q R^{2} c_{3}}{4}   \text{,}   
\end{align}
\noindent where $u_t$ and $u_n$ are the tangential and normal displacements to the axis of the rod, respectively. Thus, $u_x = u_t \sin (\varphi) + u_n \cos (\varphi)$ and $u_y = u_t \cos (\varphi) - u_n \sin (\varphi)$, where $u_x$ and $u_y$ are the $x$ and $y$ components of the displacement vector $\mathbf{u}$, respectively.

\begin{figure} [h!] 
 \centering
\subfigure[NURBS]{\includegraphics[scale=0.52]{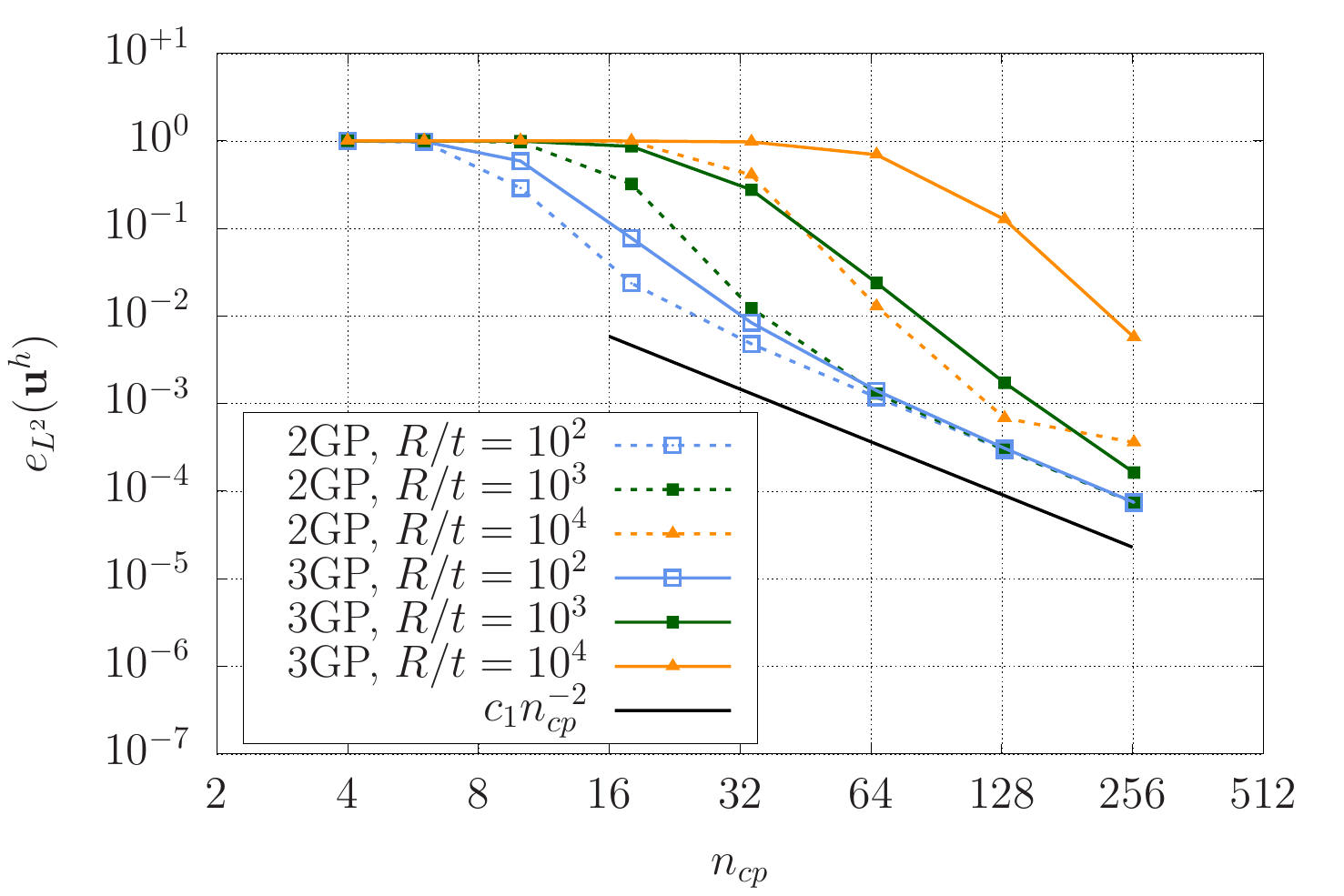}}
 \subfigure[CAS]{\includegraphics[scale=0.52]{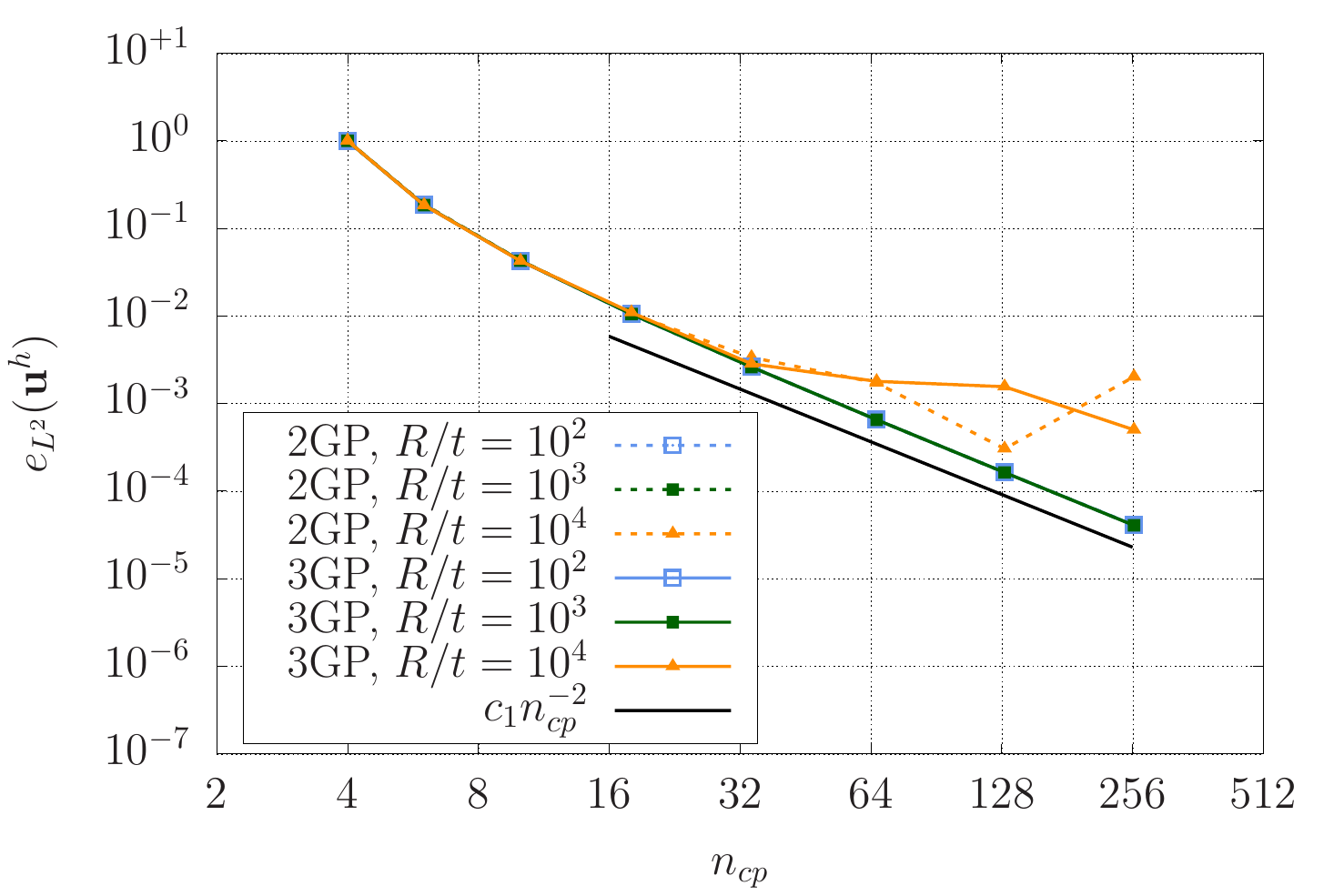}} \\
\subfigure[NURBS]{\includegraphics[scale=0.52]{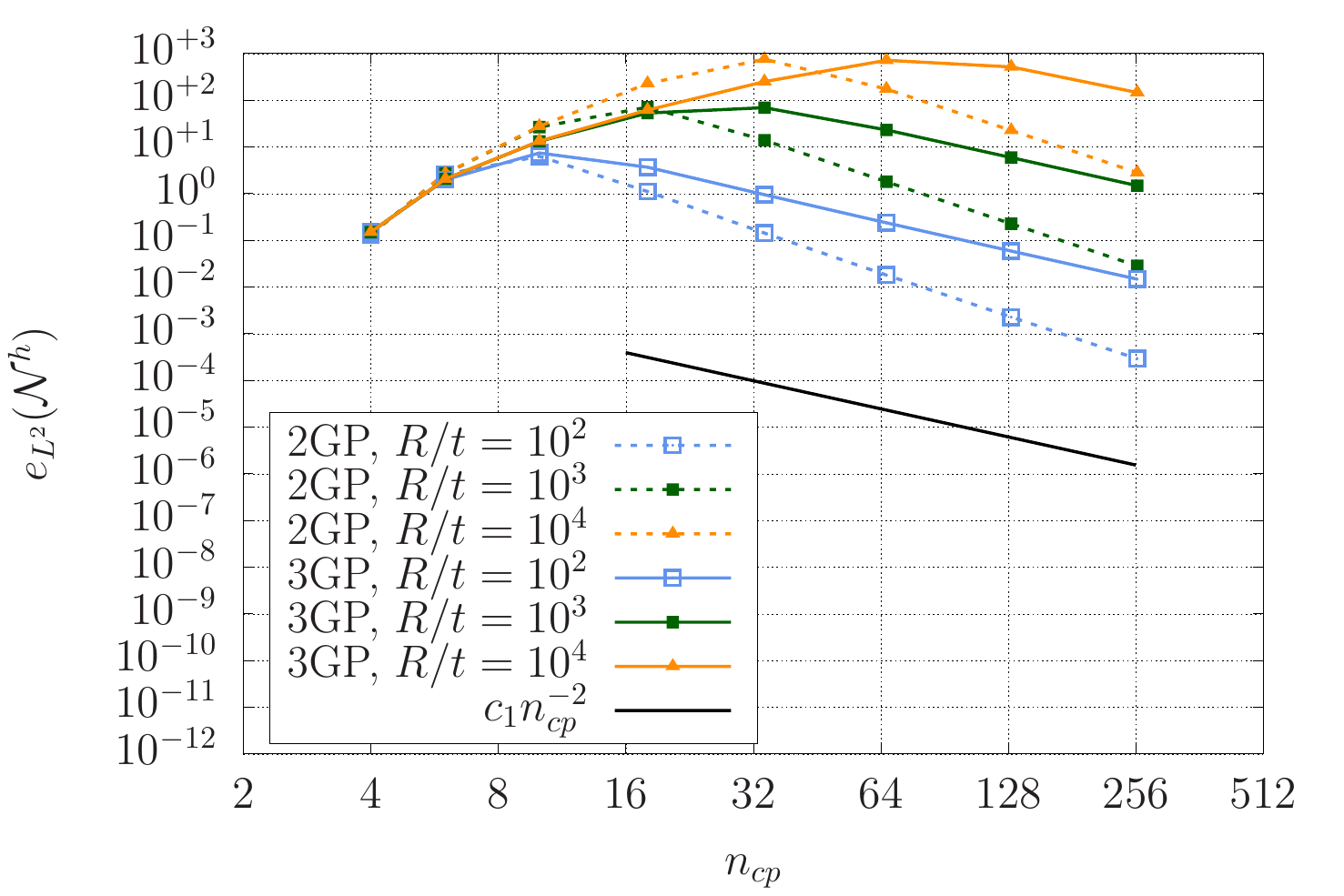}}
 \subfigure[CAS]{\includegraphics[scale=0.52]{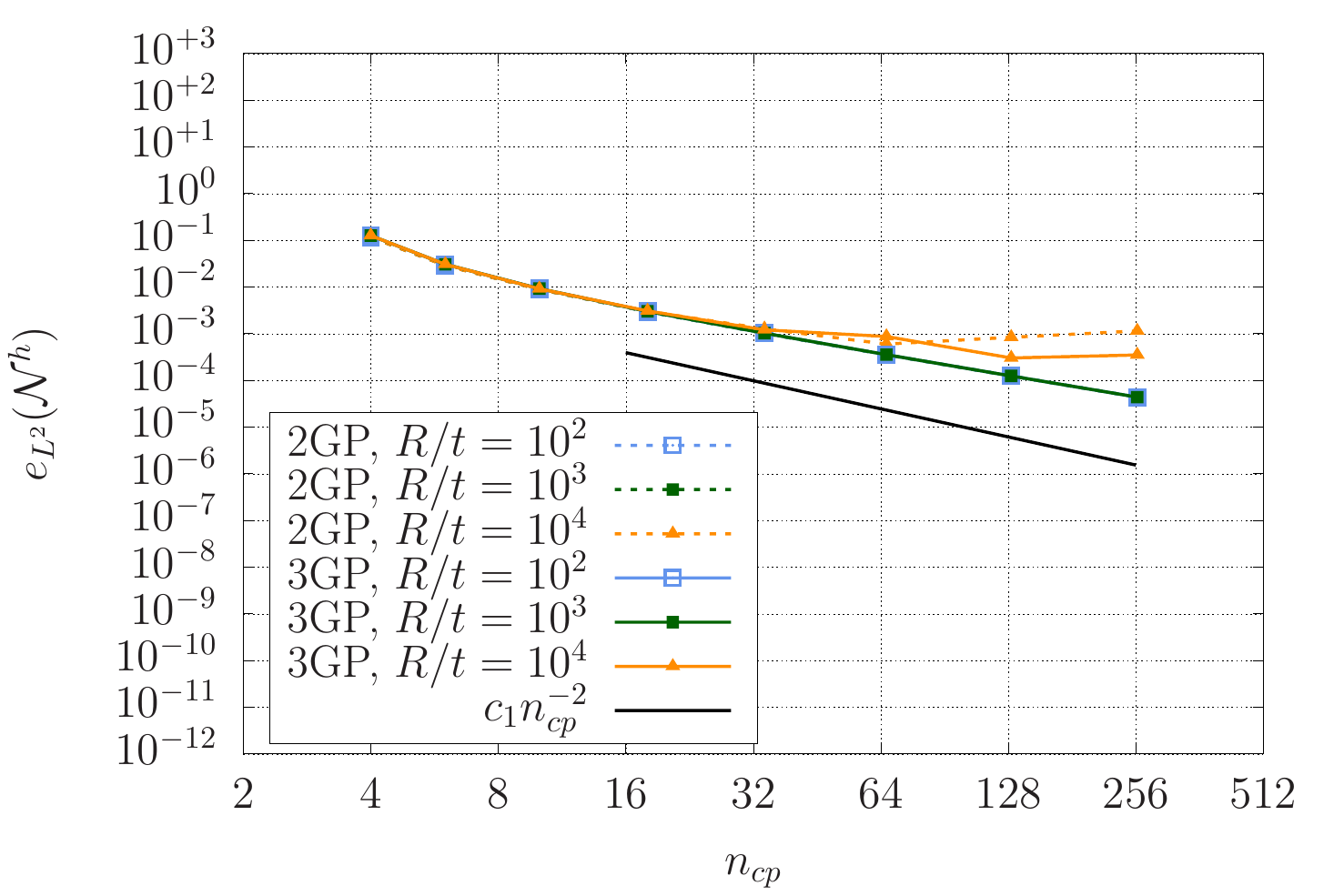}} \\
 \subfigure[NURBS]{\includegraphics[scale=0.52]{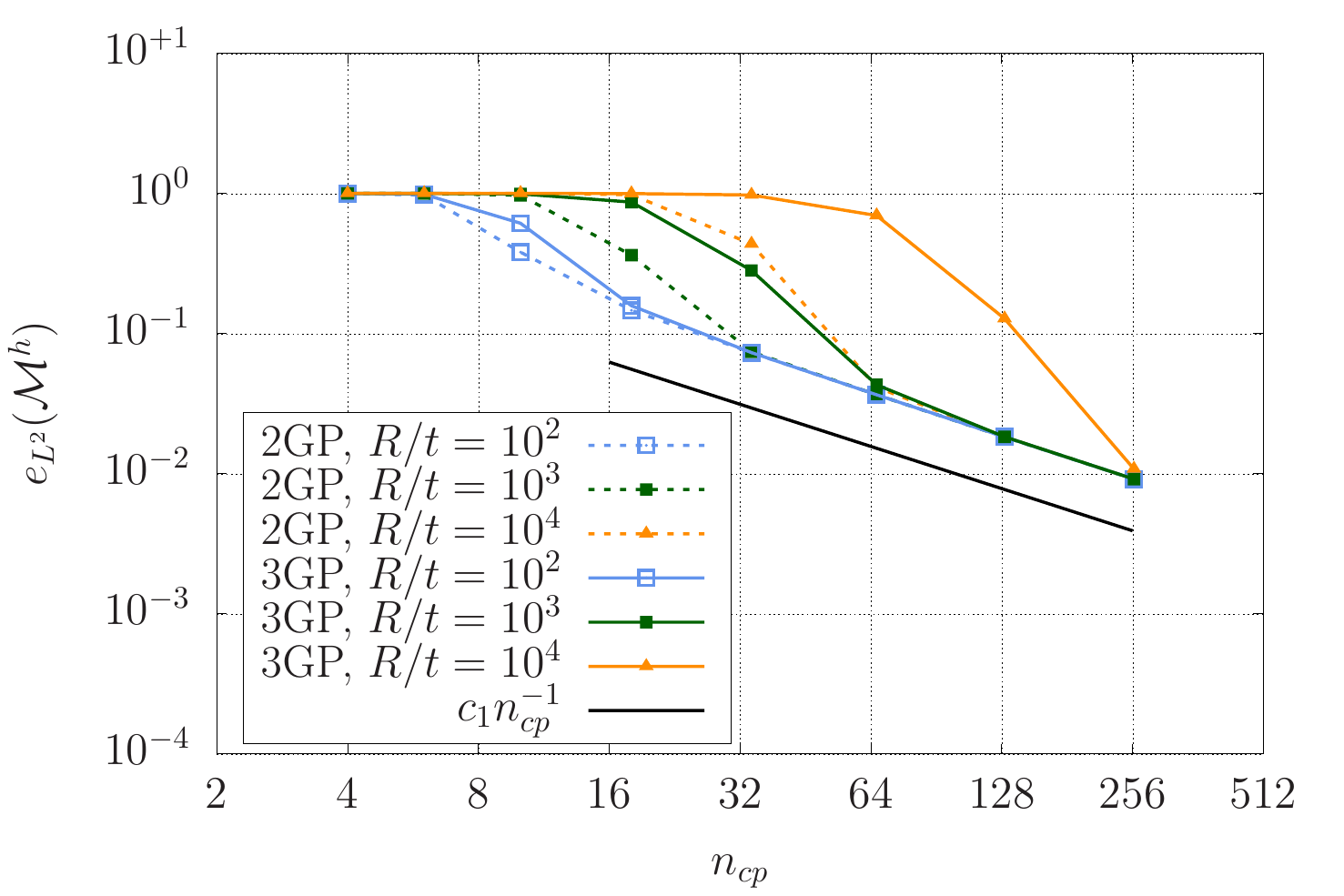}}
 \subfigure[CAS]{\includegraphics[scale=0.52]{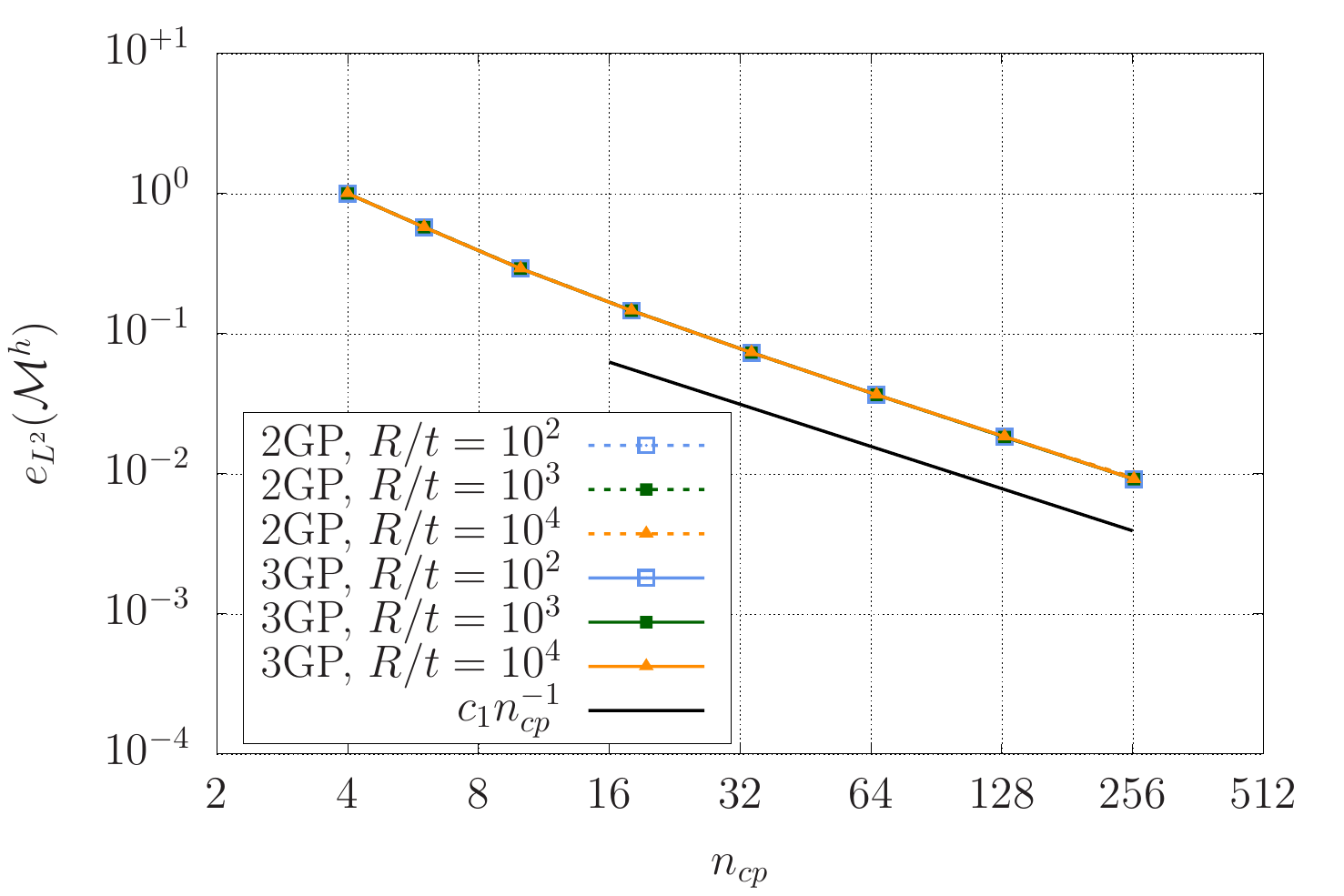}} \\
\caption{(Color online) Clamped-clamped semi-circular arch under a distributed load. Convergence comparison of NURBS elements and CAS elements using either 3 Gauss points or 2 Gauss points. For NURBS elements, the results improve when using 2 Gauss points, but locking is still present for any of the slenderness ratios considered. For CAS elements, the same level of accuracy is obtained with 2 and 3 Gauss points.}
\label{halfcircle2GP3GP}
\end{figure}

As in the preceding section, we start our convergence study with a uniform mesh composed of two quadratic elements and then perform uniform $h$-refinement seven times. Fig. \ref{halfcircleconvergence} plot the convergence in $L^2$ norm of the displacement vector, the membrane force, and the bending moment using local $\bar{B}$ elements \cite{greco2017efficient, hu2016order, antolin2020simple}, local ANS elements \cite{caseiro2014assumed, caseiro2015assumed, greco2017efficient}, NURBS elements, and CAS elements. When applied to linear plane Kirchhoff rods discretized using quadratic NURBS, local $\bar{B}$ elements perform a $L^2$ projection of the membrane strain at the element level into the space of linear Lagrange polynomials \cite{greco2017efficient, antolin2020simple} and local ANS elements collocate the membrane strain at the element level into the space of linear Lagrange polynomials using a Gauss-Legendre quadrature rule with 2 integration points as collocation points \cite{caseiro2014assumed, caseiro2015assumed, greco2017efficient}. Both local $\bar{B}$ elements and local ANS elements result in discontinuous membrane strains across element boundaries. As shown in Fig. \ref{halfcircleconvergence}, the convergence curves of CAS elements for $R/t=10^2$, $10^3$, and $10^4$ overlap (with the exception of the finer meshes for $R/t=10^4$ which slightly deteriorate) while the convergence of NURBS elements, local $\bar{B}$ elements, and local ANS elements heavily deteriorates as $R/t$ increases. Neither local $\bar{B}$ elements nor local ANS elements overcome locking for any slenderness ratio, which is consistent with the results included in \cite{greco2017efficient}. Thus, these two element types are still locking-prone discretizations. This can be easily seen by plotting the distribution of the membrane force. Even for $R/t=100$ (note that structural theories based on Kirchhoff assumptions are supposed to be used only for $R/t \geq 20$ \cite{bischoff2004models}) and using a moderate mesh resolution (16 elements), local $\bar{B}$ elements and local ANS elements have spurious oscillations whose amplitude is more than four times greater than the maximum exact membrane force of this problem as shown in Fig. \ref{halfcirclemfdistributions1}.

\begin{figure} [h!] 
 \centering
\subfigure[NURBS elements]{\includegraphics[scale=0.52]{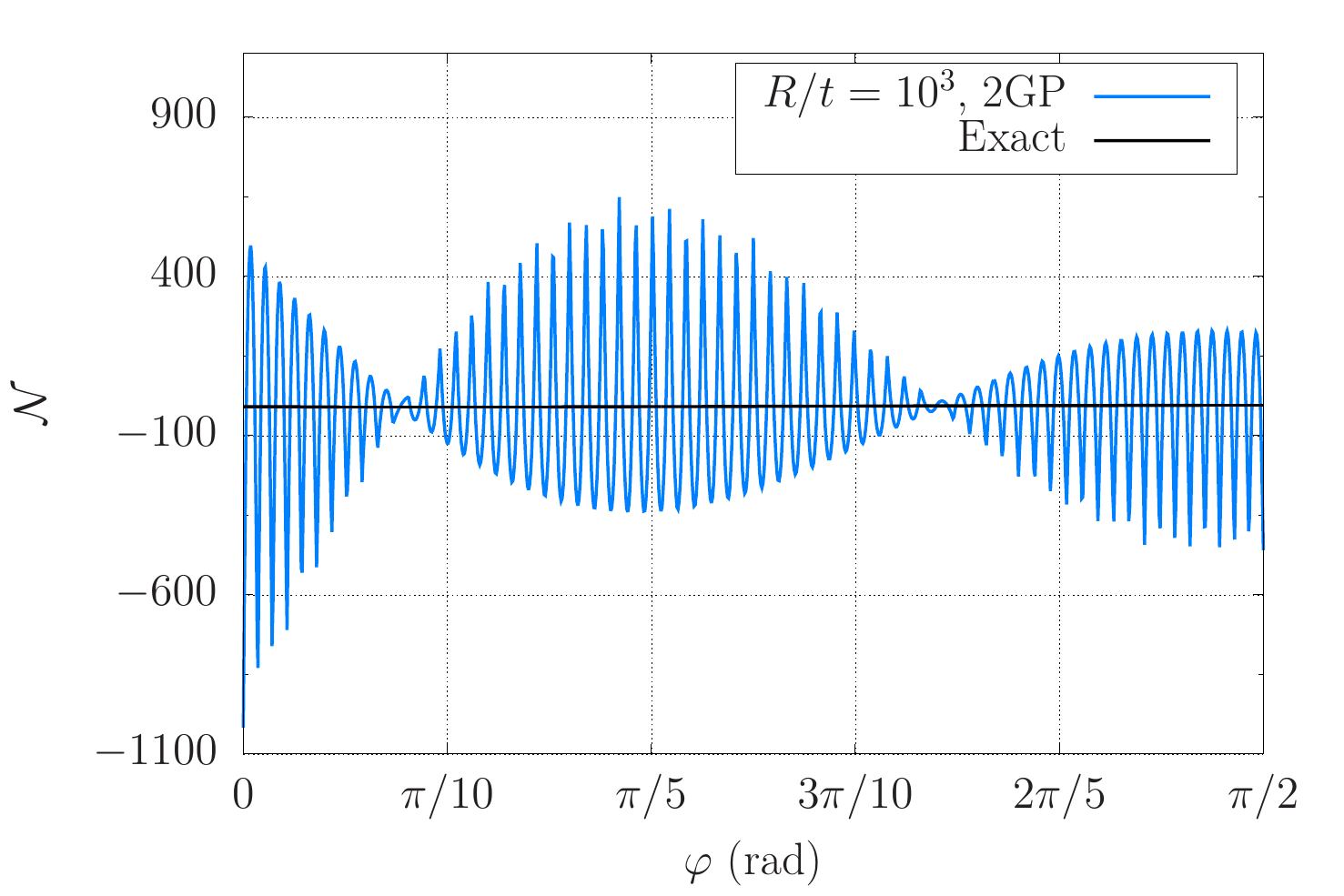}}
 \subfigure[CAS elements]{\includegraphics[scale=0.52]{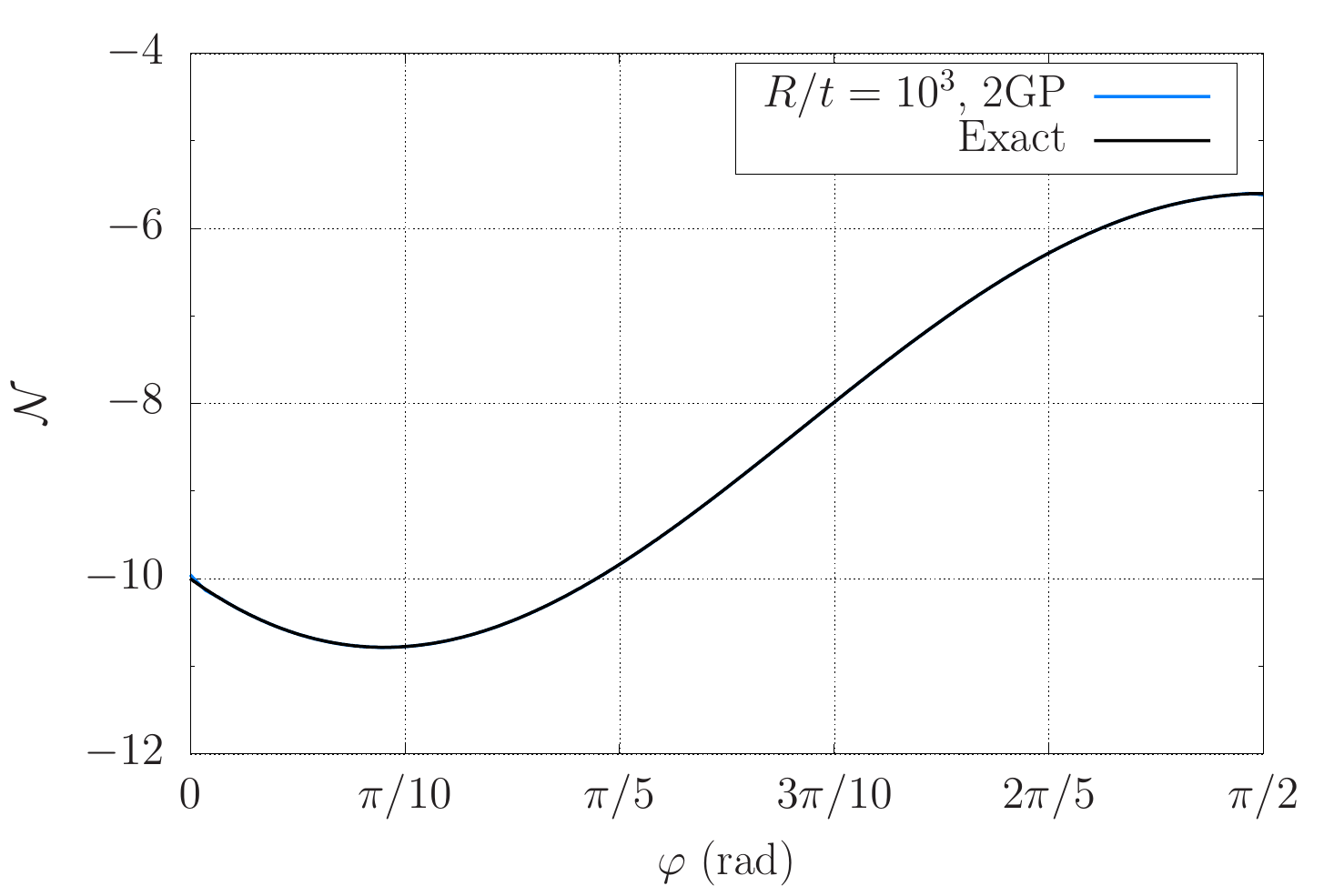}} \\
\caption{(Color online) Membrane force of the clamped-clamped semi-circular arch under a distributed load using NURBS and CAS elements with 2 Gauss points. The mesh has 64 elements and $R/t = 10^3$. The numerical solution using CAS elements overlaps with the exact solution. The numerical solution using NURBS elements has spurious oscillations whose amplitude is more than an order of magnitude larger than the maximum exact membrane force of this problem. Note the different vertical scale used in each plot.}
\label{halfcirclemfdistributions2}
\end{figure}

As shown in \cite{leonetti2019simplified}, when using reduced integration rules at the patch level, the continuity of the integration space cannot be greater than the continuity of the strains in order to exactly reproduce constant stress states which is needed for having accurate results. When discretizing fourth-order structural models with $C^1$-continuous quadratic NURBS, the strains are discontinuous across elements. Thus, the continuity of the integration space has to be discontinuous across elements in order to exactly reproduce constant stress states. In this case, the integration is no longer patch-wise but is carried out at the element level and coincides with the standard Gauss-Legendre quadrature. Therefore, using both NURBS and CAS elements, we next solve this problem using 2 Gauss points (2GP) instead of 3 Gauss points (3GP) to compute all integrals. Fig. \ref{halfcircle2GP3GP} plots the convergence in $L^2$ norm of the displacement vector, the membrane force, and the bending moment obtained with 2 and 3 Gauss points. As shown in Fig. \ref{halfcircle2GP3GP}, CAS elements result in essentially the same accuracy regardless of whether 2 Gauss points or 3 Gauss points are used. Thus, the use of 2 Gauss points is a potential option to decrease the computational time when using CAS elements. As Fig.  \ref{halfcircle2GP3GP} shows, the accuracy of NURBS elements improves when 2 Gauss points are used instead of 3 Gauss points. However, the convergence of NURBS elements still heavily deteriorates as $R/t$ increases. In other words, NURBS elements with 2 Gauss points are still a locking-prone discretization. This can be easily seen by plotting the distribution of the membrane force. Even using a fine mesh (64 elements) and a moderate $R/t$ ratio ($R/t=10^3$), NURBS elements with 2 Gauss points result in a membrane force distribution with spurious oscillations whose amplitude is more than an order of magnitude larger than the maximum exact membrane force as shown in Fig. \ref{halfcirclemfdistributions2}.

	\begin{figure} [h!] 
\centering
\includegraphics[width=8cm]{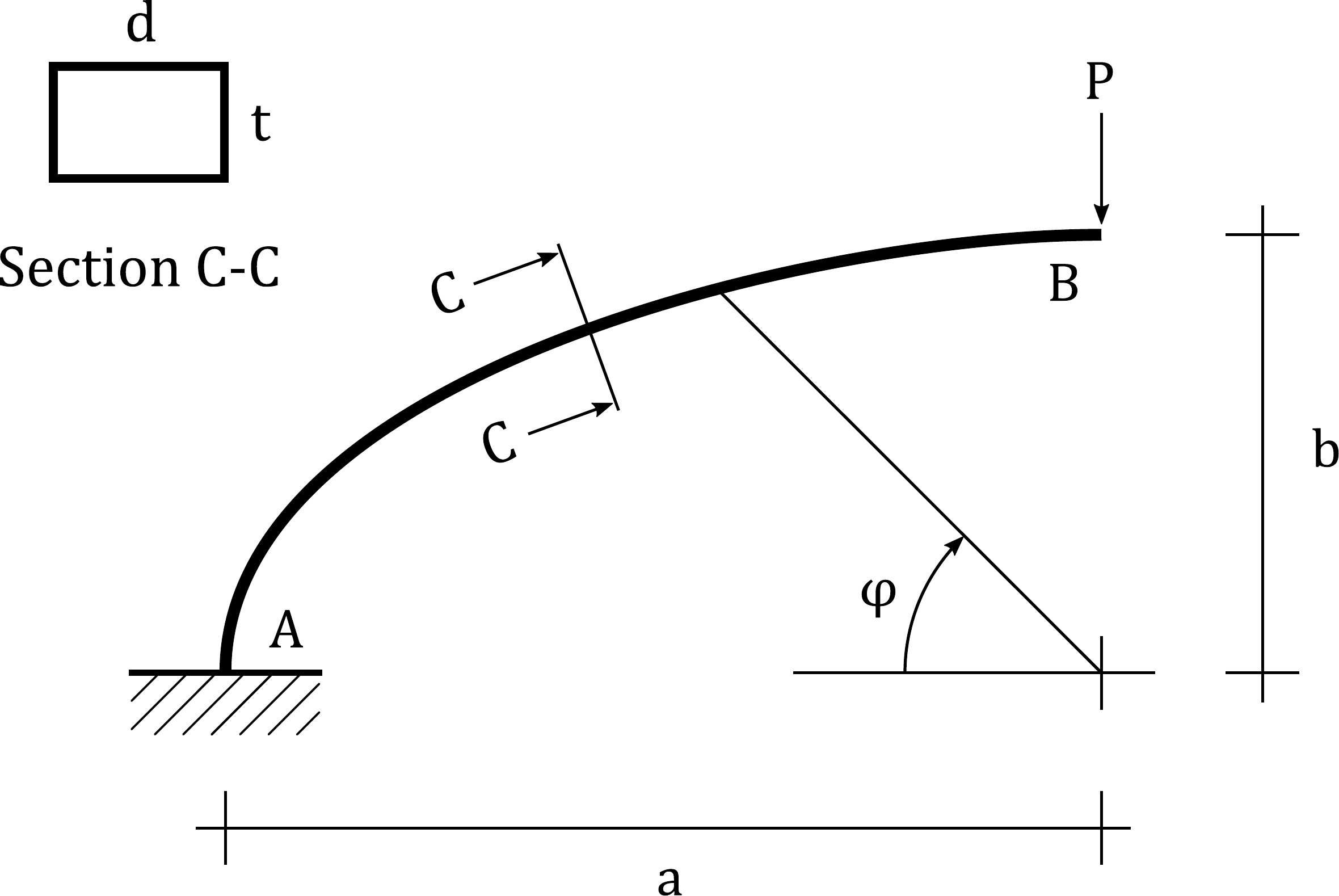}
\caption{Geometry, boundary conditions, and applied load for the clamped elliptical arch under a point load at the free end.} 
\label{archgeo}
\end{figure}

% add comment about patchwise integration

\subsection{Clamped elliptical arch under a point load at the free end}

% F =  1.0 \; \si{kN},  \quad E= 1.999 \times 10^8 \; \si{kN/m^{2}},  \quad R = 1.0 \; \si{m},  \quad b = 0.1 \; \si{m} \text{.} 

	\begin{figure} [h!] 
 \centering
\subfigure[Horizontal displacement]{\includegraphics[scale=0.53]{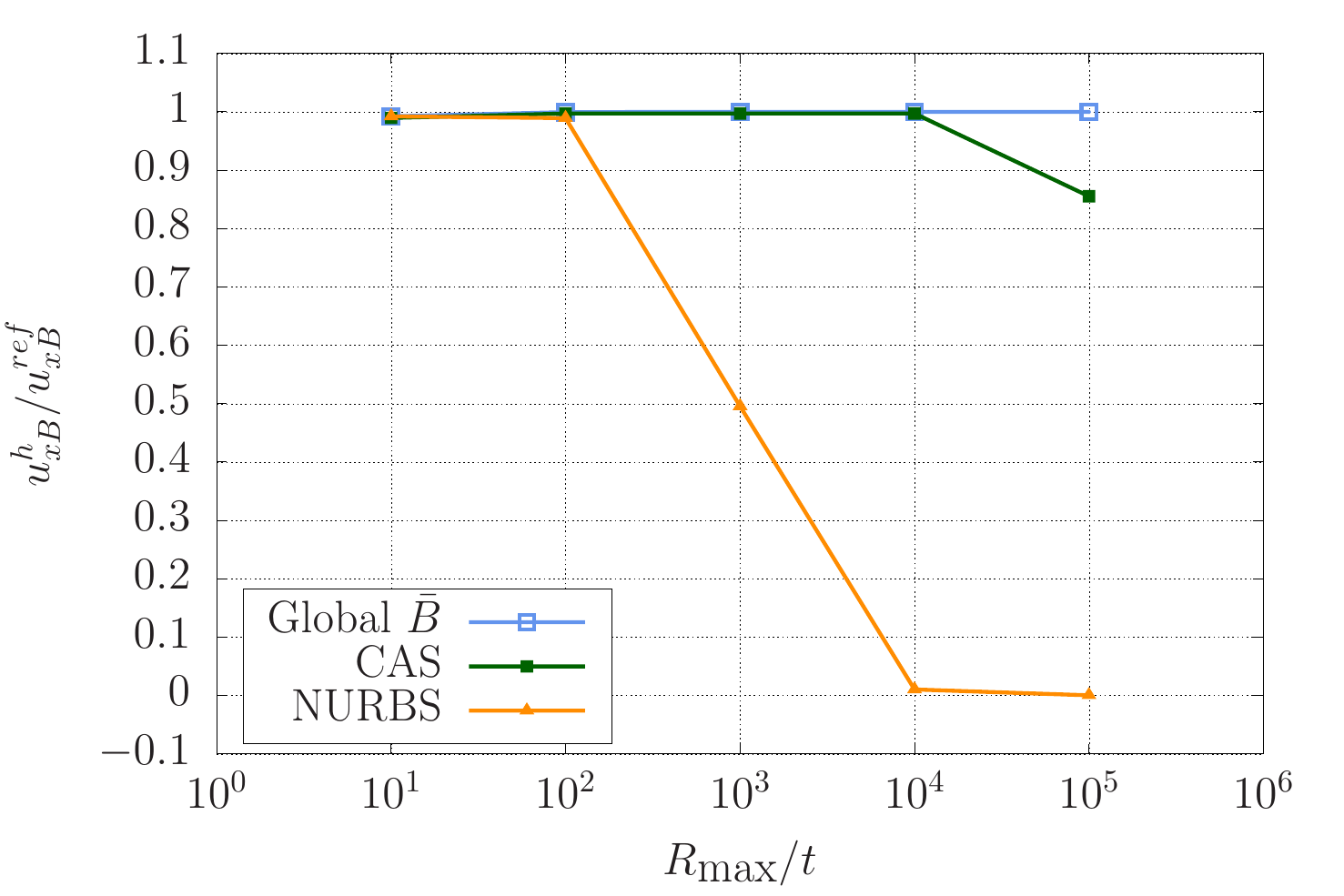}}
 \subfigure[Vertical displacement]{\includegraphics[scale=0.53]{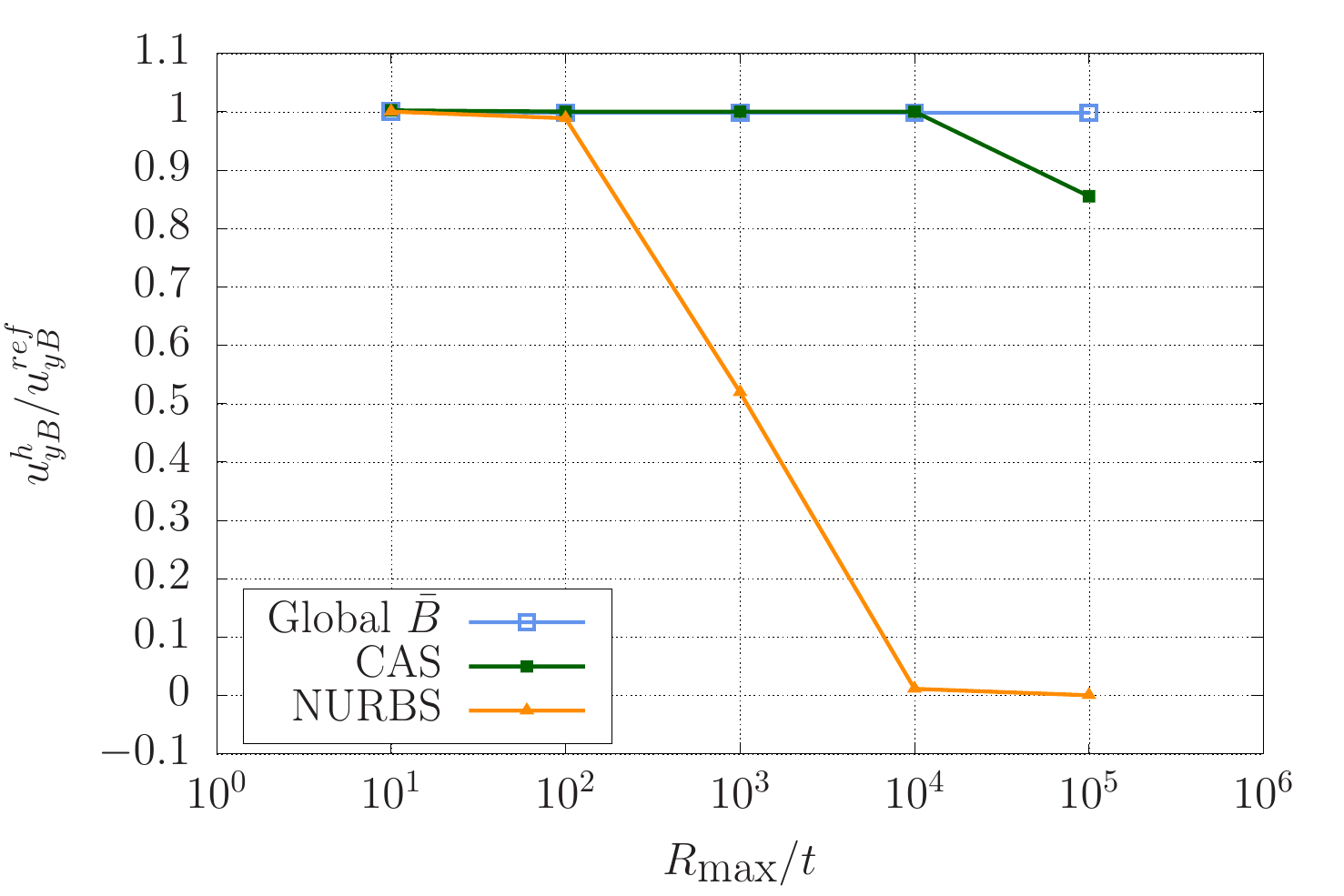}} \\
 \subfigure[Membrane force]{\includegraphics[scale=0.53]{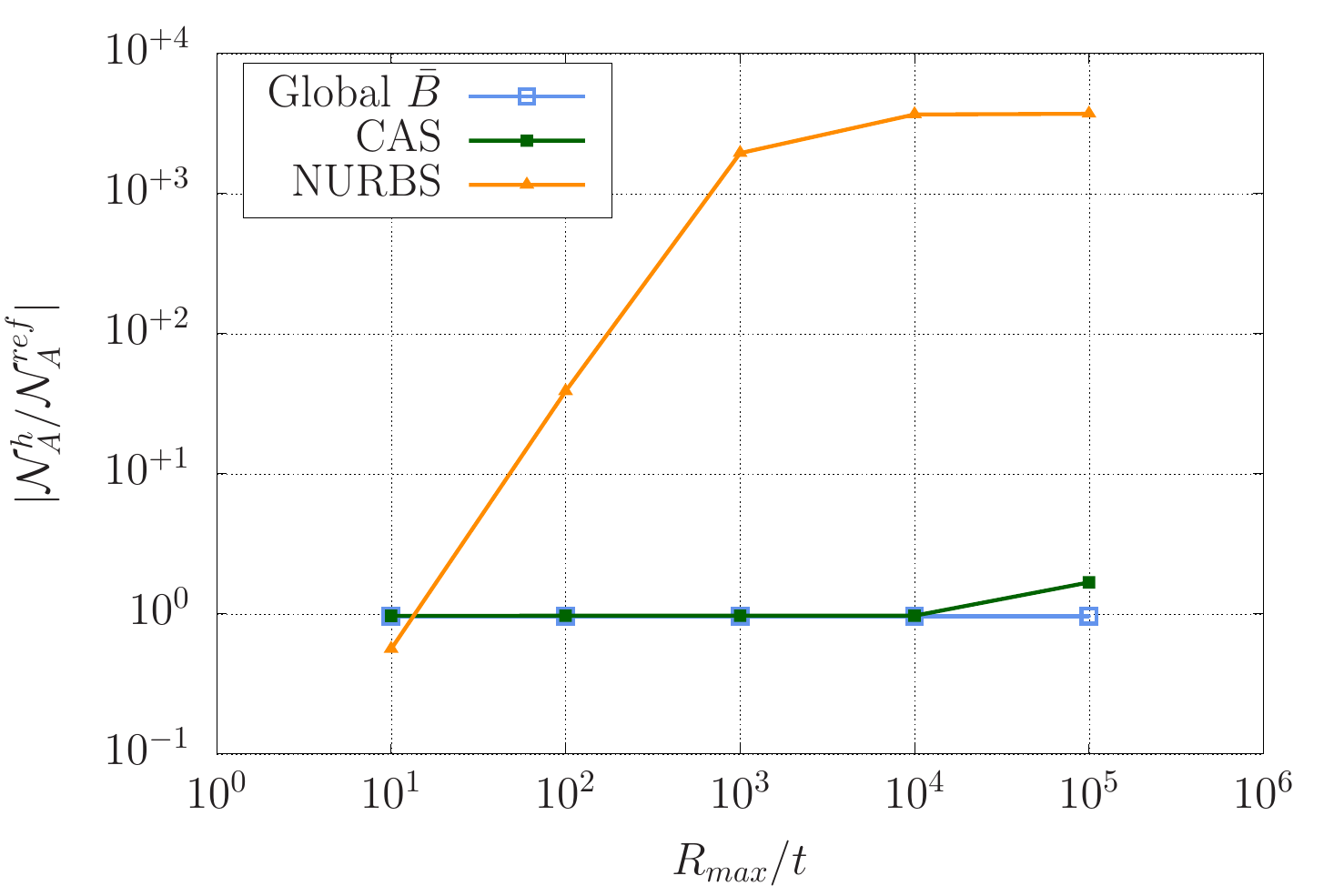}}
 \subfigure[Bending moment]{\includegraphics[scale=0.53]{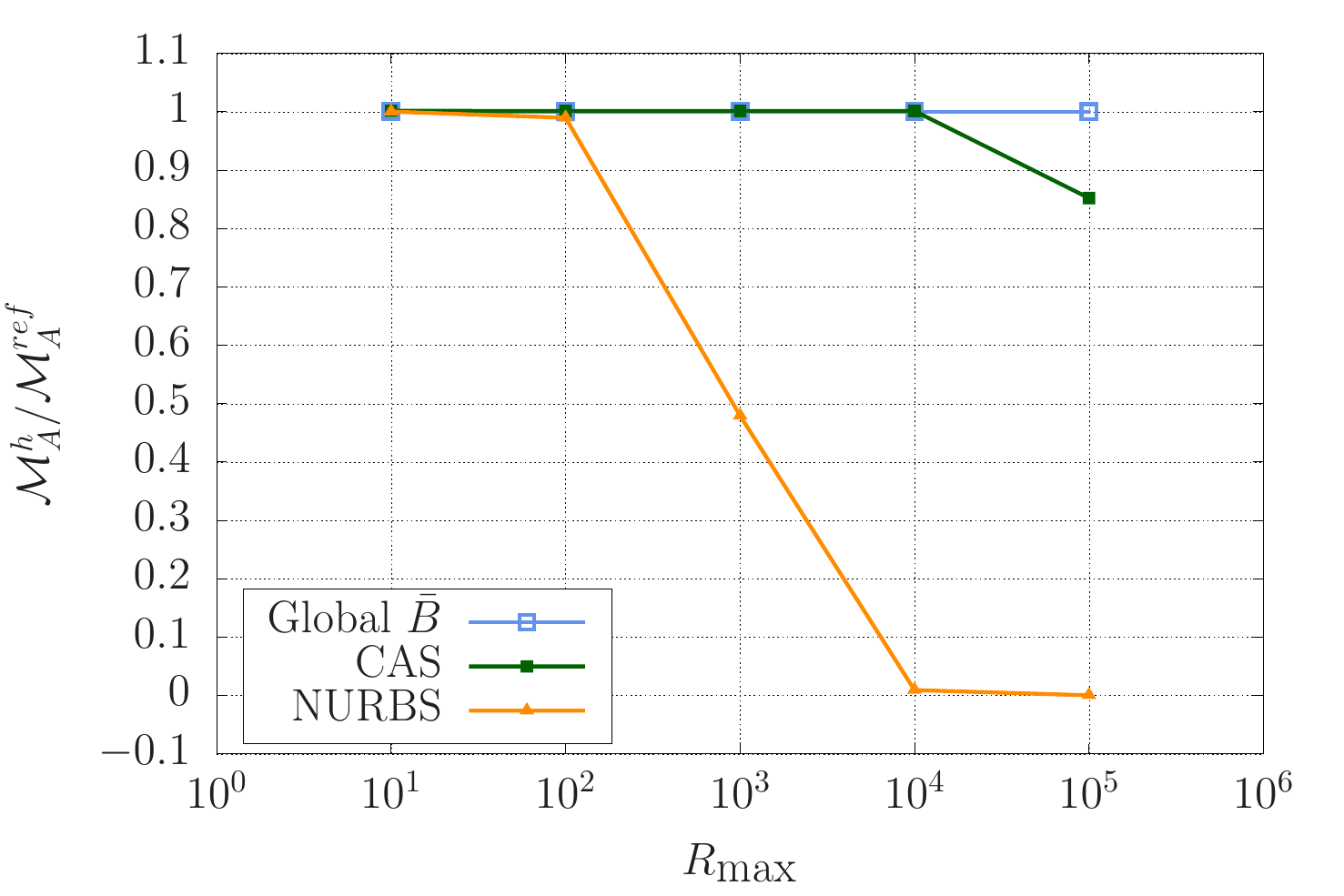}} \\
\caption{(Color online) Clamped elliptical arch under a point load at the free end. Using the global $\bar{B}$ method, NURBS elements, and CAS elements, the horizontal and vertical displacements at the free end and the membrane force and bending moment at the clamped end are plotted for different slenderness ratios. For $R_{\text{max}}/t \le 10^4$, both CAS elements and the global $\bar{B}$ method are accurate, but only the global $\bar{B}$ method is accurate for the extreme slenderness ratio of $10^5$.}
\label{ellipseconvergence}
\end{figure}

	\begin{figure} [h!] 
 \centering
\subfigure[Horizontal displacement]{\includegraphics[scale=0.53]{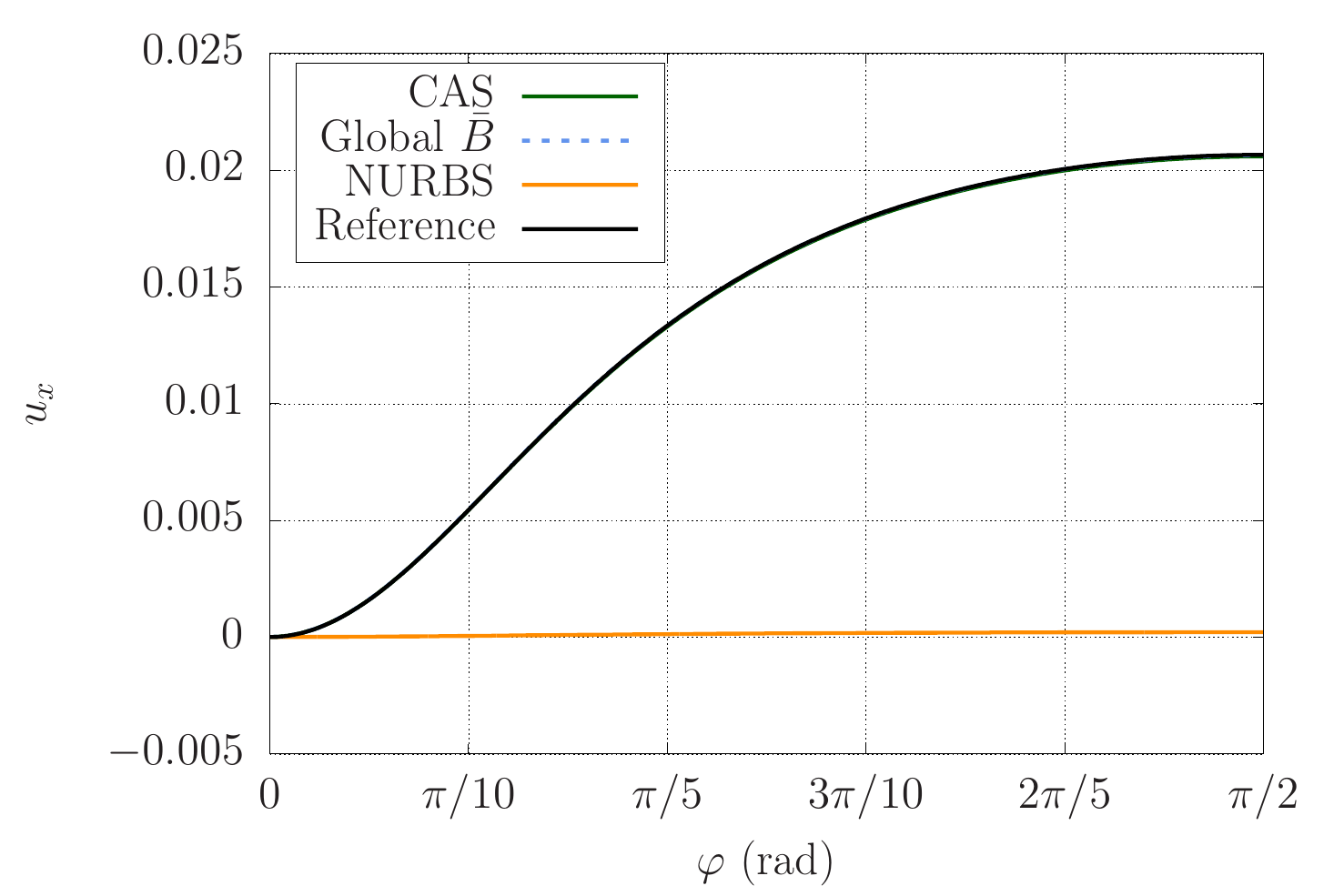}}
 \subfigure[Vertical displacement]{\includegraphics[scale=0.53]{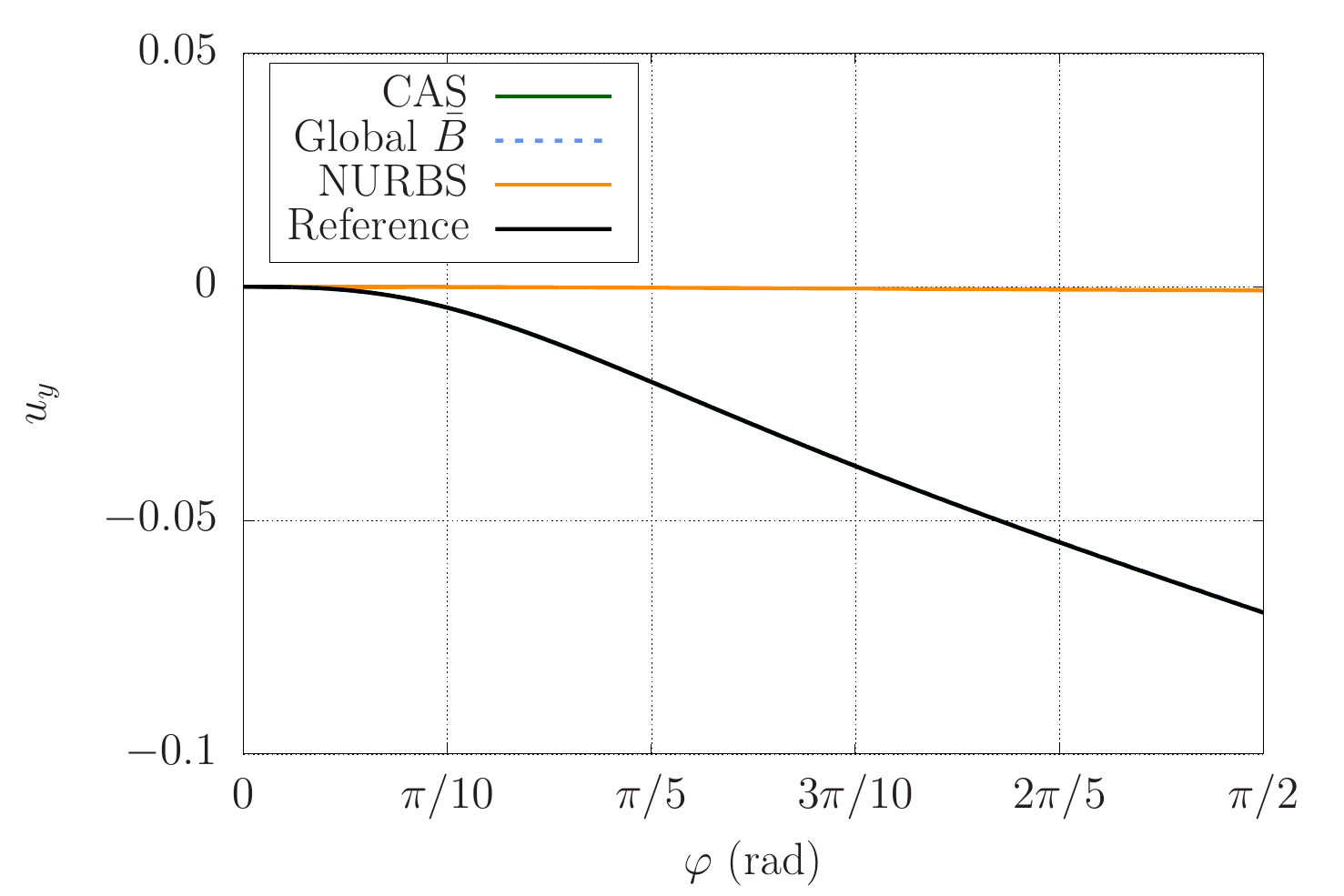}} \\
\caption{(Color online) Displacements of the clamped elliptical arch under a point load at the free end using the global $\bar{B}$ method, NURBS elements, and CAS elements. The mesh has 16 elements and $R_{\text{max}}/t = 10^4$. The numerical solutions using CAS elements and the global $\bar{B}$ method overlap with the exact solution. The numerical solution using NURBS elements locks resulting in zero displacements.}
\label{ellipseuxuy}
\end{figure}

The third numerical investigation  considers a clamped elliptical arch under a point load at the free end. The geometry, the boundary conditions, and the applied load are shown in Fig. \ref{archgeo}. The next values are used in this example
\begin{equation} 
 P =  10^7 t^3,  \quad a = 2.0,  \quad b = 1.0,  \quad E= 7.0 \times 10^{10},  \quad d = 0.1 \text{.} 
\end{equation}
The maximum and minimum radii of curvature are $R_{\text{max}} = a^2/b = 4$ and $R_{\text{min}} = b^2/a = 0.5$, respectively. In order to consider different slenderness ratios, five values are used for the thickness in this example, namely, $t=0.4$, $t=0.04$, $t=0.004$, $t=0.0004$, and $t=0.00004$. Note that for the first two thickness values there are some $R/t$ ratios for which $R/t \geq 20$ is not satisfied \cite{bischoff2004models}, but we include it here to show that not only thin structures can undergo membrane locking, but also thick structures. In the figures of this example, we use $R_{\text{max}}/t$ as slenderness ratio. Since the cross section of the rod is a rectangle, $A = td$ and $I= t^3d/12$.
 
	\begin{figure} [h!] 
 \centering
\subfigure[NURBS]{\includegraphics[scale=0.53]{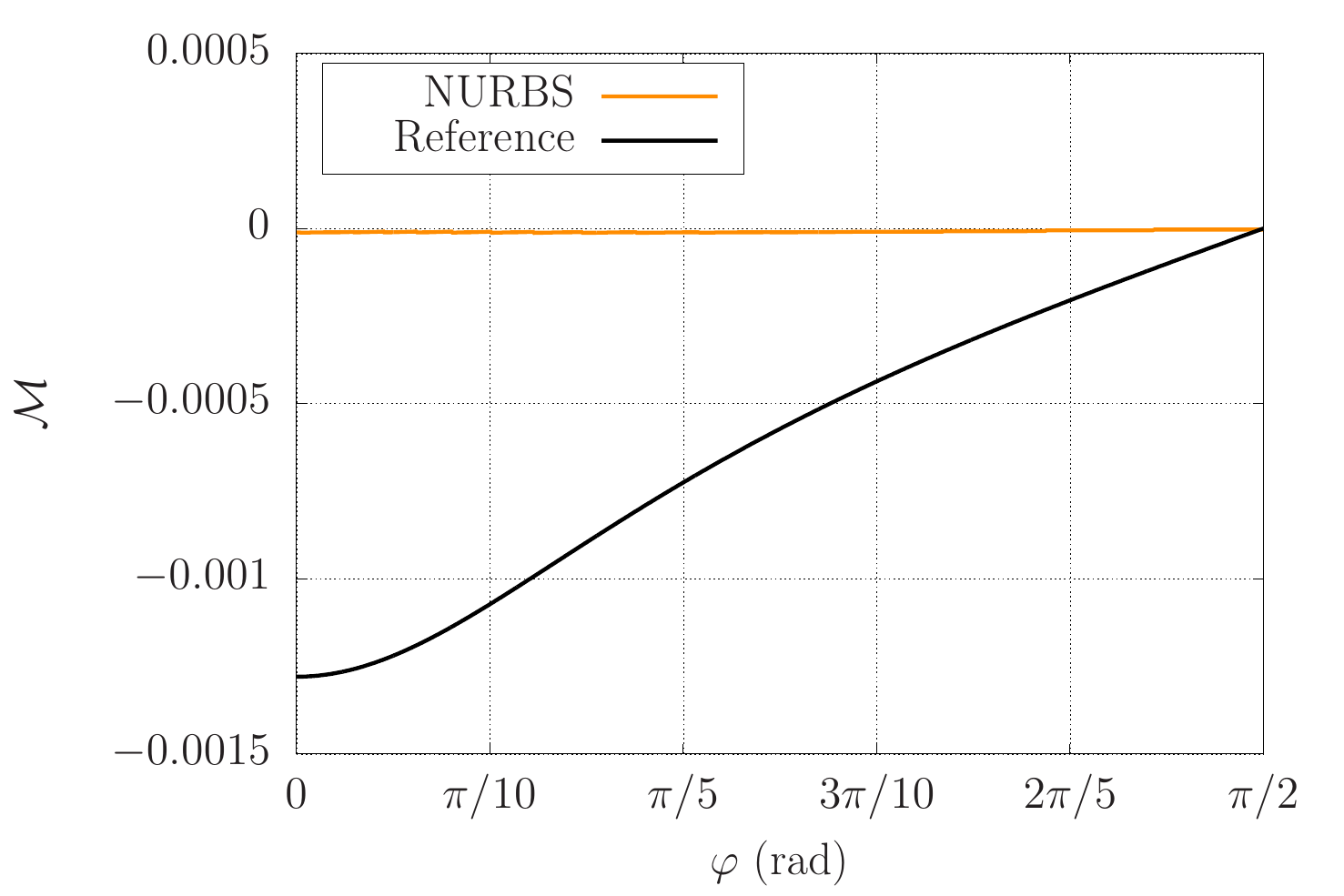}}
 \subfigure[Global $\bar{B}$ and CAS]{\includegraphics[scale=0.53]{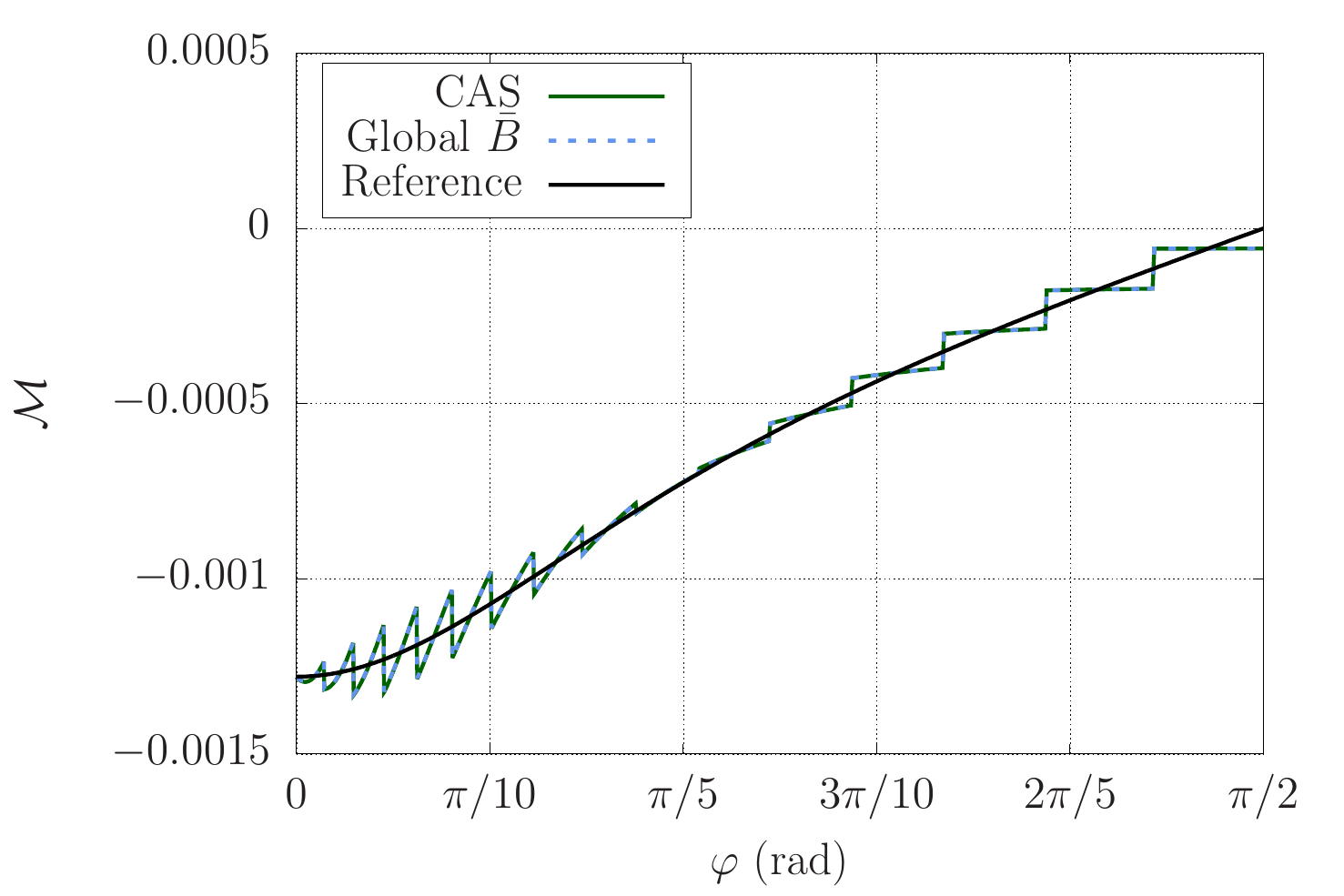}} \\
\caption{(Color online) Bending moment of the clamped elliptical arch under a point load at the free end using the global $\bar{B}$ method, NURBS elements, and CAS elements. The mesh has 16 elements and $R_{\text{max}}/t = 10^4$. The numerical solutions using CAS elements and the global $\bar{B}$ method overlap. The numerical solution using NURBS elements locks resulting in zero bending moment.}
\label{ellipsebm}
\end{figure}

	\begin{figure} [h!] 
 \centering
\subfigure[NURBS]{\includegraphics[scale=0.53]{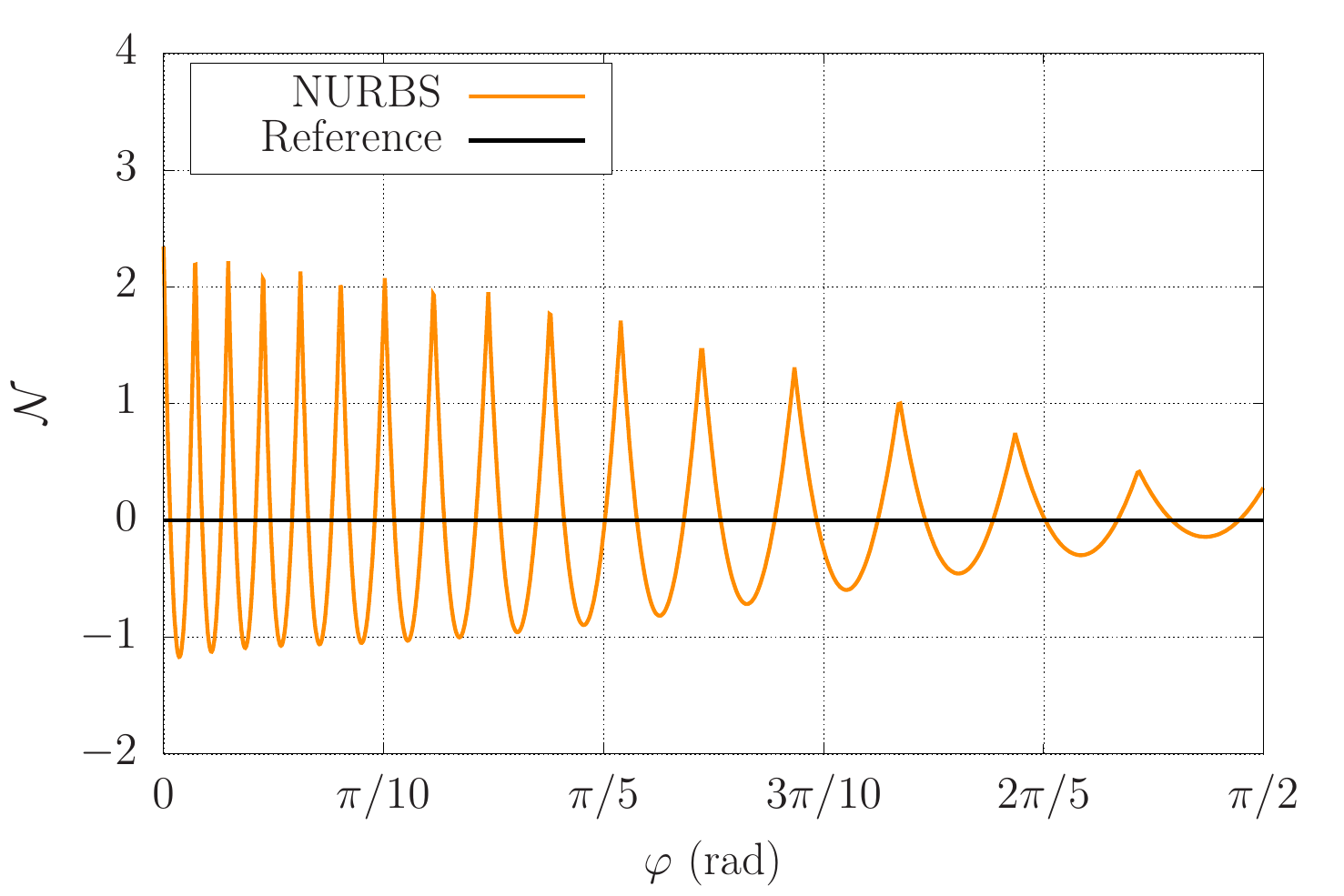}}
 \subfigure[Global $\bar{B}$ and CAS]{\includegraphics[scale=0.53]{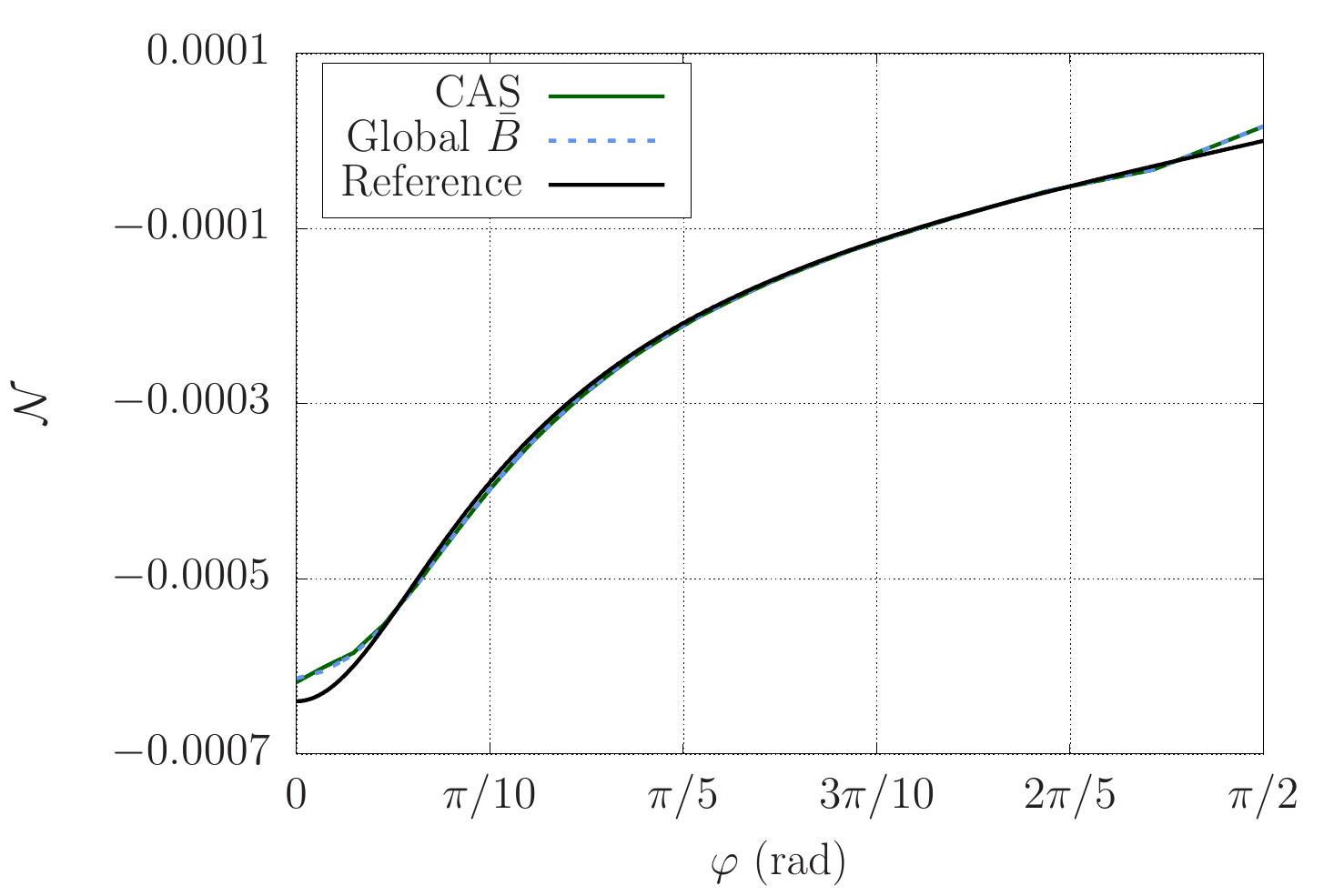}} \\
\caption{(Color online) Membrane force of the clamped elliptical arch under a point load at the free end using the global $\bar{B}$ method, NURBS elements, and CAS elements. The mesh has 16 elements and $R_{\text{max}}/t = 10^4$. The numerical solutions using CAS elements and the global $\bar{B}$ method overlap. The numerical solution using NURBS elements locks resulting in oscillations whose amplitude is more than three orders of magnitude larger than the maximum exact membrane force of this problem. Note the different vertical scale used in each plot.}
\label{ellipsemf}
\end{figure}

In this example, we fix the mesh resolution to 16 elements and investigate the accuracy of the global $\bar{B}$ method, NURBS elements, and CAS elements in obtaining the displacements at the free end and the membrane force and the bending moment at the clamped end. As reference values, we use the values obtained with 256 NURBS elements with degree $p=9$. For the extreme slenderness ratio of $R_{\text{max}}/t = 10^5$, the value of the membrane force at the clamped edge with this reference discretization was not accurate anymore. Thus, we computed this reference value applying static equilibrium instead. As shown in Fig. \ref{ellipseconvergence}, NURBS elements suffer from membrane locking for all the slenderness ratios. For $R_{\text{max}}/t = 10$, $10^2$, $10^3$, and $10^4$, both CAS elements and the global $\bar{B}$ method are locking-free, but only the global $\bar{B}$ method stays locking-free for the extreme slenderness ratio of $R_{\text{max}}/t = 10^5$.

For $R_{\text{max}}/t = 10^4$, the distributions of the horizontal and vertical displacements, the bending moment, and the membrane force are plotted in Figs. \ref{ellipseuxuy}, \ref{ellipsebm}, and \ref{ellipsemf}, respectively, using the global $\bar{B}$ method, NURBS elements, and CAS elements. As shown in Figs. \ref{ellipseuxuy}, \ref{ellipsebm}, and \ref{ellipsemf}, the numerical solution obtained using NURBS elements locks resulting in zero displacements, zero bending moment, and large-amplitude oscillations of the membrane force. In contrast, the numerical solutions obtained using the global $\bar{B}$ method and CAS elements overlap and are locking-free. Note that a small-amplitude zigzag is expected in the bending moment since it is discontinuous across element boundaries. The mean value of the bending moment in any element obtained using either CAS elements or the global $\bar{B}$ method approximates very accurately the mean reference value of the bending moment in that element.

% \subsection{Circular arch clamped at both ends with a distributed load}

\section{Conclusions and future work}

In this work, linear plane curved Kirchhoff rods are used as a model problem to investigate how to effectively remove membrane locking from quadratic NURBS-based discretizations. We develop an assumed natural strain treatment, named continuous-assumed-strain (CAS) elements, that removes membrane locking for an ample range of slenderness ratios by linearly interpolating the membrane strain with $C^0$ inter-element continuity thanks to the $C^1$ inter-element continuity of the displacement vector given by quadratic NURBS. Membrane locking brings about not only smaller displacements and bending moments than expected, but also large-amplitude spurious oscillations of membrane forces. CAS elements eliminate these spurious oscillations while NURBS elements with full and reduced integration, local $\bar{B}$ elements, and local ANS elements undergo large-amplitude spurious oscillations. In addition, the convergence of CAS elements is independent of the slenderness ratio up to $10^4$ while the convergence of NURBS elements with full and reduced integration, local $\bar{B}$ elements, and local ANS elements depends acutely on the slenderness ratio and have errors that can even increase as the mesh is refined. Finally, for a given mesh, CAS elements barely increase the computational cost with respect to the locking-prone NURBS-based discretization of the Galerkin method.

Future research directions include:

\begin{itemize}
\item Treat the shear and membrane locking of Timoshenko rods using CAS elements.
\item Extend CAS elements to the nonlinear regime.
\item Generalize CAS elements to remove locking from shell formulations.
\end{itemize}

\section*{Acknowledgements}

H. Casquero and M. Golestanian were partially supported by the NSF grant CMMI-2138187, Honda Motor Co., and Ansys Inc.

% \section*{References}

%% The Appendices part is started with the command \appendix;
%% appendix sections are then done as normal sections
%% \appendix

%% \section{}
%% \label{}

%% If you have bibdatabase file and want bibtex to generate the
%% bibitems, please use
%%
%%  \bibliographystyle{elsarticle-num} 
%%  \bibliography{<your bibdatabase>}

\bibliographystyle{elsarticle-num} 
\bibliography{./Bibliography}

%% else use the following coding to input the bibitems directly in the
%% TeX file.

\end{document}